\documentclass[floats,floatfix,showpacs,amssymb,prd,twocolumn,superscriptaddress,nolongbibliography,reprint]{revtex4-2}

\usepackage{amssymb,amsmath,verbatim,mathtools,needspace,enumitem,etoolbox,graphicx,physics,microtype,afterpage,bigints,gensymb,tabularx, comment}

\usepackage[dvipsnames, usenames]{xcolor}
\definecolor{linkcolor}{rgb}{0.0,0.3,0.5}
\definecolor{dodgerblue}{HTML}{1E90FF}
\usepackage[unicode, colorlinks=true, linkcolor=linkcolor, citecolor=linkcolor, filecolor=linkcolor,urlcolor=linkcolor, pdfusetitle]{hyperref}
\usepackage[all]{hypcap}
\usepackage[T1]{fontenc}
\usepackage[utf8]{inputenc}
\usepackage{booktabs}
\usepackage{orcidlink}
\usepackage[normalem]{ulem}
\usepackage{caption}
\usepackage{subfig}

\graphicspath{{figures/}}

\def\be{\begin{equation}}
\def\ee{\end{equation}}

\def\oms{\rm OMS}
\def\tm{\rm TM}
\def\aet{$AET$ }

\makeatletter
\newcommand*{\balancecolsandclearpage}{\close@column@grid \cleardoublepage \twocolumngrid}
\makeatother

\newcommand{\rr}[1]{{#1}}

\newcommand{\AEIp}{\affiliation{Max Planck Institute for Gravitational Physics (Albert Einstein Institute), Am M\"{u}hlenberg 1, 14476 Potsdam, Germany}}
\newcommand{\AEIh}{\affiliation{Max Planck Institute for Gravitational Physics (Albert Einstein Institute), Callinstra\ss e 38, 30167 Hannover, Germany}}
\newcommand{\leibniz}{\affiliation{Leibniz Universität Hannover, Institut für Gravitationsphysik, Callinstra\ss e 38, 30167 Hannover, Germany}}

\def\aap{{Astron.~Astrophys.}}

\def\jcap{{J. Cosmology Astropart. Phys.}}

\def\mnras{{Mon.~Not.~R.~Astron.~Soc.}}

\def\prd{{Phys.~Rev.~D}}

\def\pasp{{PASP}}

\begin{document}
\title{A flexible, GPU-accelerated approach for the joint characterization of LISA instrumental noise and Stochastic Gravitational Wave Backgrounds}
\author{Alessandro Santini$\,$\orcidlink{0000-0001-6936-8581}}
\email{alessandro.santini@aei.mpg.de}
\AEIp

\author{Martina Muratore$\,$\orcidlink{0000-0002-9630-5698}}
\AEIp

\author{Jonathan Gair$\,$\orcidlink{0000-0002-1671-3668}}
\AEIp

\author{Olaf Hartwig$\,$\orcidlink{0000-0003-2670-3815}}
\AEIp \AEIh \leibniz

\begin{abstract}

LISA data analysis represents one of the most challenging tasks ahead for %
gravitational-wave (GW) astronomy. 
Characterizing the instrument's noise properties while fitting for all the other detectable sources is a key 
requirement %
of any robust inference pipeline.
Noise estimation will also play a crucial role in searches and parameter estimation of cosmological and astrophysical stochastic signals. Previous studies have tackled this topic by assuming perfect knowledge of the spectral shape of the instrumental noise and of different possible types of GW Stochastic Backgrounds (SGWBs), usually resorting to parametrized templates. 
Recently, various works that %
employ template-agnostic methods have been presented. 
In this work, %
we take an additional step further, 
introducing flexible spectral shapes in both the instrumental noise and the stochastic signals. We account for the lack of knowledge of the exact shape of the individual contributions to the overall power spectral density %
by using splines to represent arbitrary perturbations of the noise and signal spectral densities.
We implement a data-driven Reversible Jump MCMC algorithm to fit different components simultaneously and to infer the level of flexibility required under different scenarios. %
We test this approach on simulated LISA data produced under different assumptions. We investigate the impact of this increased flexibility on the reconstruction of both the injected signal and the noise level, and we discuss the prospects for %
claiming a successful SGWB detection.
\end{abstract}

\maketitle

\section{Introduction}\label{sec: intro}

The Laser Interferometer Space Antenna (LISA)~\cite{2024arXiv240207571C} is a joint ESA-NASA space-based Gravitational-Wave (GW) detector that is expected to be launched in the mid-2030s. Operating in the millihertz region of the frequency spectrum, it will enable the observation of sources in a completely unexplored band, 
complementing ground-based GW detectors~\cite{2019PhRvX...9c1040A, 2021PhRvX..11b1053A, 2021arXiv210801045T, 2021arXiv211103606T} and Pulsar Timing Arrays~\cite{EPTA:2023xxk, NANOGrav:2023hvm, Reardon:2023gzh, Xu:2023wog} observations.
Over the mission duration, LISA is expected to observe a wide range of individually resolvable sources, including massive black-hole binary (MBHB) mergers~\cite{Klein:2015hvg}, Galactic white dwarf binaries (WDBs)~\cite{Korol:2020lpq}, extreme mass ratio inspirals (EMRIs)~\cite{Babak:2017tow}, and stellar-mass black-hole binaries (SBHBs) in their early inspiral phase~\cite{Gerosa:2019dbe, Buscicchio:2021dph, Toubiana:2022vpp}. 

In addition, LISA will be sensitive to stochastic gravitational-wave backgrounds (SGWBs) of astrophysical or cosmological origin. Astrophysical stochastic backgrounds are generated by the incoherent superposition of GW signals from a large number of individual sources. Astrophysical populations that could produce detectable stochastic backgrounds for LISA include Galactic~\cite{Karnesis:2021tsh} and extragalactic~\cite{Farmer:2003pa} WDBs, SBHBs~\cite{Babak:2023lro} and EMRIs~\cite{Barack:2004wc, Bonetti:2020jku, 2023PhRvD.108j3039P}. The WDB background in particular is expected to dominate over the expected %
instrumental noise~\cite{scirdv} between $\sim 0.5$ and $\sim 3$ mHz. SGWBs in the LISA band could also be of cosmological origin~\cite{Caprini:2015tfa}. Possible cosmological sources include First Order Phase Transitions (FOPTs), which are expected in many theories that extend physics beyond the Standard Model~\cite{Caprini:2019egz, Caprini:2019pxz, Caprini:2015zlo}, Cosmic Strings~\cite{2020JCAP...04..034A}, primordial fluctuations in the early universe~\cite{Ricciardone:2016ddg}, and Primordial Black Holes (PBHs)~\cite{Cai:2018dig, Bartolo:2018evs, LISACosmologyWorkingGroup:2022jok}.

The wide variety of sources populating the LISA band represents an exciting and unique opportunity to explore the GW Universe with unprecedented precision. However, it also poses significant data analysis challenges. 
Besides the instrumental noise and stochastic backgrounds,
which will always be present in data we will receive from the instrument, many individually resolvable sources will be long-lived and will remain in band for months or years. 
This scenario will require a global analysis of the data, where the parameters of
all the transient and stochastic components will have to be estimated simultaneously. 
The development of the necessary data analysis pipeline, referred to in the LISA community as the ``global fit'', is currently an active area of research.
Early implementations are described in Refs.~\cite{Littenberg:2023xpl,2024arXiv240504690K, Deng:2025wgk, Strub:2024kbe}.

In this work, we focus on the task of simultaneously characterizing the instrumental noise and possible SGWBs present in the data. Previous studies have tackled this problem by using relatively simple parametrized templates with $2-12$ parameters to describe the spectral shape of the instrumental noise and of the different SGWBs considered (see, e.g., Refs.~\cite{Boileau:2022ter, Adams:2013qma, Hartwig:2023pft}).
\rr{Past works also include a templated analysis of the WDB background anisotropy~\cite{Criswell:2024hfn}.}
Nevertheless, we expect that the instrumental noise entering the LISA data streams will be significantly more complex and hard to fully model analytically.
The LISA-Pathfinder (LPF) mission~\cite{PhysRevLett.116.231101, 2024PhRvD.110d2004A} has already shown
a similar behaviour, with a significant fraction of the total measured noise that could not be explained by on-board measurements of individual noise components.  
As the goal of LPF was to test the behavior of (part of) the key LISA hardware, we can already expect the same to happen for LISA, which is a significantly more complex system than LPF. Therefore, we must be prepared to account for uncertainties in the noise models we will ultimately use in data analysis pipelines, which will come at the cost of affecting the detection and characterisation of SGWBs~\cite{2024PhRvD.109d2001M}. 
Moreover, the different SGWBs described are expected to have different spectral shapes. Fitting them would require the inclusion of models for all of them in our pipelines, which is not only impractical but would also require a complete SGWB template bank. Since our knowledge of the exact spectral shape of the different SGWBs is limited, we cannot compile a complete catalogue of all the possible sources and associated spectral shapes. For this reason, robust and flexible signal reconstruction algorithms are of pivotal importance to successfully extract the information carried by these persistent signals. 

Some previous works have explored possible solutions to the two problems. In Refs.~\cite{2019JCAP...11..017C, Flauger:2020qyi}, the authors propose an automated pipeline based on a binned power-law signal reconstruction, where the frequency band is divided into bins. In each bin, the data is fitted with a power-law signal plus a model of the instrumental noise. The latter is parametrized and common across all the bins. 
Conversely, in Ref.~\cite{2023JCAP...04..066B}, the authors propose a flexible model for the instrumental noise, where the spectral shape is described by a set of splines, maintaining a parametrized description of the SGWB signals. 
Finally, in Ref.~\cite{2024PhRvD.109h3029P}, the authors propose a flexible model for both the instrumental noise and the SGWB signals, where their weakly-parametric models rely on expectation values of Gaussian processes~\cite{10.7551/mitpress/3206.001.0001}. 
In all these works, the authors assume the perfect knowledge of how noise and SGWBs are converted in the time-delay interferometry (TDI) variables~\cite{Tinto:2004wu} used to represent, and analyze, the data. A description of time-delay interferometry can be found in Sec.~\ref{sec:noise_and_tdi}.

With the methodology proposed in this paper, we take an additional step further, lifting the crucial assumption of knowing exactly how the different noise sources propagate in the chosen TDI basis. This is done by introducing spline components directly at the TDI level. 
Our model builds on the work of Ref.~\cite{2024PhRvD.109d2001M}, where the authors include realizations of cubic splines in the noise model to account for possible deviations from the baseline spectral density. 
Those splines had a fixed number of knots placed at fixed frequencies, while their amplitudes were left free to vary. %
There, the goal was to assess the impact of the noise knowledge uncertainties on the measurement precision of the parameters of SGWBs with different shapes. This was done relying on the widely employed Fisher information matrix formalism~\cite{Cutler:1994ys, Vallisneri:2007ev}. Here, we extend the idea to a full Bayesian analysis pipeline, capable of estimating not only the amplitudes of the spline knots but also the required number and location, as well as the parameters of the injected noise model and SGWB signals. 

This work is part of a broader, collaborative effort involving the authors of Refs.~\cite{2023JCAP...04..066B, 2019JCAP...11..017C, Flauger:2020qyi, 2024PhRvD.109h3029P}. There, our goal is to compare the performance of different pipelines when analyzing the same datasets. Those results will be presented separately. 
 
The remainder of this paper explores the following questions:
\begin{itemize}
	\item Can we successfully use flexible, data-driven approaches to model noise and signal components at the same time?
	\item Could a flexible instrumental noise model account for the totality of the data, mistakenly absorbing also stochastic contributions from GW origin? %
	 In this case, what would be our chances of confidently claiming an SGWB detection?
	\end{itemize}
We describe the properties of the instrumental noise and signal considered in this work in Sec.~\ref{sec: data}. We introduce the details of our inference pipeline in Sec.~\ref{sec: method}, while in Sec.~\ref{sec: results}, we present our results. Finally, we draw our conclusions and explore possible future research directions in Sec.~\ref{sec: conclusions}.

\section{Gravitational Wave signals and noise models}\label{sec: data}

In this section, we report the properties of the instrumental noise and SGWBs considered in this work. We then combine them to describe the datasets simulated for the analysis. 
\rr{In particular, we describe in Secs.~\ref{sec:noise_and_filters} and~\ref{sec:sgwb} the properties of the parametrized noise and SGWB templates, respectively, that we use as baselines in our flexible model.}  

\subsection{Time-delay Interferometry}\label{sec:noise_and_tdi}

The LISA constellation will be composed of three identical spacecraft forming a quasi-equilateral triangle. 
The arm length between the three satellites is $2.5$ million kilometres, and they are interconnected by six active laser links. 
Because of the motion of the satellites, the constellation presents unequal and time-varying arm lengths~\cite{2017arXiv170200786A}, which makes the suppression of laser frequency noise in the measurement channels highly non-trivial~\cite{Tinto:1994kg, Armstrong_1999}.
To mitigate this, a post-processing method called time-delay interferometry (TDI) is employed~\cite{Tinto:2004wu}. This technique works by combining raw phasemeter data on the ground and applying suitable time delays to simulate an interferometer with equal arm lengths, thereby effectively suppressing by eight orders of magnitude the laser frequency noise.
TDI combinations can be of different generations. First-generation TDI variables can suppress laser frequency noise for a constellation with static unequal arm lengths~\cite{Shaddock:2004ua}, while second-generation variables can also be used for time-varying arm lengths~\cite{Shaddock:2003dj, Vallisneri:2005ji}.
Other dominant noise contributions that have to be subtracted from the data are the clock and the tilt-to-length (TTL) noises~\cite{Paczkowski:2022nrt, Hartig:2023ofu, Wanner:2024eoa}.

In this work, we assume that the aforementioned noises have already been suppressed~\cite{Hartig_2022, PhysRevApplied.14.014030, PhysRevD.103.123027}. %
We also adopt the idealized configuration of LISA forming a perfect equilateral triangle, which allows us to assume static and equal arm lengths~\cite{PhysRevD.105.062006}. This simplification is valid in the context of analyzing secondary noise contributions, i.e., those not eliminated by the initial noise reduction pipelines~\cite{Muratore:2021rwq, Hartwig:2021dlc, Staab:2023idg, Reinhardt:2025}
—and for modeling the instrument’s response to GWs under the assumptions introduced in this work. We still employ second-generation TDI variables to lay the ground for future, more realistic investigations.
We assume a noise model made of two components: the test mass (TM) acceleration noise and the optical metrology system (OMS) noise~\cite{2023PhRvD.107h2004M}. %

While many different TDI combinations can be (and have been) constructed~\cite{Armstrong_1999, muratore2020revisitation, Hartwig:2021dlc}, the Michelson variables $X$, $Y$, and $Z$ are widely employed in the community~\cite{ldc, radler_doc, sangria_doc, spritz_doc}.
These can be further combined to form the three noise-orthogonal $A,\, E, \, T$ channels~\cite{PhysRevD.105.062006, PhysRevD.66.122002}. In principle, this TDI basis does not present cross-correlations between the different channels, significantly simplifying GW likelihoods and speeding up their evaluation. However, this orthogonality holds only under the assumptions of equal arm lengths, uncorrelated noise contributions, and identical OMS and TM noises in all the links.
This TDI basis is constructed as a linear combination of the Michelson channels:
\begin{equation*}
{A} = \frac{{Z} - {X}}{\sqrt{2}}\;, \quad  {E} = \frac{{X} - 2 {Y} + {Z}}{\sqrt{6}}, \quad T = \frac{ X + Y +Z}{\sqrt{3} }.
\end{equation*}
On the other hand, the second-generation TDI variable $X_2$ is defined as:
\begin{align}
{X_2}  = & (1-D_{12}D_{21} - D_{12}D_{21}D_{13}D_{31}
\\ &
\quad +D_{13}D_{31}D_{12}D_{21}D_{12}D_{21})\times (\eta_{13} + D_{13} \eta_{ \,31}) \nonumber
\\ &- (1-D_{13}D_{31} - D_{13}D_{31}D_{12}D_{21}
\nonumber
\\ &
\quad +D_{12}D_{21}D_{13}D_{31}D_{13}D_{31})\times (\eta_{12} + D_{12} \eta_{ \,21}). \nonumber
\label{eq:tdi2-definition}
\end{align}
\rr{In the above, the quantities $\eta_{ij}(t)$ represent the single-link measurements between spacecrafts~\cite{Hartwig:2021dlc}, describing the %
relative phase difference between the laser beam emitted from spacecraft \( j \) at time \( t - \tau_{ij} \) and the laser local to spacecraft \( i \) at time \( t \). Here, $\tau_{ij} = L_{ij} / c$ represents the light travel time between spacecrafts $j$ (the emitter) and $i$ (the receiver).
$D_{ij}$ is, instead,  an operator that applies a constant delay equal to $\tau_{ij}$, with:
\begin{equation}
	D_{ij} x(t) = x(t - \tau_{ij})\;.
\end{equation} 
}
In the frequency domain, its effect can be expressed as a multiplication by the phase factor $\mathcal{F}(D_{ij}) = e^{-i 2 \pi f \tau_{ij}}$.
In this work, we assume equal and constant arm lengths, so that $L_{ij} = L = 2.5 \times 10^9 \, \rm m$ for all the $i,j$ indices.
\rr{The other two TDI variables, $Y_2$ and $Z_2$, are defined similarly to $X_2$ by cyclic permutations of the three satellites.}

\subsection{Parametrized noise model}\label{sec:noise_and_filters}
Denoting the time series of the TM and OMS noises in the single link $\eta_{ij}$ by $x^g_{ij}(t)$ and $x^m_{ij}(t)$ respectively, the noise in a single-link measurement can be expressed as  
\begin{equation}
\tilde{\eta}^N_{ij}(f) = \tilde{x}^g_{ji}(f)\, e^{-i f \tau_{ij}} + \tilde{x}^g_{ij}(f) + \tilde{x}^m_{ij}(f), \label{eq:link}
\end{equation}
where \( \tilde{\eta}^N_{ij}(f) \) denotes the total noise in $\eta_{ij}(t)$.  
This expression shows that each single-link measurement includes TM noise contributions from both spacecraft involved, leading to correlations between measurements taken at either end of the same arm. Specifically, the TM noise introduces a non-zero cross-correlation between the two corresponding link measurements \cite{2024PhRvD.109d2001M}.

In this work, we assume a noise curve~\cite{ldc, 2021arXiv210801167B} such that the TM acceleration component is described by:
\begin{align}
&\ev{\tilde{x}^g_{ij}(f) \tilde{x}^{g\, *}_{lm}(f')} = \frac{1}{2} \delta_{il}\delta_{jm} \delta(f-f') S_{\text{TM}}(f)\\
 &S_{\text{TM}}(f) = A_{\rm TM}^2 \left[ 1 + \left( \frac{0.4\, \rm{mHz}}{f} \right)^2 \right] \left[1 + \left(\frac{f}{8\, \rm{mHz}}\right)^4 \right] \nonumber \\
  & \qquad \qquad \quad \times \left(\frac{1}{2\pi f c}\right)^2 \! .\nonumber
\end{align}
We set the test mass noise amplitude to $A_{\rm TM} = 2.4 \times 10^{-15}\,\rm m \, s^{-2} \, Hz^{-1/2}$~\cite{ldc}. %
Conversely, the OMS component reads:
\begin{align}
\ev{\tilde{x}^m_{ij}(f) \tilde{x}^{m \, *}_{lm}(f')} &= \frac{1}{2} \delta_{il}\delta_{jm} \delta(f-f') S_{{\rm OMS}}(f)\\
S_{{\rm OMS}}(f)  &= A_{\rm OMS}^2 \left[ 1 + \left( \frac{2 \, \rm mHz}{f} \right)^4 \right] \left(\frac{2\pi f}{c}\right)^2\nonumber 
\label{eq:readout}.
\end{align}
We set the OMS noise amplitude to $A_{\rm OMS} = 7.9 \times 10^{-12} \, \rm m \, Hz^{-1/2}$~\cite{ldc}.
\rr{Asterisks} represent complex conjugates, and we use the notation $\ev{\cdot}$ to indicate expectation values. $S_{\rm TM}$ and $S_{\rm OMS}$ represent the one-sided Power Spectral Densities (PSDs) of the two components. We assume that the PSDs of all six test mass noise terms and, separately, all six OMS noise terms are the same, and the noise components are independent. These assumptions mean that the noises in the $A$, $E$ and $T$ channels will be uncorrelated.

The $A, \, E, \text{and } T$ one-sided power spectral densities are built applying the TDI transfer functions to the noise model:

\begin{subequations}
\label{eq:noise_tdi}
	\begin{align}
		S_{n, AA}(f) &= S_{n, EE}(f)  \\
			&= 32 \sin^2\left( \frac{2 \pi f L}{c} \right) \sin^4\left( \frac{2 \pi f L}{c} \right) \notag \\ 
			&\Bigg\{ \left[2 + \cos \left( \frac{2 \pi f L}{c}\right) \right] S_{\oms}(f) \notag \\
			&+  4\left[ 1 + \cos\left( \frac{2 \pi f L}{c}\right) + \cos^2\left( \frac{2 \pi f L}{c}\right) \right] S_{\tm}(f)  \Bigg\}, \notag
	\end{align}
	\begin{align}
	S_{n, TT}(f) &= 64 \sin^2\left( \frac{2 \pi f L}{c} \right) \sin^4\left( \frac{2 \pi f L}{c} \right)\\
	&\Bigg\{ \left[1 - \cos \left( \frac{2 \pi f L}{c}\right) \right] S_{\oms}(f) \notag \\
	&+ 2\left[1 - \cos \left( \frac{2 \pi f L}{c}\right) \right]^2 S_{\tm}(f) \Bigg\},  \notag
	\end{align}
	\begin{align}
	S_{n, AE}(f) &= S_{n, AT}(f) = S_{n, EA}(f) \\
	&= S_{n, ET}(f) = S_{n, TA}(f) = S_{n, TE}(f) = 0 \notag
	\end{align}
\end{subequations}

In the following sections, we refer to  Eq.~\eqref{eq:noise_tdi} as the template for the noise.

\subsection{Parametrized SGWB model}\label{sec:sgwb}
Stochastic backgrounds in the LISA band can have a large variety of spectral shapes, depending on the physical processes that generate them~\cite{Caprini_2018}.
For the %
purposes of this work, we only consider SGWB \rr{templates} whose energy density can be described by a power-law:
\begin{equation}\label{eq:pl}
h^2 \Omega(f) = \frac{1}{\rho_c} \frac{\text{d} \rho_{\rm GW}(f) }{\text{d} \log f}= A \, \left(\frac{f}{f_{\rm knee}}\right)^n.
\end{equation} %
Here, $\rho_{\rm GW}(f)$ is the GW energy density at frequency $f$, $h$ is the dimensionless Hubble parameter~\cite{2020A&A...641A...6P} $H_0 = h \times 100 \, \mathrm{Km \,s^{-1} \, Mpc^{-1}} \simeq h \times 3.24 \times 10^{-18} \, \mathrm{Hz}$, $\rho_c=3 H_0^2/(8 \pi G h^2)$ is the critical energy density, $f_{\rm knee}$ is a reference frequency and $A$ and $n$ are the amplitude and spectral slope of the background. The values of $A$ and $n$ are determined by the physical processes generating the background~\cite{Babak:2023lro, 2023MNRAS.526.4378L}.
As an example of an SGWB that can be described by Eq.~\eqref{eq:pl} in the LISA frequency band, we consider here the concrete case of the SGWB produced by cosmic strings loops~\cite{Caprini:2015tfa, 2020JCAP...04..034A}.
For the purposes of the present work, we can rewrite Eq.~\eqref{eq:pl} in the cosmic string case as: 
\begin{equation}
		h^2 \Omega_{\rm cs} = 0.55 \times 10^{-11} \left( \frac{G\mu}{10^{-13}} \right).
		\label{eq: cs}
\end{equation}
Here, $G \mu \sim 10^{-6} (\eta / 10^{16} \, \rm GeV)^2$ is the string tension and $\eta$ is a characteristic energy scale. 
For the cosmic string background considered here we assume a flat plateau characterized by $n=0$. 
The overall spectral shape of the background is expected to be more complex, being a function also of the string loop size and the string reconnection probability~\cite{Caprini:2015tfa, 2020JCAP...04..034A, Binetruy:2012ze}.
While the location and amplitude of the flat part of the spectrum is affected by the aforementioned parameters, the simple description given in Eq.~\eqref{eq: cs} is sufficient for the purposes of this work, as we are interested in assessing our pipeline's response to the introduction of flexible models when trying to infer the presence of a SGWB in the data. The analysis of a more comprehensive set of signal models is left for future work.

Finally, we express the SGWB energy density in terms of the one-sided signal power spectral density in the TDI channels as
\begin{equation}
\label{eq: Sh}
\boldsymbol{S_h}(f) = \boldsymbol{\mathcal{R}}(f) \, h^2 \Omega(f) \, \frac{3 (H_0 / h)^2}{4 \pi^2 f^3},
\end{equation}
where 
$\boldsymbol{\mathcal{R}}(f)$ denotes the sky-averaged LISA response function. Here, bold symbols represent matrices in the chosen TDI basis.
Derivations of the LISA response to stochastic signals can be found in Refs.~\cite{2024PhRvD.109h3029P,2023JCAP...04..066B, 2024PhRvD.109d2001M}. 
In this work, we follow the derivation and implementation described in Ref.~\cite{2024PhRvD.109d2001M} and use the associated \texttt{Mathematica} code~\footnote{\url{https://github.com/martinaAEI/noise_knowledge_uncertainty}}.
We first numerically evaluate $\boldsymbol{\mathcal{R}}(f)$ on a grid of frequencies logarithmically spaced in the range $[5\times10^{-6}, \, 0.1] \, \rm Hz$.
We then interpolate these data to evaluate the response function at the frequencies of interest.
The aforementioned code can compute the response function for different LISA constellation configurations (e.g., for both equal and unequal arm lengths).
Therefore, adopting this setup allows our inference pipeline to be easily adapted to different scenarios by switching the grids used to build the interpolants.

From the one-sided PSDs of noise and signal in the chosen TDI basis, it is straightforward to define a signal-to-noise ratio (SNR) of a SGWB as
\begin{equation}\label{eq: snr}
	\text{SNR} = \sqrt{T_{\rm obs} \sum_{i, \,j} \int_{0}^{\infty} \text{d} f \, \frac{S_{h, \, ij}^2}{S_{n, \, ij}^2}},
\end{equation} 
with $T_{\rm obs}$ the LISA observation time~\cite{Thrane:2013oya}. This is not strictly equivalent to the definition of signal-to-noise ratio for deterministic sources, as it does not correspond to the output of a search statistic. However, it is commonly used to compare the detectability of different SGWBs with LISA.%

In the remainder of this paper, we refer to  Eq.~\eqref{eq: Sh} as the template for the signals.

\subsection{Data generation}\label{sec: generation}
We test our pipeline on two different setups to study its performance under different conditions.
In the first case, we generate data consistent with a \rr{parametrized} stochastic background from cosmic strings just outside the reach of SKA~\cite{2015aska.confE..37J},
choosing $G\mu = 10^{-13}$ \rr{in Eq.~\eqref{eq: cs}}.
We inject the signal in a realization of LISA instrumental noise generated according to the \rr{parametrized} LISA PSD introduced in Eq.~\eqref{eq:noise_tdi}.
Conversely, in the second case, we analyze a noise realization that is produced from a PSD whose spectral shape differs from what we expect from the instrument. This is described in Sec.~\ref{sec: perturbed noise}.

For all the analyzed datasets, we generate the noise-orthogonal \aet TDI variables directly in the frequency domain under the assumptions of stationarity and Gaussianity. LISA instrumental noise will neither be stationary nor Gaussian (see, e.g., Ref.~\cite{2020PhRvD.102h4062E}), but a more realistic description of it is out of the scope of this work.
We generate data according to the models described in the previous sections, 
working with second-generation TDI variables and assuming constant, equal arm lengths $L = 2.5 \times 10^9 \, \rm m$. This assumption, despite being highly simplifying, allows us to streamline the description and generation of the simulated datasets and, therefore, to focus this work on presenting the details of our inference methodology.

We generate data assuming an observation time of $T_{\rm obs}=1$ yr and a sampling rate of $2$ seconds, or equivalently a sampling frequency $f_s = 0.5\, \text{Hz}$. We then restrict the analyzed frequency range in the interval $f \in \left[ 10^{-4}, \, 2.9\times 10^{-2} \right] \, \text{Hz}$. Although this narrower range limits our constraining power, we can avoid artifacts due to the zeros introduced in the LISA noise transfer function~\cite{QuangNam:2022gjz}.
For these parameters, the aforementioned SGWB has SNR equal to $262$.

\section{Methodology}

\label{sec: method}
\subsection{Power spectral density breakdown}
In this work, we are interested only in stochastic contributions to the overall LISA data streams.
Following~\cite{2023JCAP...04..066B} and~\cite{2024arXiv240504690K}, for a single data stream we can separate its deterministic and stochastic signal components and express it as
\begin{align}
\label{eq: subtraction}
\tilde{d}(f) &= \sum_{i,j} \tilde{h}_{ij}(f) + \sum_i \tilde{y}_i^{\rm GW} (f) + \tilde{n}(f) \\
			 &= \tilde{H}(f) + \tilde{y}(f), \nonumber
\end{align}
where $\tilde{d}(f)$ represents the frequency domain data, $\tilde{h}_{ij}(f)$ the resolvable sources, $\tilde{y}_i^{\rm GW} (f)$ the stochastic contributions due to gravitational waves and $\tilde{n}(f)$ the instrumental noise. %
We use the index $i$ to refer to the different source types (MBHBs, EMRIs, GWDs, SBHBs) and the index $j$ to label the individually resolved sources belonging to the same type. 
In this work, we do not include any noise artifacts (gaps, glitches), and we assume the perfect subtraction of all the resolvable sources $\tilde{H}(f) = \sum_{i,j} \tilde{h}_{ij}(f)$, so that we can consider only the total stochastic contribution $\tilde{y}(f) = \sum_i \tilde{y}_i^{\rm GW} (f) + \tilde{n}(f)$. \\

We perform our analysis using multiple channels. Thus, Eq.~\eqref{eq: subtraction} applies to each of them. We can write the data vector of the Fourier transformed TDI variables as $\boldsymbol{\tilde{\rm d}} \equiv (\tilde{A}, \, \tilde{E}, \, \tilde{T})^T$ and then, at each frequency, we can define the $3\times3$ periodogram matrix $\boldsymbol{\rm P}(f) \equiv \boldsymbol{\tilde{\rm d}}(f)\boldsymbol{\tilde{\rm d}}(f)^\dagger$. \rr{Daggers represent conjugate transposes.} 
Following~\cite{2023JCAP...04..066B}, we include in the Fourier variables the normalization factor $\sqrt{2 / (N\, f_s)}$, with $N = T_{\rm obs} f_s$ the length of the time series. In this way, the periodogram is directly comparable to the one-sided PSD. 

Since signal and noise are independent processes, the data covariance matrix $\mathbf{C}_d(f) = \left< \boldsymbol{\tilde{\rm d}} \boldsymbol{\tilde{\rm d}}^\dagger \right>(f)$ can be rewritten as the sum of the instrumental noise and SGWBs covariances.
On the TDI level, this reads
\begin{equation}
\mathbf{C}_{AET}(f) = \frac{1}{2}
\begin{bmatrix}
S_{AA} (f) & 0 & 0 \\
0 & S_{EE} (f) & 0 \\
0 & 0 & S_{TT} (f) \\
\end{bmatrix},
\end{equation}
with
\begin{equation}
\label{eq: psd_tot}
S_{ii}(f) = S_{n, ii}(f) + S_{h, ii}(f), \quad i \in \{A,\,E,\,T\}.
\end{equation}
Here $S_{n, ii}(f)$ and $S_{h, ii}(f)$ represent respectively the noise and background contribution to the overall one-sided PSD in each channel. 
We account for the non-perfect knowledge of their spectral shape, \rr{factoring the overall PSD as the product of a parametrized template $S_{n/h}^{\rm template}(f | \boldsymbol{\lambda}_{n/h})$ and a perturbation $\mathcal{P}_{n/h}(f | \boldsymbol{\gamma}_{n/h})$. We perform our inference algorithm on the parameters of both the templates ($\boldsymbol{\lambda}=\{\boldsymbol{\lambda}_n, \, \boldsymbol{\lambda}_h\}$) and the perturbations ($\boldsymbol{\gamma}=\{\boldsymbol{\gamma}_n, \, \boldsymbol{\gamma}_h\}$). We represent the entire parameter space with $\boldsymbol{\theta}$: $ \boldsymbol{\theta} =\{\boldsymbol{\theta}_n, \, \boldsymbol{\theta}_h\} = \{ \boldsymbol{\lambda}, \, \boldsymbol{\gamma}\} = \{ \boldsymbol{\lambda}_n, \, \boldsymbol{\lambda}_h, \, \boldsymbol{\gamma}_{n, ii}, \, \boldsymbol{\gamma}_h\}$. Therefore, each component reads:
}
\begin{subequations}
\label{eq: splines}
\begin{align}
S_{n, ii}(f | \boldsymbol{\theta}_n) &= S_{n, ii}^{\rm template}(f | \boldsymbol{\lambda}_n) \cdot 10^{\mathcal{P}_{n, ii}(f|\boldsymbol{\gamma}_{n, ii})}, \label{eq: noise_pert}\\
S_{h, ii}(f | \boldsymbol{\theta}_h) &= S_{h, ii}^{\rm template}(f | \boldsymbol{\lambda}_h) \cdot 10^{\mathcal{P}_{h}(f|\boldsymbol{\gamma}_h)}. \label{eq: sum_b}
\end{align}
\end{subequations}
This factorization results in two correlated terms, but we choose it to allow the templates to catch the overall frequency evolution of the PSD and let eventual deviations from it be absorbed into $\mathcal{P}_{n/h}(f)$. 
In this work, we assume three independent perturbations for the noise contributions across the three TDI channels, whereas the background spectral shape is perturbed in a consistent (i.e., shared) way.
Hence, the presence of the indices ${ii}$ only in the exponent of Eq.~\eqref{eq: noise_pert}. This assumption, which translates into using four different perturbations,
is based on the hypothesis that the transfer function of the signal will be well-known.
While the same holds for any noise source entering the measurement chain, 
we cannot know a-priori the full set of noises that will be present in the data stream. In addition, we will not have access to all the physical parameters that constitute the noises we expect to be there 
(see, e.g., Refs.~\cite{2024arXiv240207571C} and~\cite{2023PhRvD.108h2004Q}). 
Thus, we decided to be fully generic only in the noise contribution to the overall PSD in each channel.
This approach has a caveat: the \aet basis is noise-orthogonal only under the assumption of a perfect equilateral constellation where all the secondary noise sources are exactly equal in the three spacecrafts~\cite{2023PhRvD.107l3531H},
thus using independent splines in the three channels may seem a violation of this assumption. For a complete analysis, we should also include perturbations in the off-diagonal components of the noise matrix, which would increase the number of perturbations for the noise contributions by another three terms. The off-diagonal terms would also account for cross-correlations between the noise and SGWB components at high frequencies~\cite{Hartwig:2023pft, 2024PhRvD.109d2001M}.
Cross correlations can be as large as $\sim10\,\%$ of the diagonal terms~\cite{Hartwig:2023pft}. These can have an impact on the distinguishability of instrumental noise and SGWBs, but from the results of Ref.~\cite{2024PhRvD.109d2001M} we do not expect our conclusions to be heavily affected. 
For completeness, this should be validated against a full analysis in the Michelson TDI basis, where the (hermitian \rr{, positive-definite}) covariance matrix $\mathbf{C}_{XYZ}(f)$ is dense with $6$ independent components, \rr{since it is possible to construct a positive definite matrix whose elements are functions of $6$ independent, flexible terms~\cite{2024PhRvD.109d2001M}.} This requests the evaluation of $6$ splines at every likelihood call only for the noise contribution, significantly increasing the computational cost. 
Here, we neglect off-diagonal terms for computational simplicity.
Having generated the frequency domain data directly in the \aet basis, we know the noise-orthogonal assumption holds. 
We leave a full covariance treatment of the three noise channels for future analysis. 

We use as perturbations $\mathcal{P}_{n/h}(f)$ realizations of Akima splines~\cite{10.1145/321607.321609}, 
\rr{which although only being $C^1$ differentiable, were introduced to produce smoother fits to data in the presence of rapidly varying second derivatives with respect to other widely used $C^2$ Cubic splines implementations~\cite{2020SciPy-NMeth, Katz:2021yft, Chapman-Bird:2025xtd}. A description of the Akima splines algorithm can be found in App.~\ref{sec: akima_impl}.}
To be as generic as possible, we do not fix the number or the position of the spline knots beforehand. Instead, 
we use a Reversible Jump Markov Chain Montecarlo (RJ-MCMC) scheme~\cite{10.1093/biomet/82.4.711} to dynamically infer the most probable number of knots. 
We use the parallel tempered ensemble sampler \texttt{Eryn}~\cite{2013PASP..125..306F, 2023MNRAS.526.4814K, michael_katz_2023_7705496} for this.
Our knot placement scheme can be described as follows:
\begin{itemize}
	\item We fix the frequency position of the first and last spline knots to the extrema of the frequency range considered. We label these two points as edges, and we only fit for their amplitudes. This is crucial to always make the splines cover the whole frequency range.
	\item For the in-between knots, we sample both the position $f_k$ and the amplitude $w_k$. The sampler proposes a change in their number at each step. Following \texttt{Eryn}'s jargon, this is done through a \emph{between-models} move, proposing the addition of a new knot to the active ones (i.e., the ones currently used to evaluate the splines) or removing one. When the addition of a new knot is proposed, its parameters are drawn from the prior. We refer the reader to Refs.~\cite{2023JCAP...04..066B, 2023MNRAS.526.4814K, 2024arXiv240504690K, muratore2025pipelinesearchingfittinginstrumental} for additional details about the \texttt{Eryn} reversible jump implementation and its recent applications in LISA data analysis.
\end{itemize}

We set the minimum and maximum number of in-between knots to $0$ and $10$, respectively. %
Since our Akima splines implementation uses boundary conditions that require at least $4$ points to be computed,
we rely %
on a linear interpolant for cases with less than two in-between knots. 

With the nomenclature introduced in this section and the templates presented in Eqs.~\eqref{eq:noise_tdi} and~\eqref{eq: Sh},
we can summarize the parameters of our model as follows:
\begin{itemize}
	\item $\boldsymbol{\lambda}_n = \{A_{\rm TM}, \, A_{\rm OMS}\}$: the two noise amplitudes;
	\item $\boldsymbol{\lambda}_h = \{A, \, n\}$: the amplitude and the spectral index of the signal power law;
	\item $\boldsymbol{\gamma}_{\alpha} = \{w_0^\alpha, \, \{f_k, \, w_k \}_{k=1,..., n-1}^\alpha,\, w_n^\alpha\}$: the amplitudes at the edges,
	and the position
	and amplitude of each in-between knot added through reversible jump, for each independent spline component $\alpha$.
\end{itemize}

Given the properties presented in this section, it is clear that our inference model includes components without a ``physical'' origin (the splines) and does not leverage the different noise and signal TDI transfer functions to discriminate between the two contributions. 
Although these may sound like disadvantages, we believe they represent the core features of our work.
In fact, the generality we introduced enables analyses in which the totality of the stochastic contributions are fitted as a whole in a global analysis that also includes deterministic sources. 
The total stochastic power found in this way could be broken down into individual contributions in a post-processing step.
We leave the assessment of the feasibility of this approach to future investigations.

\subsection{Hardware acceleration}\label{sec: gpu}
In our setup, we must evaluate many potentially different splines at each MCMC step. All the walkers of our ensemble can have different numbers of spline knots with different positions and amplitudes $\left(f_k, \, w_k \right)$. Therefore, we implement a version of Akima splines~\cite{cudakima_2024_13919394}, which leverages parallelization and GPU (Graphics Processing Units) acceleration to speed up their evaluation substantially. While the \texttt{Cupy} library~\cite{cupy_learningsys2017} already offers a GPU implementation of Akima splines, this is mostly suitable for scenarios where the number and the position of the interpolation knots are fixed. We tailor our code to handle the variability caused by the reversible jump setup.  

Additionally, we employ the \texttt{JAX} library~\cite{jax2018github} extensively in our codebase, leveraging its built-in GPU support, vectorization, and just-in-time (JIT) compilation.

\subsection{MCMC setup}\label{sec: mcmc}
We run \texttt{Eryn} using 20 different temperatures with 20 walkers each. We initialise our walkers, drawing their position in the parameter space randomly from the prior. The only exception is for the spline knots, whose initial amplitudes are drawn from a multivariate normal centred on zero $\mathcal{N}(\boldsymbol{\mu}=\mathbf{0}, \, \boldsymbol{\Sigma}=\boldsymbol{\rm I} \times 10^{-8})$. This choice follows our interpretation of the splines as ``small'' perturbations of parametrised templates.
For each of our parameter estimation (PE) runs, we monitor the convergence of the samples by making sure that %
the chains are longer than $200 \times \max(\hat{\tau})$, with $\hat{\tau}$ being the average autocorrelation time across all the model parameters~\cite{2010CAMCS...5...65G}. %
We discard (at least) the first $5 \times \max(\hat{\tau})$ samples to account for the burn-in phase,
and we thin the chains by a factor of $0.5 \times \min(\hat{\tau})$.

The large number of parallel temperatures helps to increase the acceptance rate of our runs,
and allows us to estimate the evidence $Z$ through the stepping-stone method~\cite{10.1093/sysbio/syq085}. In the standard Bayesian framework, the evidence, or marginal likelihood, is a normalization constant equal to the integral over the parameter space of the unnormalized posterior distribution:%
\be\label{eq: bayes}
p (\boldsymbol{\theta} | \, d) = \frac{\mathcal{L} (d \, | \, \boldsymbol{\theta})\, \pi(\boldsymbol{\theta})}{Z},%
\quad%
Z = \int \text{d} \boldsymbol{\theta} \, \mathcal{L} (d \, | \, \boldsymbol{\theta})\, \pi(\boldsymbol{\theta}),
\ee
where $d$ represents the observed data, $\boldsymbol{\theta}$ the model parameters, $\pi(\boldsymbol{\theta})$ their prior and $\mathcal{L}$ the assumed likelihood. We report our prior distributions in Tab.~\ref{tab:priors}.
The evidence is usually expensive to compute, and being a normalization constant, it can safely be disregarded in parameter estimation studies. %
However, it is required to perform Bayesian model selection. The ratio of the evidence computed for two competing models, $M_1$ and $M_2$, is called the Bayes factor, $\mathcal{B}_{12} = Z_1 / Z_2$. A large Bayes factor indicates that model $M_1$ describes the observed data better than $M_2$. %
Examples of the use of Bayes factor's for model selection in the field of GW astronomy can be found in Refs.~\cite{2011PhRvD..83h2002D, 2021PhRvD.104h3027T, 2024PhRvD.109j4019T}, \rr{while applications of the stepping-stone method in this field can be found in Refs.~\cite{Maturana-Russel:2018yos, 2025MNRAS.540.3818Z}}.
Following~\cite{10.1093/sysbio/syq085}, the stepping-stone approximation for the evidence can be expressed as:
\begin{align}
\label{eq: ss}
\log{\hat{Z}} =& \sum_{k=1}^{K} \log \sum_{i=1}^{n} \mathcal{L}(d \, | \, \boldsymbol{\theta}^i_{\beta_{ \,k-1}}, \, M)^{\beta_k - \beta_{ \,k-1}} \\
			& - K \, \log n, \notag
\end{align}
where $K$ is the number of temperature $T_k$ in the ladder, $\beta_k = 1 / T_k$ are the inverse temperatures, $M$ is the model the evidence is computed for, and $\boldsymbol{\theta}^i_{\beta_{ \,k-1}}$ are $n$ draws from the annealed posterior at temperature $T_{k-1}$.%
For low $\beta$ values, the likelihood is suppressed by the tempering. When $\beta$ tends to $1$, the tempered posterior approaches the target distribution. To cover these behaviours and to resolve the turnover point between the two, we place 5 chains with $\beta$ logarithmically equispaced between $10^{-20}$ and $10^{-12}$, 10 between $10^{-12}$ and $10^{-5}$, and 5 between $10^{-5}$ and $1$.

To further reduce the computational cost of our runs, 
\rr{we do not sample using the totality of the frequency domain data and the Whittle likelihood~\cite{faaafff7-3364-33b0-82ea-e107a78593d3}.
This represents the standard in the context of GW astronomy, and it has already been widely implemented in the literature (see, e.g., Ref.~\cite{2024PhRvD.109h3029P}, and Ref.~\cite{Franciolini:2025leq} for a more comprehensive SGWB likelihoods review).}
Here, we average the periodograms over $N_{\rm bins}$ consecutive and non-overlapping frequency bins, and we sample using a complex Wishart likelihood, as described in Ref.~\cite{2023JCAP...04..066B}.
We define the averaged periodogram in a segment $i$ as
\begin{equation}
\boldsymbol{\rm P}(f_i^{\rm bin}) \equiv \frac{1}{n_i} \sum_{k = i - \frac{n_i}{2}}^{i + \frac{n_i}{2}}\boldsymbol{\tilde{\rm d}}(f_k)\boldsymbol{\tilde{\rm d}}(f_k)^\dagger,
\end{equation}
where $n_i$ is the number of frequency samples in a bin centred on $f_i^{\rm bin}$. Introducing the number of Degrees of Freedom (DoFs) in each bin $\nu(f_i^{\rm bin}) = n_i$, the matrix $\boldsymbol{\rm Y}(f) = \nu(f) \boldsymbol{\rm P}(f)$ would follow the aforementioned complex Wishart distribution \rr{if $\boldsymbol{\tilde{\rm d}}(f_k)$ were uncorrelated between frequency bins}. When working with time domain data, we apply a window function to the timeseries before converting it into the frequency domain. This would introduce correlations between frequency bins, effectively reducing the number of DoFs $\nu(f)$. Since we generate our data directly in the frequency domain, this step is not necessary.
Keeping only the terms that depend on the model parameters, for each bin we can write:
\begin{equation}
	\log p(\boldsymbol{\rm Y}(f) | \, \boldsymbol{\theta}) = -\text{tr}\left( \boldsymbol{\rm{C}}_d^{-1} \boldsymbol{\rm {Y}}(f)  \right) - \nu(f) \log |\boldsymbol{\rm{C}}_d(f)|,
\end{equation} 
where $\rm{tr}(\cdot)$ denotes the trace operator, and $|\cdot|$ represents the determinant of a matrix.
Combining all the bins together, our likelihood reads out:
\begin{equation}
	\log \mathcal{L}(\boldsymbol{\rm Y} | \, \boldsymbol{\theta}) = \sum_{i=0}^{N_{\rm bins} - 1} \log p(\boldsymbol{\rm Y}(f_i^{\rm bin}) | \, \boldsymbol{\theta}).
\end{equation}
\rr{In the limit of $n_i = 1$ for all $i$, having only one frequency sample per bin, the complex Wishart likelihood converges to the Whittle likelihood~\cite{2023JCAP...04..066B}.}
For completeness, we include in our code the possibility of using directly the latter with the totality of the frequency domain data.
We have checked that the two methods yield consistent results.

\begin{table}
\caption{Prior distributions used for the MCMC runs presented in this work. $A$ and $n$ represent the amplitude and slope of the power-law in Eq.~\eqref{eq:pl}.
We use uniform distributions for most of the sampling parameters.
The prior distribution assumed for the weight of each spline knot and edge is a normal distribution centered on $\mu=0$. %
In each channel and for each contribution, we assume the same priors on $f_k$ and $w_k$. 
}\label{tab:priors}
\centering
\begin{tabular}{ccc}
\toprule
Parameter & \qquad & Prior distribution \\	
\midrule
$A_{\rm TM} / 10^{-15} \, [\rm m \, s^{-2} \, Hz^{-1/2}]$ & \qquad & $U[10^{-2},\, 10^2]$ \\
$A_{\rm OMS} / 10^{-12} \, [\rm m \, Hz^{-1/2}]$ & \qquad & $U[10^{-2},\, 10^2]$ \\
$\log_{10}A$ & \qquad & $U[-20, \, -10]$ \\
$n$ & \qquad & $U[-5, \, 5]$ \\
$\log_{10} f_k$ & \qquad & $U[\log_{10} f_{\min}, \, \log_{10} f_{\max}]$ \\
$w_k$ & \qquad & $\mathcal{N}(\mu=0, \, \sigma=0.1)$ \\
\bottomrule
\end{tabular}
\end{table}

\subsection{Two-step inference}\label{sec: two_steps}
Our last acceleration step consists of splitting the inference into two consecutive steps. There is a degeneracy in the model in that changes to the model template parameters can be mimicked by changes to the spline parameters. We address this degeneracy by first running an initial MCMC with only the modelled noise parameters and the background templates.
This corresponds to enforcing all the spline knots to have $w_k=0$ in Eq.~\eqref{eq: splines}. 
The resulting posterior distributions are used to produce the results labelled as ``template recovery'' in Figs.~\ref{fig: cs_psd} and~\ref{fig: noiseperturbed_psd},
and to fix the template parameters in a second MCMC run.
In the second run, we let the splines free to explore the parameter space, while the template parameters $\boldsymbol{\lambda}$ are fixed to the medians of the aforementioned posteriors.
If compared to running one single MCMC where all the parameters $\boldsymbol{\theta}$ are updated together, this separation dramatically improves the convergence of the chains, 
as we remove the degeneracy between the template parameters and the splines. 
We can do this because in our model the splines are not used to describe the overall spectral shape of the PSD, but only to account for (small) deviations from the templates. We can interpret the recovered template model parameters as the best fit model, and the spline parameters as deviations from that best fit. This applies to both the noise and the signal analyzed. 
In general, if the spline model was used to describe the totality of the stochastic contribution as one individual component, the priors would be increased to be as large as needed. Instead, 
we interpret the splines as deviations from the best-fit model; therefore, we use tight zero-mean Gaussians as priors on $w_k$.

We have checked that when fixing the template parameters to different values sampled from the posterior inferred in step 1, the splines parameters change accordingly, resulting in consistent recovered PSDs. %

\rr{This two-steps approach, combined with the hardware acceleration and averaging procedure described in Secs.~\ref{sec: gpu} and~\ref{sec: mcmc}, allows our runs to converge and accumulate enough samples within a day on an NVIDIA A100-SXM4 GPU despite the large number of parallel walkers and the dimensionality of the parameter space.}

\section{Results}\label{sec: results}
In this section, we present the results of our pipeline on the different datasets described in Sec.~\ref{sec: data}. 
We compare the results obtained using only the templates with those obtained using splines to account for deviations from them.
The analysis in Sec.~\ref{sec: perturbed noise} of a datastream generated from a different PSD with respect to the template used in the recovery
will allow us to test the robustness of our pipeline in scenarios where the analyzed data do not match the baseline parametrized templates.
We leave the analysis of SGWBs whose spectral shape differs from our baseline power law template for future work.

\subsection{Cosmic strings}
\begin{figure*}[ht]
\centering
\subfloat[][\emph{Templates recovery}\label{fig: cs_templates}]
{\includegraphics[width=0.45\textwidth]{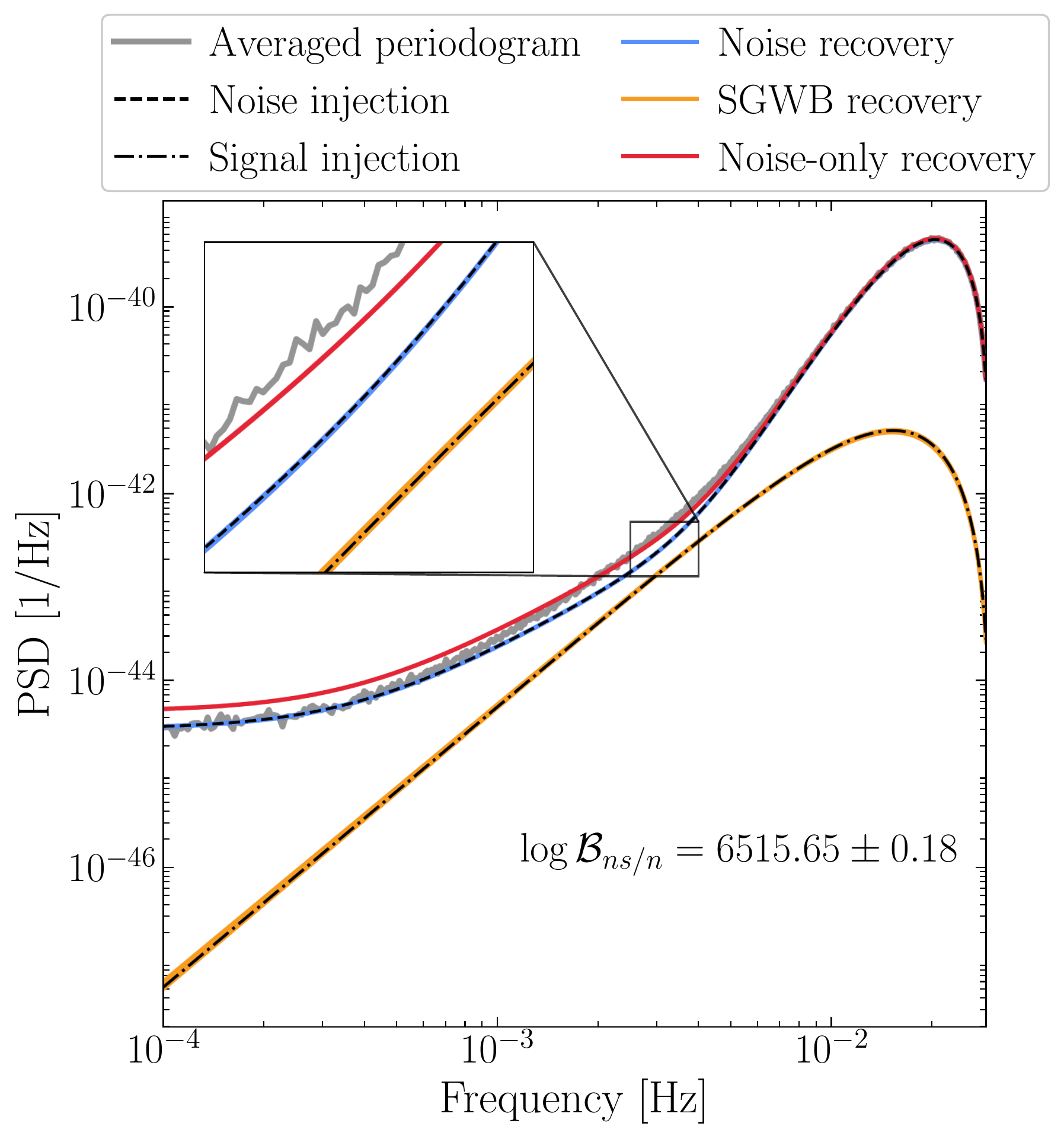}}\quad
\subfloat[][\emph{Splines recovery}\label{fig: cs_splines}]
{\includegraphics[width=0.45\textwidth]{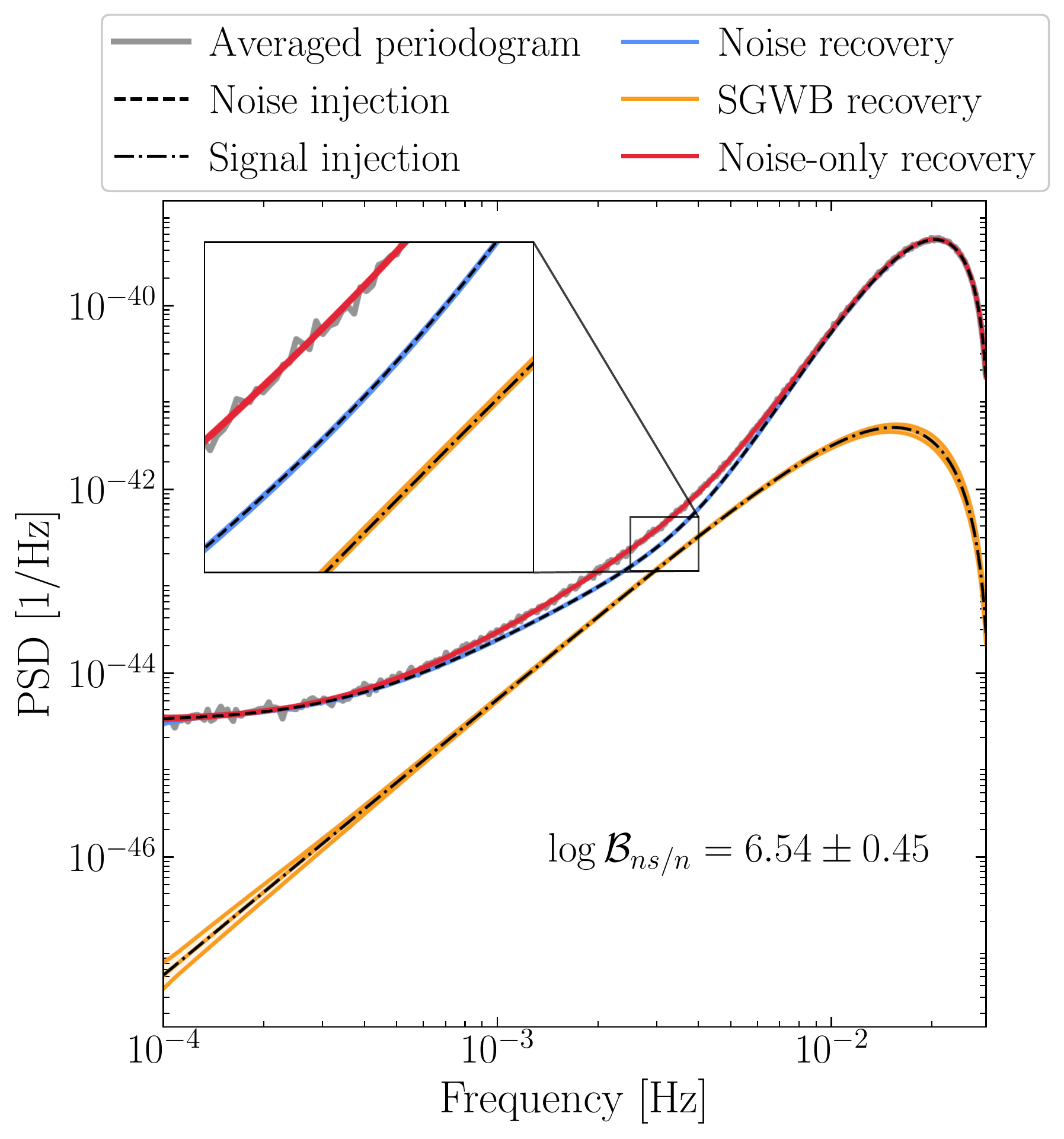}}
\caption{%
Reconstructed posterior distribution of the different contributions to the total PSD in the $A$ channel. 
The black dashed (dash-dotted) line represents the PSD used to generate the noise (SGWB from cosmic strings) dataset.
The averaged data periodogram is shown as a grey line.
Blue (orange) lines and filled contours represent medians and $90\, \%$ credible regions for the reconstructed noise (signal) PSD when both contributions are present in the model. 
Red lines and contours refer to the case where only the noise has been used to fit the totality of the data.
Insets show a zoom-in on the region $f \in [2.5, \, 4]\, \rm mHz$.
\emph{Left panel:} Result of the analysis performed using only the baseline templates
\emph{Right panel:} Result of the analysis performed introducing splines in the model. 
}\label{fig: cs_psd}
\end{figure*}
First, we present results for the cosmic string-like SGWB.
Our reconstructed posterior distributions on the PSDs used for the data generation are shown in Fig.~\ref{fig: cs_psd}, together with the respective injections. As anticipated in Sec.~\ref{sec: two_steps}, we report in the two panels the results obtained with only the templates (left) and with the inclusion of the splines (right).
In each panel, we report the results obtained under two separate hypotheses: first, we include both the noise and signal components in the model fitted to the data, %
shown as the blue and orange curves in the plot. 
Next, we perform the inference using only the noise component, shown in red.
 The two models read respectively like the totality and the first term of Eq.~\eqref{eq: psd_tot}.

While this is not the case for the analysis performed only with templates,
when we introduce splines in the model, both approaches can successfully account for the totality of the data.
Figure~\ref{fig: cs_templates} clearly shows that the red curve,
representing the recovery performed using only the noise template in Eq.~\eqref{eq:noise_tdi},
fails to model the totality of the data.
This is particularly evident at low frequencies, where the reconstruction significantly overestimates the data content.
The main driver for this is the inclusion of the $T$ channel in the analysis. $T$ is mostly insensitive to GWs (especially at low frequency)~\cite{Muratore:2021uqj}, so it can, in principle, be used as a noise monitor.
This holds especially in the idealized equal arm length configuration assumed here.
Its noise PSD is dominated by the OMS component, which controls the $A$ and $E$ PSD only at high frequency. Since $T$ is largely unaffected by the presence of a signal, when the latter is present in the data but not included in the model, $A_{\rm OMS}$ is better constrained than $A_{\rm TM}$. While the median of the $A_{\rm TM}$ posterior is more than $167$ standard deviations ($\sigma$) away from the corresponding injected value, the bias for $A_{\rm OMS}$ is of ``only'' $8\sigma$. The poor estimate of $A_{\rm TM}$ dominates the error at low frequency.

The situation is different for the splines analysis in Fig.~\ref{fig: cs_splines}. 
Under the first hypothesis, the reconstructed $90\%$ credible regions, represented as filled contours,
entirely encompass the injected PSDs. 
In the second, the reconstruction perfectly overlays with the averaged data periodogram shown in grey. %
In both panels of Fig.~\ref{fig: cs_psd}, the $90\%$ credible regions for the noise recovery under both assumptions are tight enough to be mostly indistinguishable from the median of the posterior distributions. 
We can compute the Bayes factor between the two hypotheses to assess which is better supported by the data. %
This is a pivotal question posed by resorting to flexible methods, as lifting the assumption of knowing the spectral shape of the different contributions inevitably reduces our constraining power. Flexible noise models can accommodate the excess power present in the data because of a stochastic signal, making it harder to detect an SGWB and constrain its parameters.
Labeling the first hypothesis
as ``noise+signal'' ($ns$) and the second as ``noise'' ($n$), we compute the respective evidences using Eq.~\eqref{eq: ss} and the natural logarithm of the Bayes factor $\log \mathcal{B}_{ns/n} = \log Z_{ns} - \log Z_n$. We find a log-Bayes factor of $\log\mathcal {B}_{ns/n} = 6.54\pm0.44$, which, according to the criterion outlined in Ref.~\cite{doi:10.1080/01621459.1995.10476572}, translates to ``very strong'' evidence for the presence of a stochastic signal in the data. 
Thus, in this case, our flexible parametrization allows us to confidently discriminate between the presence or absence of an SGWB. Although the splines can absorb the SGWB, they do so by varying considerably more parameters, so the more parsimonious description of an SGWB is still preferred. 
Nevertheless, we want to stress that the introduction of the splines reduced the log-Bayes factor by three orders of magnitude, as shown in Fig.~\ref{fig: cs_psd}. Therefore, the increased flexibility introduced in the noise (and signal) model(s) could make detecting SGWBs with SNR lower than the one considered much more challenging, as also shown in Ref.~\cite{2023JCAP...04..066B}. This is particularly relevant for SGWBs that are quieter than our example, but that would still be detectable in principle in a templates-only analyses.

The interpretation of the absolute value of the Bayes factor in the context of SGWB analyses is an active area of research with the authors of Refs.~\cite{2023JCAP...04..066B, 2019JCAP...11..017C, Flauger:2020qyi, 2024PhRvD.109h3029P}. Moreover, it is always influenced by the prior distributions assumed during the analysis. Nevertheless, here it still provides a valuable metric to assess the impact of model-agnostic analyses on the detectability of these signals. We also refer the reader to Ref.~\cite{Pozzoli:2024hkt}, where the impact of the inclusion of model uncertainties in the Bayes factor computation was investigated.
\rr{
	For comparison, we also apply the Deviance Information Criterion (DIC)~\cite{https://doi.org/10.1111/1467-9868.00353, MR2027492}, an alternative model selection tool that does not require the computation of the evidence. 
	We define the deviance as $D(\boldsymbol{\theta}) = -2 \log \mathcal{L}(d | \, \boldsymbol{\theta}) + C$, where $C$ is a constant that cancels out in all model comparisons.
	Using the notation $\overline{x}$ to indicate the mean over the posterior distribution of the quantity $x$, 
	we can define the effective number of model parameters as $p_D = \overline{D(\boldsymbol{\theta})} - D(\overline{\boldsymbol{\theta}})$.
	From these quantities, the DIC is computed as
	\begin{equation}
		\text{DIC} = \overline{D(\boldsymbol{\theta})} \, + \, p_D.
	\end{equation}
	Models with smaller DIC values should be preferred. In particular, Ref.~\cite{doi:10.1080/01621459.1995.10476572} suggests that the absolute value of the difference in DIC between two models, $|\Delta \text{DIC}|$, should roughly follow the same scale of twice the natural logarithm of the Bayes.
	For the template analysis, we find $\Delta \text{DIC} = \text{DIC}_{ns} - \text{DIC}_{n} = -1.3 \times 10^4$, while in the splines analysis we obtain $\Delta \text{DIC} = -30$. This agrees with our Bayes factor results, favouring the ``noise+signal'' hypothesis in both cases. Once again, the inclusion of splines significantly reduces the strength of the support for the presence of a signal. However, when analysing the ``noise+signal'' data, we find $\text{DIC}_{ns}$ to be (slightly) smaller in the templates-only analysis, while the splines case shows a (slightly) larger evidence than the templates-only case. Therefore, under our inference setup, the DIC seems (slightly) more penalizing towards increased model complexity than the Bayes factor.
}
 
Given the stochastic nature of the datasets, we would expect that 
flexible models are capable of achieving larger likelihoods than the analytic templates,
even if the latter ones match the spectral shape used during the data generation.
This is a consequence of the flexible models being able to better accommodate the random fluctuations inherent to stochastic processes.
Both the noise and signal datasets simulated here are random realizations whose PSD is described by Eqs.~\eqref{eq:noise_tdi} and~\eqref{eq: Sh} only on average. Therefore, we expect the presence of local fluctuations that deviate from the templates.
Given the local nature of the splines, they can account for these fluctuations, while the templates cannot. 
In principle, without an upper bound on the number of knots, one could reach larger and larger likelihood values by adding knots until the recovered PSD exactly matches the data periodogram everywhere. This effect is balanced by the increased prior volume associated with the additional parameters. The fact that only a small number of interior knots are added when the data is generated from a PSD consistent with the template family gives us confidence that our priors on the spline parameters have been chosen well.

\begin{figure}[t]
	\centering
	\includegraphics[width=\linewidth]{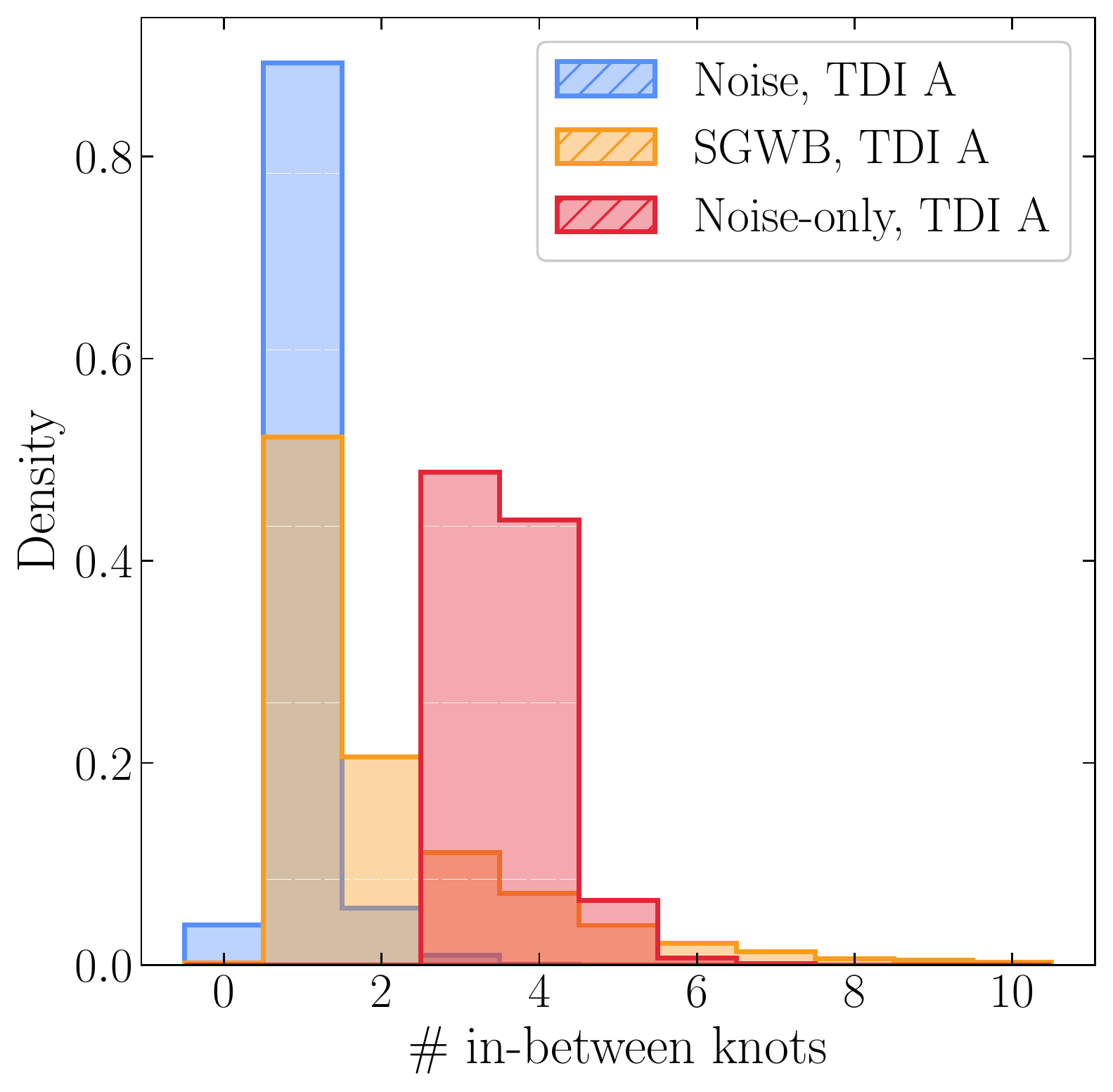}
	\caption{
	Posteriors on the number of in-between spline knots for the datastream containing noise and an SGWB due to cosmic strings. Blue (orange) bars refer to the noise in the $A$ channel (signal) contribution. Red bars refer to the analysis performed only with the noise component. 
	}\label{fig: cs_leaves}
\end{figure}

\begin{figure}[t]
	\centering
	\includegraphics[width=\linewidth]{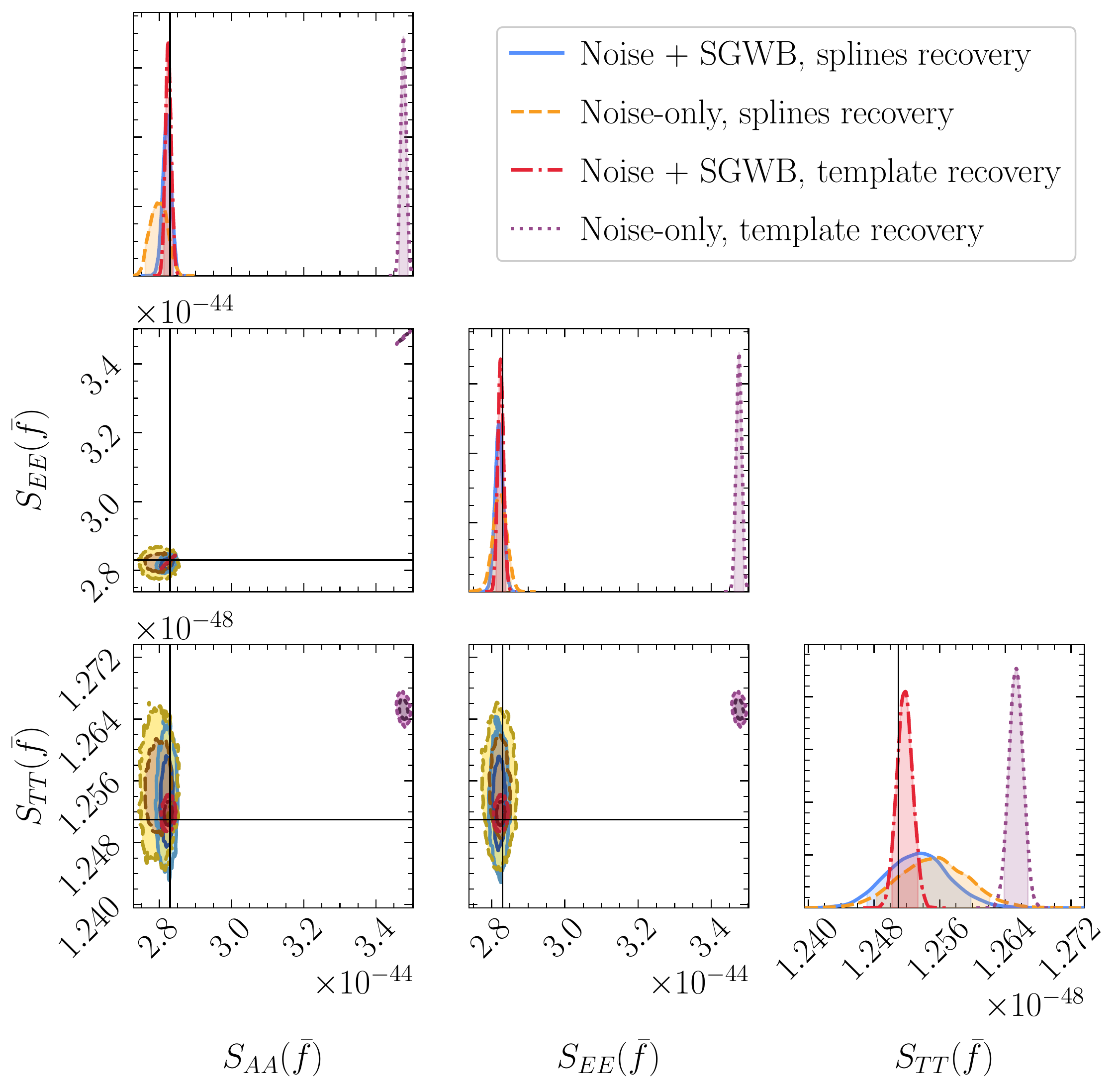}
	\caption{
	Posterior distribution of reconstructed total PSD in the three TDI channels at a reference frequency of $\bar{f}=10^{-3} \, \rm Hz$,
	for the datastream containing noise and an SGWB due to cosmic strings. Black solid lines represent the values evaluated at the injected parameters. 
	2D contours represent the $1-\sigma$ and $2-\sigma$ regions, while the 1D shaded areas represent $95\%$ credible regions.
	}\label{fig: cs_corner}
\end{figure}
We report the posterior on the number of in-between spline knots in the $A$ channel in Fig.~\ref{fig: cs_leaves}.
The posteriors in $A$ and $E$ always look the same, since the PSD in the two channels has the same shape. Conversely, given its low sensitivity to GWs, the $T$ channel posteriors always converge to one single knot. 
Colors reflect the labels already described for Fig.~\ref{fig: cs_psd}. 
Under the ``noise+signal'' hypothesis, 
all the spline perturbations converge to having only one in-between knot. 
On the other hand, given the loud nature of the signal, 
the recovery under the noise-only hypothesis requires between a total of 5 and 6 knots to account for the totality of the power present in the data stream. %
This behavior reflects our expectations. 
Since the simulated data are consistent with the templates used during the inference stage,
we do not have to rely on the spline component to catch deviations from the parametrized baseline. 
Nevertheless, since one extra knot is still added by the reversible jump algorithm, 
we believe that its role is to account for the aforementioned random fluctuations.
The different posteriors help interpret the Bayes factors previously presented. In fact, even if our flexible model under the ``noise-only'' hypothesis can provide a good description of the data, it requires a larger number of parameters than the model including both noise and signal. 
Usually, we would like our analyses to implement the concept of ``Occam's razor,'' i.e., if two models describe the data equally well, the more complex one is disfavoured. It's often said that the Bayes always automatically includes it, but this holds only in cases where suitable priors are chosen (see, e.g., Ref~\cite{10.5555/971143}). The Bayes factors found here demonstrate that our priors are chosen appropriately, since we prefer the simplest description of the data in this case, which is the noise model template plus an SGWB. 

Figure~\ref{fig: cs_psd} shows the $90\%$ credible regions getting wider when splines are included in the model.
The effect is more pronounced for the SGWB contribution, while it is barely visible for the remaining components.
To better highlight this behaviour,
we show in Fig.~\ref{fig: cs_corner} the posterior distribution of the reconstructed total PSD in the three TDI channels at a reference frequency of $\bar{f}=10^{-3} \, \rm Hz$. We note that the colorscheme in this Figure is different to the one used in Figs.~\ref{fig: cs_psd} and~\ref{fig: cs_leaves}.
This Figure allows us to compare how the location and width of the PSD posteriors change with the assumed model.
The tightest contours are obtained by the two template only analysis.
The noise-only recovery in purple is clearly biased in all three channels.
Including the splines in the model has the effect of broadening the posteriors for both the ``noise+signal'' and the ``noise'' recoveries,
and to shift the latter enough to be in agreement with the injection.
\rr{Both here and in Fig.~\ref{fig: cs_splines}, the credible intervals for the splines analysis account only for the uncertainty in the spline parameters, while the template parameters remain fixed. The same applies to the evidence and DIC calculations in the splines analyses. While being a caveat, the posteriors in Fig.~\ref{fig: cs_corner} suggest that the uncertainty in the spline coefficients translates in larger contours than the uncertainty in the template parameters alone. Therefore, we do not expect our conclusions to change significantly.}

\subsection{Perturbed noise spectrum}\label{sec: perturbed noise}
\begin{figure}
\centering
\includegraphics[width=\linewidth]{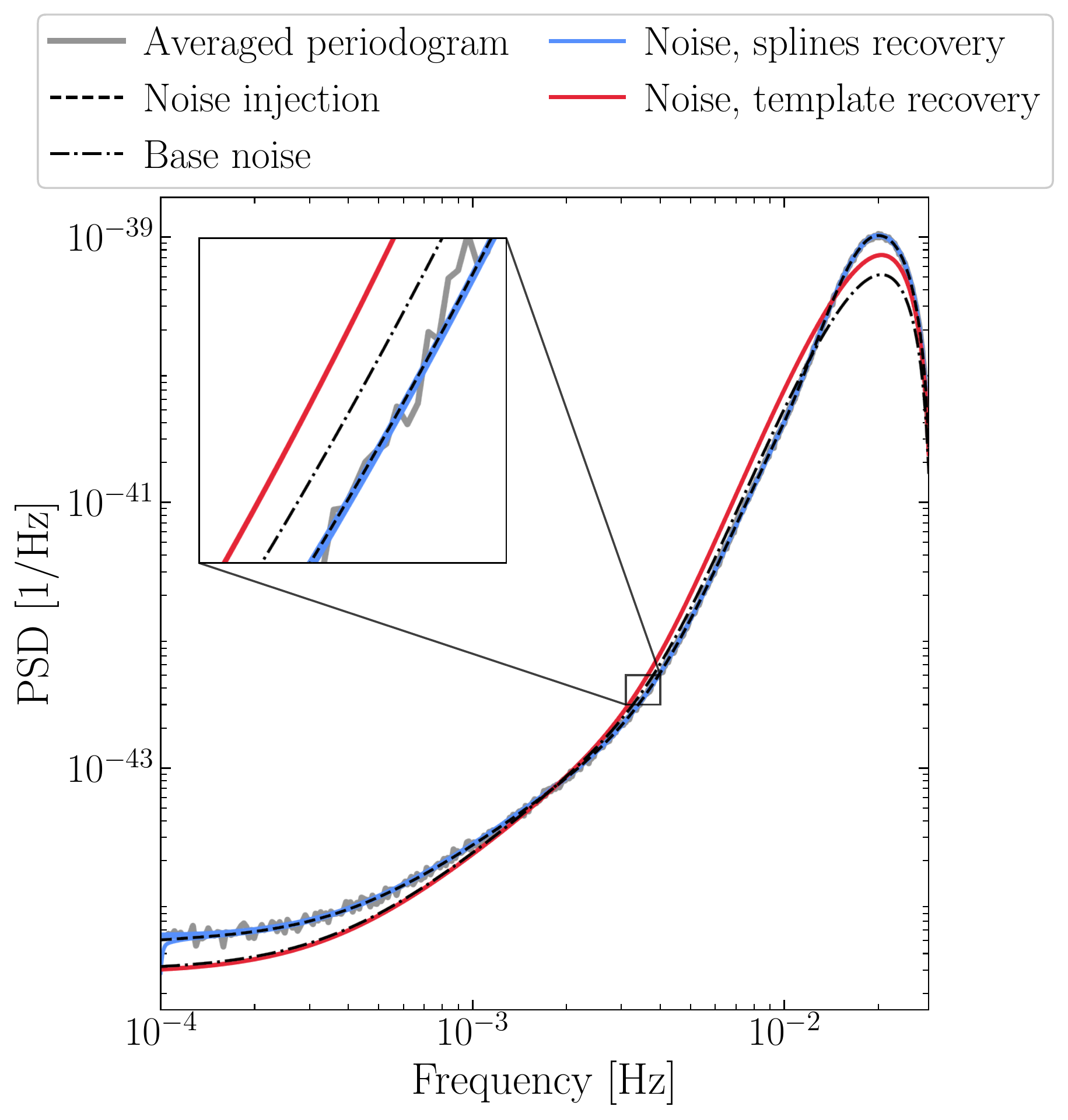}
\caption{%
Reconstructed posterior distribution of the reconstruction of the total PSD in the $A$ channel. 
The black dashed (dash-dotted) line represents the ``perturbed'' PSD used to generate the noise dataset (the base noise PSD before applying the spline perturbation).
The averaged data periodogram is shown as a grey line.
Blue lines and filled contours represent the median and $90\%$ credible interval for the reconstructed noise PSD. 
The inset shows a zoom-in on the region $f \in [3.1,\,4]\, \rm mHz$.
}\label{fig: noiseperturbed_psd}
\end{figure}

Finally, we present our results for the ``perturbed'' noise PSD.
In this case, we have generated the data according to a PSD of the same shape as Eq.~\eqref{eq: noise_pert} with $\mathcal{P}_{n, ii}(f)$ being splines with knot positions $\log_{10}f_k = \left\{-7.5, \,-4, \, -2.01,\, -1.71,\, -1.54, \, -0.6 \right\}$ and amplitudes $w_k = \left\{0.0,\, 0.2,\, -0.1,\, 0.3,\, 0.2,\, 0.0 \right\}$. 
We assume the same knot positions and amplitudes for all the channels, and we include the leftmost and rightmost points to cover the whole frequency range implied by our assumed observation period and cadence.
The base, unperturbed noise PSD and the one used for the injection are shown as the black dash-dotted and dashed lines in Fig.~\ref{fig: noiseperturbed_psd}.
Red and blue lines and contours represent the median and $90\%$ credible interval for the reconstructed PSD for the template and spline recoveries, respectively. 
While the first approach fails to successfully represent the power present in the data, the spline analysis is able to recover the injected PSD despite it being different from the assumed baseline.
This gives us confidence that the results of the previous section do not strongly depend on having the right template as the baseline of the flexible models,
as long as the differences in the spectral shape are small enough to be compatible with our priors.

\begin{figure}
\centering
\includegraphics[width=\linewidth]{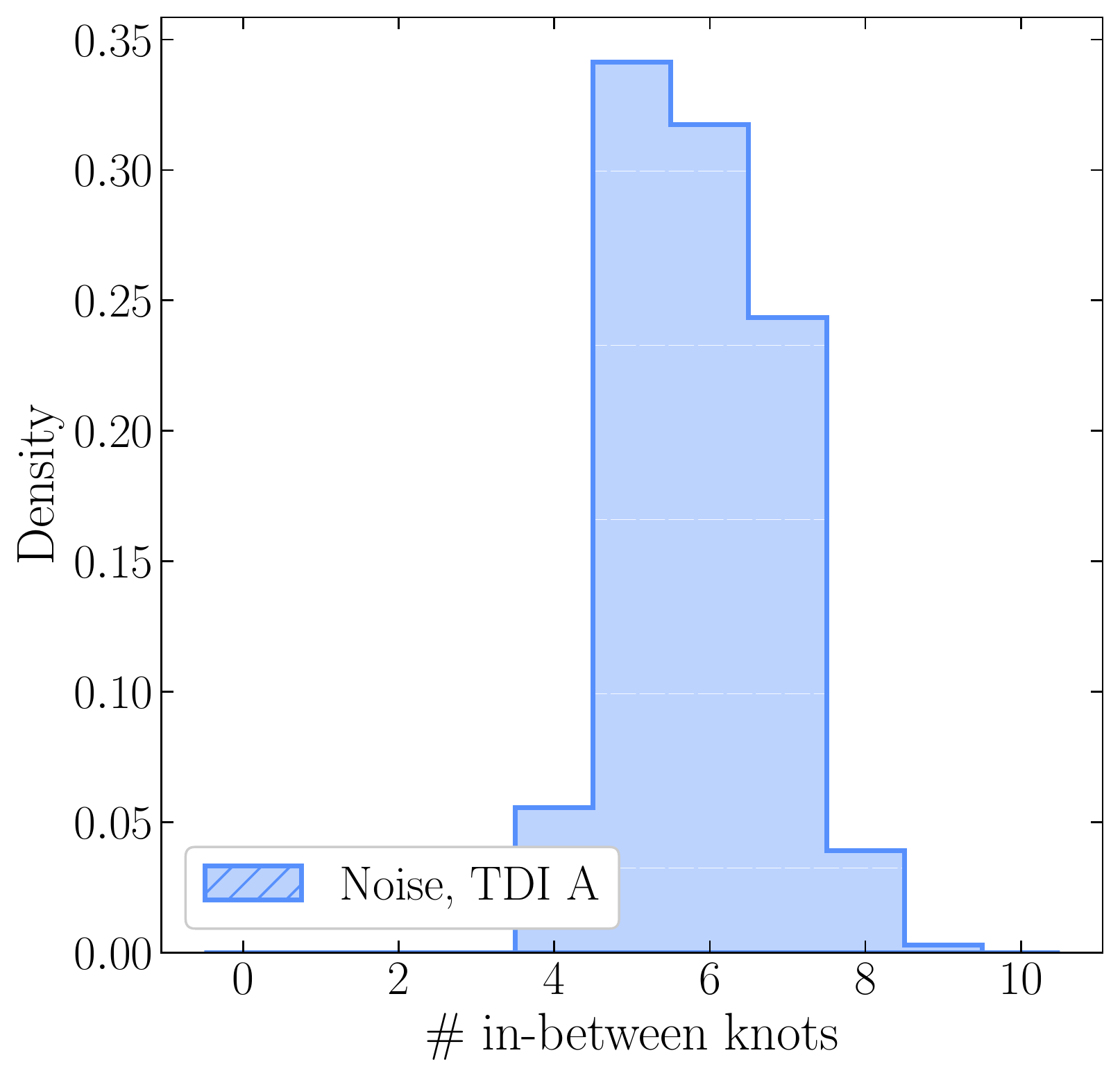}
\caption{
Posteriors on the number of in-between spline knots for the datastream containing only noise generated from the ``perturbed'' PSD.
}\label{fig: noiseperturbed_leaves}
\end{figure}

We report the posterior on the number of spline knots required to account for the totality of the data in Fig.~\ref{fig: noiseperturbed_leaves}.
In this case, we need a larger number of knots to model the simulated data, a 
result that is consistent with our expectations.
Notably, the posterior does not converge to the number of knots used when generating the data, which was $4$.
This is a consequence of the intrinsic correlation between the baseline templates and the spline components of the model.
Therefore, the recovered parameters of the two components are free to be different from those used during the data generation as long as the total PSD remains the same.

\begin{figure}
	\centering
	\includegraphics[width=\linewidth]{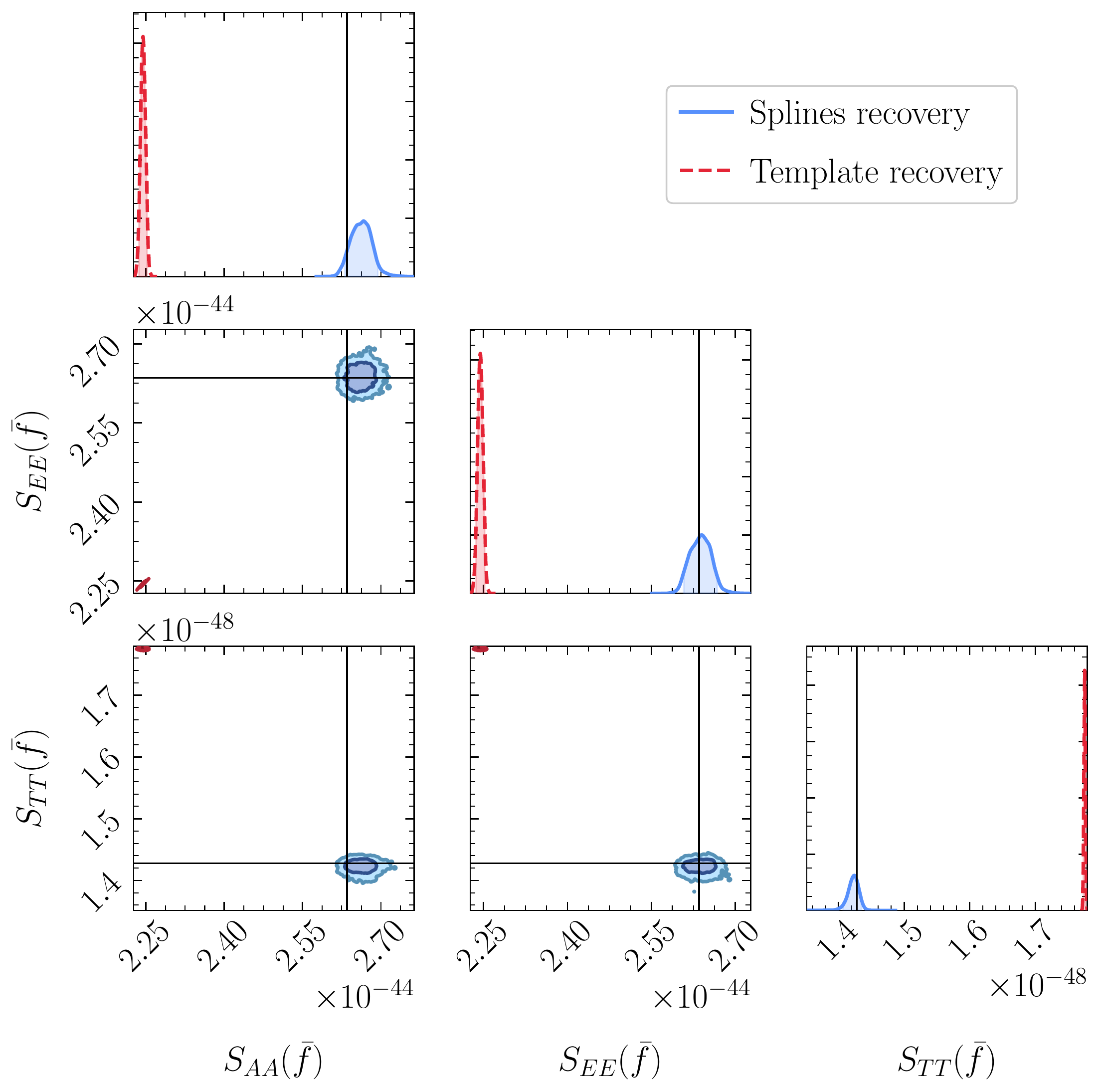}
	\caption{
	Posterior distribution of reconstructed total PSD in the three TDI channels at a reference frequency of $\bar{f}=10^{-3} \, \rm Hz$,
	for the datastream containing only ``perturbed'' noise. Black solid lines represent the values evaluated at the injected parameters. 
	2D contours represent the $1-\sigma$ and $2-\sigma$ regions, while the 1D shaded areas represent $95\%$ credible regions.
	}\label{fig: noise_corner}
\end{figure}

Finally, we show in Fig.~\ref{fig: noise_corner} the posterior distribution of the reconstructed total PSD in the three TDI channels at the reference frequency $\bar{f}=10^{-3} \, \rm Hz$.
Once again, the inclusion of splines in the model has the effect of broadening the posterior distributions,
and shifting them enough to encompass the injected values.

\section{conclusions\label{sec: conclusions}}
In this paper, we have presented a novel code designed to simultaneously extract the properties of the instrumental noise and stochastic backgrounds from LISA data under model-agnostic assumptions. 
For both components, we provide flexible models that do not necessarily have to follow rigid spectral shapes. 
The LPF mission has already shown that analytical models could not account for the totality of the observed instrumental noise~\cite{PhysRevLett.116.231101, 2024PhRvD.110d2004A}, and we expect the same to hold for LISA.
Conversely, we ultimately want SGWB models that are ready to capture unexpected features in the signal spectral shape.
Hence, the requirement of flexible models for both stochastic contributions, which have to be able to account for the unmodelled contribution that will enter our measurement chain. %

Our fit is based on the inclusion of Akima splines. In each TDI channel, we apply them %
to the state-of-the-art noise and signal models~\cite{2024arXiv240207571C} and run our inference scheme on the parameters of both.
We introduce the splines directly at the TDI level, after the LISA response and transfer functions, to account for any arbitrary contribution. This level of flexibility comes at a price, as it reduces our capability of discriminating stochastic signals from pure noise. We investigate the behaviour of our pipeline in two different scenarios.

We first test our setup against a simulated dataset containing a realization of instrumental noise and a stochastic background compatible with Cosmic Strings. 
In this case, both the noise and the signal injections fall inside the parameter space of our baseline templates alone.
Relying on a detection criterion based on the Bayes factor, we find that our inference scheme can successfully disentangle the signal and the noise, reconstructing both. We find a log-Bayes factor of $\log\mathcal {B}_{ns/n} = 6.54\pm0.44$ in favour of the presence of a signal in the data. 
Nevertheless, the extra flexibility introduced comes at the price of losing constraining power. 
Running our pipeline with only rigid templates produces a much larger Bayes factor ($\log\mathcal {B}_{ns/n} = 6515.65 \pm 0.18$) and tighter posteriors on the total PSDs with respect to the model with the splines, as shown in Figs.~\ref{fig: cs_psd} and~\ref{fig: cs_corner}.
For the power-law signal considered, having an SNR of $262$, we find a decrease in the log-Bayes factor of $3$ orders of magnitude. This is a consequence of our flexible noise model's ability to account for the totality of power contained in the data stream, at the price of a larger parameter space dimensionality.

Next, we assess the performance of our model when the analyzed noise data do not match the assumed baseline template. 
Even in this case, our flexible noise model successfully recovers the spectral shape of the analyzed data, but it needs a large number of spline knots to do so.

While we work under a set of simplifying assumptions, this paper represents a further step forward towards an efficient, general, and agnostic setup able to address the challenges posed by stochastic signals in LISA. 
From here, natural future research directions include (i) extending our approach to the full covariance matrix, (ii) applying our methodology to datasets of increasing realism, including the future LISA Data Challenges~\cite{ldc}, (iii) analyzing an SGWB whose PSD spectral shape does not match the analytic template assumed as baseline during the inference,
and (iv) leveraging our agnostic setup to fit the totality of the data as a whole and discriminate between noise and signals a posteriori.
\\

\section*{Code availability}
The library used to perform the analysis presented in this paper is already available at~\url{https://github.com/asantini29/lisa-ps}~\cite{lisaps}. This package is still under early development, and it is likely to undergo significant changes in the future. We also release our parallel, GPU-accelerated implementation of Akima splines at~\url{https://github.com/asantini29/CudAkima}~\cite{cudakima_2024_13919394}.
Besides the few packages already cited, this work made extensive use of \texttt{Numpy}~\cite{harris2020array}, \texttt{Matplotlib}~\cite{Hunter:2007}, \texttt{Numba}~\cite{lam2015numba}, and \texttt{Pysco}~\cite{pysco_2024_13930440}.

\acknowledgments
We thank Alexandre Toubiana, Lorenzo Speri, and the LISA Cosmology Working Group members for fruitful discussions. In particular, the authors acknowledge the collaboration of Germano Nardini, Chiara Caprini, Jean-Baptiste Bayle, Nikolaos Karnesis, Quentin Baghi, Riccardo Buscicchio, Federico Pozzoli, Mauro Pieroni, and Dam Quang Nam. 
A.S., M.M., and O.H. gratefully acknowledge the support of the German Space Agency, DLR. The work is supported by the Federal Ministry for Economic Affairs and Climate Action based on a resolution of the German Bundestag (Project Ref. No. FKZ 50 OQ 2301).

\appendix

\section{Akima splines implementation}\label{sec: akima_impl}
\rr{%
Given a set of knot points with coordinates ${(x_i, \, y_i)}_{i = 1, \dots, n}$, Akima splines construct a piecewise function $S(x)$ such that $S(x_i) = y_i$ for all $i$. In those points, its slope $s_i$ is a function of $(x_j, \, y_j)$ with $j \in \{i-2, \dots, i+2\}$. We can define the slope of the line connecting two consecutive knots $(x_i, \, y_i)$ and $(x_{i+1}, \, y_{i+1})$ as
\begin{equation}
m_i = \frac{y_{i+1} - y_i}{x_{i+1} - x_i}, \quad i = 0, \dots,\, n-1.
\end{equation}
Defining the quantity
\begin{equation}
w_i = |m_{i+1} - m_i|,
\end{equation}
we can construct the spline slopes as
\begin{equation}
	s_i =
		\begin{dcases}
			\frac{m_{i+1} + m_i}{2} \quad &\text{if } w_i + w_{i-2} = 0, \\
			\frac{w_i\, m_{i-1} + w_{i-2} \, m_i}{w_i + w_{i-2}} \quad &\text{otherwise}.
		\end{dcases}
\end{equation}
For each knot $(x_i, y_i)$ the coefficients $s_i$ require 4 extra points to be computed, therefore we need to introduce boundary conditions for the first and last two knots. In our implementation~\cite{cudakima_2024_13919394}, we follow Ref.~\cite{2020SciPy-NMeth}:
\begin{subequations}
	\begin{align}
		s_1 &= \frac{3m_1 - m_2}{2};\\
		s_2 &= 
			\begin{dcases}
				\frac{m_3 + m_2}{2} \quad &\text{if } w_2 + w_1 = 0\\
				\frac{w_2\, m_1 + w_1 \, m_2}{w_2 + w_1} &\text{otherwise};
			\end{dcases}\\
		s_{n-1} &= 
			\begin{dcases}
				\frac{m_{n-3} + m_{n-2}}{2} \quad \text{if } w_{n-3} + w_{n-2} = 0\\
				\frac{w_{n-3}\, m_{n-2} + w_{n-2} \, m_{n-3}}{w_{n-3} + w_{n-2}}  \quad \text{otherwise};
			\end{dcases}\\
		s_{n} &= \frac{3m_{n-1} - m_{n-2}}{2}.
	\end{align}
\end{subequations}
Since the spline is defined as a piecewise cubic polynomial, in each interval $x \in [x_i, \, x_{i+1})$ we have:
\begin{equation}
S(x) = a_i + b_i (x - x_i) + c_i (x - x_i)^2 + d_i (x - x_i)^3.
\end{equation}
The coefficients are determined by imposing the continuity of $S(x)$ and its first derivative $S'(x)$ at each knot:
\begin{subequations}
	\begin{align}
		S(x_i) = y_i, &\quad S(x_{i+1}) = y_{i+1}\\
		S'(x_i) = s_i, &\quad S'(x_{i+1}) = s_{i+1}.
	\end{align}
\end{subequations}
Therefore, the polynomial coefficients read:
\begin{subequations}
	\begin{align}
		a_i &= y_i, \\
		b_i &= s_i, \\
		c_i &= \frac{3m_i - 2s_i - s_{i+1}}{x_{i+1} - x_i}, \\
		d_i &= \frac{s_i + s_{i+1} - 2m_i}{(x_{i+1} - x_i)^2}.
	\end{align}
\end{subequations}
Since there are no conditions on the second derivative of $S(x)$, this is not necessarily continuous. 
}
\bibliography{reference}

\begin{thebibliography}{118}%
\makeatletter
\providecommand \@ifxundefined [1]{%
 \@ifx{#1\undefined}
}%
\providecommand \@ifnum [1]{%
 \ifnum #1\expandafter \@firstoftwo
 \else \expandafter \@secondoftwo
 \fi
}%
\providecommand \@ifx [1]{%
 \ifx #1\expandafter \@firstoftwo
 \else \expandafter \@secondoftwo
 \fi
}%
\providecommand \natexlab [1]{#1}%
\providecommand \enquote  [1]{``#1''}%
\providecommand \bibnamefont  [1]{#1}%
\providecommand \bibfnamefont [1]{#1}%
\providecommand \citenamefont [1]{#1}%
\providecommand \href@noop [0]{\@secondoftwo}%
\providecommand \href [0]{\begingroup \@sanitize@url \@href}%
\providecommand \@href[1]{\@@startlink{#1}\@@href}%
\providecommand \@@href[1]{\endgroup#1\@@endlink}%
\providecommand \@sanitize@url [0]{\catcode `\\12\catcode `\$12\catcode
  `\&12\catcode `\#12\catcode `\^12\catcode `\_12\catcode `\%12\relax}%
\providecommand \@@startlink[1]{}%
\providecommand \@@endlink[0]{}%
\providecommand \url  [0]{\begingroup\@sanitize@url \@url }%
\providecommand \@url [1]{\endgroup\@href {#1}{\urlprefix }}%
\providecommand \urlprefix  [0]{URL }%
\providecommand \Eprint [0]{\href }%
\providecommand \doibase [0]{https://doi.org/}%
\providecommand \selectlanguage [0]{\@gobble}%
\providecommand \bibinfo  [0]{\@secondoftwo}%
\providecommand \bibfield  [0]{\@secondoftwo}%
\providecommand \translation [1]{[#1]}%
\providecommand \BibitemOpen [0]{}%
\providecommand \bibitemStop [0]{}%
\providecommand \bibitemNoStop [0]{.\EOS\space}%
\providecommand \EOS [0]{\spacefactor3000\relax}%
\providecommand \BibitemShut  [1]{\csname bibitem#1\endcsname}%
\let\auto@bib@innerbib\@empty
%</preamble>
\bibitem [{\citenamefont {{Colpi}}\ \emph {et~al.}(2024)\citenamefont {{Colpi}}
  \emph {et~al.}}]{2024arXiv240207571C}%
  \BibitemOpen
  \bibfield  {author} {\bibinfo {author} {\bibfnamefont {M.}~\bibnamefont
  {{Colpi}}} \emph {et~al.},\ }\href
  {https://doi.org/10.48550/arXiv.2402.07571} {\bibfield  {journal} {\bibinfo
  {journal} {arXiv e-prints}\ ,\ \bibinfo {eid} {arXiv:2402.07571}} (\bibinfo
  {year} {2024})},\ \Eprint {https://arxiv.org/abs/2402.07571}
  {arXiv:2402.07571 [astro-ph.CO]} \BibitemShut {NoStop}%
\bibitem [{\citenamefont {{Abbott}}\ \emph {et~al.}(2019)\citenamefont
  {{Abbott}} \emph {et~al.}}]{2019PhRvX...9c1040A}%
  \BibitemOpen
  \bibfield  {author} {\bibinfo {author} {\bibfnamefont {B.~P.}\ \bibnamefont
  {{Abbott}}} \emph {et~al.},\ }\href
  {https://doi.org/10.1103/PhysRevX.9.031040} {\bibfield  {journal} {\bibinfo
  {journal} {Physical Review X}\ }\textbf {\bibinfo {volume} {9}},\ \bibinfo
  {eid} {031040} (\bibinfo {year} {2019})},\ \Eprint
  {https://arxiv.org/abs/1811.12907} {arXiv:1811.12907 [astro-ph.HE]}
  \BibitemShut {NoStop}%
\bibitem [{\citenamefont {{Abbott}}\ \emph {et~al.}(2021)\citenamefont
  {{Abbott}} \emph {et~al.}}]{2021PhRvX..11b1053A}%
  \BibitemOpen
  \bibfield  {author} {\bibinfo {author} {\bibfnamefont {R.}~\bibnamefont
  {{Abbott}}} \emph {et~al.},\ }\href
  {https://doi.org/10.1103/PhysRevX.11.021053} {\bibfield  {journal} {\bibinfo
  {journal} {Physical Review X}\ }\textbf {\bibinfo {volume} {11}},\ \bibinfo
  {eid} {021053} (\bibinfo {year} {2021})},\ \Eprint
  {https://arxiv.org/abs/2010.14527} {arXiv:2010.14527 [gr-qc]} \BibitemShut
  {NoStop}%
\bibitem [{\citenamefont {{Abbott}}\ \emph {et~al.}(2024)\citenamefont
  {{Abbott}} \emph {et~al.}}]{2021arXiv210801045T}%
  \BibitemOpen
  \bibfield  {author} {\bibinfo {author} {\bibfnamefont {R.}~\bibnamefont
  {{Abbott}}} \emph {et~al.},\ }\href
  {https://doi.org/10.1103/PhysRevD.109.022001} {\bibfield  {journal} {\bibinfo
   {journal} {\prd}\ }\textbf {\bibinfo {volume} {109}},\ \bibinfo {eid}
  {022001} (\bibinfo {year} {2024})},\ \Eprint
  {https://arxiv.org/abs/2108.01045} {arXiv:2108.01045 [gr-qc]} \BibitemShut
  {NoStop}%
\bibitem [{\citenamefont {{Abbott}}\ \emph {et~al.}(2023)\citenamefont
  {{Abbott}} \emph {et~al.}}]{2021arXiv211103606T}%
  \BibitemOpen
  \bibfield  {author} {\bibinfo {author} {\bibfnamefont {R.}~\bibnamefont
  {{Abbott}}} \emph {et~al.},\ }\href
  {https://doi.org/10.1103/PhysRevX.13.041039} {\bibfield  {journal} {\bibinfo
  {journal} {Physical Review X}\ }\textbf {\bibinfo {volume} {13}},\ \bibinfo
  {eid} {041039} (\bibinfo {year} {2023})},\ \Eprint
  {https://arxiv.org/abs/2111.03606} {arXiv:2111.03606 [gr-qc]} \BibitemShut
  {NoStop}%
\bibitem [{\citenamefont {Antoniadis}\ \emph {et~al.}(2024)\citenamefont
  {Antoniadis} \emph {et~al.}}]{EPTA:2023xxk}%
  \BibitemOpen
  \bibfield  {author} {\bibinfo {author} {\bibfnamefont {J.}~\bibnamefont
  {Antoniadis}} \emph {et~al.} (\bibinfo {collaboration} {EPTA, InPTA}),\
  }\href {https://doi.org/10.1051/0004-6361/202347433} {\bibfield  {journal}
  {\bibinfo  {journal} {Astron. Astrophys.}\ }\textbf {\bibinfo {volume}
  {685}},\ \bibinfo {pages} {A94} (\bibinfo {year} {2024})},\ \Eprint
  {https://arxiv.org/abs/2306.16227} {arXiv:2306.16227 [astro-ph.CO]}
  \BibitemShut {NoStop}%
\bibitem [{\citenamefont {Afzal}\ \emph {et~al.}(2023)\citenamefont {Afzal}
  \emph {et~al.}}]{NANOGrav:2023hvm}%
  \BibitemOpen
  \bibfield  {author} {\bibinfo {author} {\bibfnamefont {A.}~\bibnamefont
  {Afzal}} \emph {et~al.} (\bibinfo {collaboration} {NANOGrav}),\ }\href
  {https://doi.org/10.3847/2041-8213/acdc91} {\bibfield  {journal} {\bibinfo
  {journal} {Astrophys. J. Lett.}\ }\textbf {\bibinfo {volume} {951}},\
  \bibinfo {pages} {L11} (\bibinfo {year} {2023})},\ \bibinfo {note} {[Erratum:
  Astrophys.J.Lett. 971, L27 (2024), Erratum: Astrophys.J. 971, L27 (2024)]},\
  \Eprint {https://arxiv.org/abs/2306.16219} {arXiv:2306.16219 [astro-ph.HE]}
  \BibitemShut {NoStop}%
\bibitem [{\citenamefont {Reardon}\ \emph {et~al.}(2023)\citenamefont {Reardon}
  \emph {et~al.}}]{Reardon:2023gzh}%
  \BibitemOpen
  \bibfield  {author} {\bibinfo {author} {\bibfnamefont {D.~J.}\ \bibnamefont
  {Reardon}} \emph {et~al.},\ }\href {https://doi.org/10.3847/2041-8213/acdd02}
  {\bibfield  {journal} {\bibinfo  {journal} {Astrophys. J. Lett.}\ }\textbf
  {\bibinfo {volume} {951}},\ \bibinfo {pages} {L6} (\bibinfo {year} {2023})},\
  \Eprint {https://arxiv.org/abs/2306.16215} {arXiv:2306.16215 [astro-ph.HE]}
  \BibitemShut {NoStop}%
\bibitem [{\citenamefont {Xu}\ \emph {et~al.}(2023)\citenamefont {Xu} \emph
  {et~al.}}]{Xu:2023wog}%
  \BibitemOpen
  \bibfield  {author} {\bibinfo {author} {\bibfnamefont {H.}~\bibnamefont {Xu}}
  \emph {et~al.},\ }\href {https://doi.org/10.1088/1674-4527/acdfa5} {\bibfield
   {journal} {\bibinfo  {journal} {Res. Astron. Astrophys.}\ }\textbf {\bibinfo
  {volume} {23}},\ \bibinfo {pages} {075024} (\bibinfo {year} {2023})},\
  \Eprint {https://arxiv.org/abs/2306.16216} {arXiv:2306.16216 [astro-ph.HE]}
  \BibitemShut {NoStop}%
\bibitem [{\citenamefont {Klein}\ \emph {et~al.}(2016)\citenamefont {Klein}
  \emph {et~al.}}]{Klein:2015hvg}%
  \BibitemOpen
  \bibfield  {author} {\bibinfo {author} {\bibfnamefont {A.}~\bibnamefont
  {Klein}} \emph {et~al.},\ }\href {https://doi.org/10.1103/PhysRevD.93.024003}
  {\bibfield  {journal} {\bibinfo  {journal} {Phys. Rev. D}\ }\textbf {\bibinfo
  {volume} {93}},\ \bibinfo {pages} {024003} (\bibinfo {year} {2016})},\
  \Eprint {https://arxiv.org/abs/1511.05581} {arXiv:1511.05581 [gr-qc]}
  \BibitemShut {NoStop}%
\bibitem [{\citenamefont {Korol}\ \emph {et~al.}(2020)\citenamefont {Korol}
  \emph {et~al.}}]{Korol:2020lpq}%
  \BibitemOpen
  \bibfield  {author} {\bibinfo {author} {\bibfnamefont {V.}~\bibnamefont
  {Korol}} \emph {et~al.},\ }\href
  {https://doi.org/10.1051/0004-6361/202037764} {\bibfield  {journal} {\bibinfo
   {journal} {Astron. Astrophys.}\ }\textbf {\bibinfo {volume} {638}},\
  \bibinfo {pages} {A153} (\bibinfo {year} {2020})},\ \Eprint
  {https://arxiv.org/abs/2002.10462} {arXiv:2002.10462 [astro-ph.GA]}
  \BibitemShut {NoStop}%
\bibitem [{\citenamefont {Babak}\ \emph {et~al.}(2017)\citenamefont {Babak},
  \citenamefont {Gair}, \citenamefont {Sesana}, \citenamefont {Barausse},
  \citenamefont {Sopuerta}, \citenamefont {Berry}, \citenamefont {Berti},
  \citenamefont {Amaro-Seoane}, \citenamefont {Petiteau},\ and\ \citenamefont
  {Klein}}]{Babak:2017tow}%
  \BibitemOpen
  \bibfield  {author} {\bibinfo {author} {\bibfnamefont {S.}~\bibnamefont
  {Babak}}, \bibinfo {author} {\bibfnamefont {J.}~\bibnamefont {Gair}},
  \bibinfo {author} {\bibfnamefont {A.}~\bibnamefont {Sesana}}, \bibinfo
  {author} {\bibfnamefont {E.}~\bibnamefont {Barausse}}, \bibinfo {author}
  {\bibfnamefont {C.~F.}\ \bibnamefont {Sopuerta}}, \bibinfo {author}
  {\bibfnamefont {C.~P.~L.}\ \bibnamefont {Berry}}, \bibinfo {author}
  {\bibfnamefont {E.}~\bibnamefont {Berti}}, \bibinfo {author} {\bibfnamefont
  {P.}~\bibnamefont {Amaro-Seoane}}, \bibinfo {author} {\bibfnamefont
  {A.}~\bibnamefont {Petiteau}},\ and\ \bibinfo {author} {\bibfnamefont
  {A.}~\bibnamefont {Klein}},\ }\href
  {https://doi.org/10.1103/PhysRevD.95.103012} {\bibfield  {journal} {\bibinfo
  {journal} {Phys. Rev. D}\ }\textbf {\bibinfo {volume} {95}},\ \bibinfo
  {pages} {103012} (\bibinfo {year} {2017})},\ \Eprint
  {https://arxiv.org/abs/1703.09722} {arXiv:1703.09722 [gr-qc]} \BibitemShut
  {NoStop}%
\bibitem [{\citenamefont {Gerosa}\ \emph {et~al.}(2019)\citenamefont {Gerosa},
  \citenamefont {Ma}, \citenamefont {Wong}, \citenamefont {Berti},
  \citenamefont {O'Shaughnessy}, \citenamefont {Chen},\ and\ \citenamefont
  {Belczynski}}]{Gerosa:2019dbe}%
  \BibitemOpen
  \bibfield  {author} {\bibinfo {author} {\bibfnamefont {D.}~\bibnamefont
  {Gerosa}}, \bibinfo {author} {\bibfnamefont {S.}~\bibnamefont {Ma}}, \bibinfo
  {author} {\bibfnamefont {K.~W.~K.}\ \bibnamefont {Wong}}, \bibinfo {author}
  {\bibfnamefont {E.}~\bibnamefont {Berti}}, \bibinfo {author} {\bibfnamefont
  {R.}~\bibnamefont {O'Shaughnessy}}, \bibinfo {author} {\bibfnamefont
  {Y.}~\bibnamefont {Chen}},\ and\ \bibinfo {author} {\bibfnamefont
  {K.}~\bibnamefont {Belczynski}},\ }\href
  {https://doi.org/10.1103/PhysRevD.99.103004} {\bibfield  {journal} {\bibinfo
  {journal} {Phys. Rev. D}\ }\textbf {\bibinfo {volume} {99}},\ \bibinfo
  {pages} {103004} (\bibinfo {year} {2019})},\ \Eprint
  {https://arxiv.org/abs/1902.00021} {arXiv:1902.00021 [astro-ph.HE]}
  \BibitemShut {NoStop}%
\bibitem [{\citenamefont {Buscicchio}\ \emph {et~al.}(2021)\citenamefont
  {Buscicchio}, \citenamefont {Klein}, \citenamefont {Roebber}, \citenamefont
  {Moore}, \citenamefont {Gerosa}, \citenamefont {Finch},\ and\ \citenamefont
  {Vecchio}}]{Buscicchio:2021dph}%
  \BibitemOpen
  \bibfield  {author} {\bibinfo {author} {\bibfnamefont {R.}~\bibnamefont
  {Buscicchio}}, \bibinfo {author} {\bibfnamefont {A.}~\bibnamefont {Klein}},
  \bibinfo {author} {\bibfnamefont {E.}~\bibnamefont {Roebber}}, \bibinfo
  {author} {\bibfnamefont {C.~J.}\ \bibnamefont {Moore}}, \bibinfo {author}
  {\bibfnamefont {D.}~\bibnamefont {Gerosa}}, \bibinfo {author} {\bibfnamefont
  {E.}~\bibnamefont {Finch}},\ and\ \bibinfo {author} {\bibfnamefont
  {A.}~\bibnamefont {Vecchio}},\ }\href
  {https://doi.org/10.1103/PhysRevD.104.044065} {\bibfield  {journal} {\bibinfo
   {journal} {Phys. Rev. D}\ }\textbf {\bibinfo {volume} {104}},\ \bibinfo
  {pages} {044065} (\bibinfo {year} {2021})},\ \Eprint
  {https://arxiv.org/abs/2106.05259} {arXiv:2106.05259 [astro-ph.HE]}
  \BibitemShut {NoStop}%
\bibitem [{\citenamefont {Toubiana}\ \emph {et~al.}(2022)\citenamefont
  {Toubiana}, \citenamefont {Babak}, \citenamefont {Marsat},\ and\
  \citenamefont {Ossokine}}]{Toubiana:2022vpp}%
  \BibitemOpen
  \bibfield  {author} {\bibinfo {author} {\bibfnamefont {A.}~\bibnamefont
  {Toubiana}}, \bibinfo {author} {\bibfnamefont {S.}~\bibnamefont {Babak}},
  \bibinfo {author} {\bibfnamefont {S.}~\bibnamefont {Marsat}},\ and\ \bibinfo
  {author} {\bibfnamefont {S.}~\bibnamefont {Ossokine}},\ }\href
  {https://doi.org/10.1103/PhysRevD.106.104034} {\bibfield  {journal} {\bibinfo
   {journal} {Phys. Rev. D}\ }\textbf {\bibinfo {volume} {106}},\ \bibinfo
  {pages} {104034} (\bibinfo {year} {2022})},\ \Eprint
  {https://arxiv.org/abs/2206.12439} {arXiv:2206.12439 [gr-qc]} \BibitemShut
  {NoStop}%
\bibitem [{\citenamefont {Karnesis}\ \emph {et~al.}(2021)\citenamefont
  {Karnesis}, \citenamefont {Babak}, \citenamefont {Pieroni}, \citenamefont
  {Cornish},\ and\ \citenamefont {Littenberg}}]{Karnesis:2021tsh}%
  \BibitemOpen
  \bibfield  {author} {\bibinfo {author} {\bibfnamefont {N.}~\bibnamefont
  {Karnesis}}, \bibinfo {author} {\bibfnamefont {S.}~\bibnamefont {Babak}},
  \bibinfo {author} {\bibfnamefont {M.}~\bibnamefont {Pieroni}}, \bibinfo
  {author} {\bibfnamefont {N.}~\bibnamefont {Cornish}},\ and\ \bibinfo {author}
  {\bibfnamefont {T.}~\bibnamefont {Littenberg}},\ }\href
  {https://doi.org/10.1103/PhysRevD.104.043019} {\bibfield  {journal} {\bibinfo
   {journal} {Phys. Rev. D}\ }\textbf {\bibinfo {volume} {104}},\ \bibinfo
  {pages} {043019} (\bibinfo {year} {2021})},\ \Eprint
  {https://arxiv.org/abs/2103.14598} {arXiv:2103.14598 [astro-ph.IM]}
  \BibitemShut {NoStop}%
\bibitem [{\citenamefont {Farmer}\ and\ \citenamefont
  {Phinney}(2003)}]{Farmer:2003pa}%
  \BibitemOpen
  \bibfield  {author} {\bibinfo {author} {\bibfnamefont {A.~J.}\ \bibnamefont
  {Farmer}}\ and\ \bibinfo {author} {\bibfnamefont {E.~S.}\ \bibnamefont
  {Phinney}},\ }\href {https://doi.org/10.1111/j.1365-2966.2003.07176.x}
  {\bibfield  {journal} {\bibinfo  {journal} {Mon. Not. Roy. Astron. Soc.}\
  }\textbf {\bibinfo {volume} {346}},\ \bibinfo {pages} {1197} (\bibinfo {year}
  {2003})},\ \Eprint {https://arxiv.org/abs/astro-ph/0304393}
  {arXiv:astro-ph/0304393} \BibitemShut {NoStop}%
\bibitem [{\citenamefont {Babak}\ \emph {et~al.}(2023)\citenamefont {Babak},
  \citenamefont {Caprini}, \citenamefont {Figueroa}, \citenamefont {Karnesis},
  \citenamefont {Marcoccia}, \citenamefont {Nardini}, \citenamefont {Pieroni},
  \citenamefont {Ricciardone}, \citenamefont {Sesana},\ and\ \citenamefont
  {Torrado}}]{Babak:2023lro}%
  \BibitemOpen
  \bibfield  {author} {\bibinfo {author} {\bibfnamefont {S.}~\bibnamefont
  {Babak}}, \bibinfo {author} {\bibfnamefont {C.}~\bibnamefont {Caprini}},
  \bibinfo {author} {\bibfnamefont {D.~G.}\ \bibnamefont {Figueroa}}, \bibinfo
  {author} {\bibfnamefont {N.}~\bibnamefont {Karnesis}}, \bibinfo {author}
  {\bibfnamefont {P.}~\bibnamefont {Marcoccia}}, \bibinfo {author}
  {\bibfnamefont {G.}~\bibnamefont {Nardini}}, \bibinfo {author} {\bibfnamefont
  {M.}~\bibnamefont {Pieroni}}, \bibinfo {author} {\bibfnamefont
  {A.}~\bibnamefont {Ricciardone}}, \bibinfo {author} {\bibfnamefont
  {A.}~\bibnamefont {Sesana}},\ and\ \bibinfo {author} {\bibfnamefont
  {J.}~\bibnamefont {Torrado}},\ }\href
  {https://doi.org/10.1088/1475-7516/2023/08/034} {\bibfield  {journal}
  {\bibinfo  {journal} {JCAP}\ }\textbf {\bibinfo {volume} {08}},\ \bibinfo
  {pages} {034}},\ \Eprint {https://arxiv.org/abs/2304.06368} {arXiv:2304.06368
  [astro-ph.CO]} \BibitemShut {NoStop}%
\bibitem [{\citenamefont {Barack}\ and\ \citenamefont
  {Cutler}(2004)}]{Barack:2004wc}%
  \BibitemOpen
  \bibfield  {author} {\bibinfo {author} {\bibfnamefont {L.}~\bibnamefont
  {Barack}}\ and\ \bibinfo {author} {\bibfnamefont {C.}~\bibnamefont
  {Cutler}},\ }\href {https://doi.org/10.1103/PhysRevD.70.122002} {\bibfield
  {journal} {\bibinfo  {journal} {Phys. Rev. D}\ }\textbf {\bibinfo {volume}
  {70}},\ \bibinfo {pages} {122002} (\bibinfo {year} {2004})},\ \Eprint
  {https://arxiv.org/abs/gr-qc/0409010} {arXiv:gr-qc/0409010} \BibitemShut
  {NoStop}%
\bibitem [{\citenamefont {Bonetti}\ and\ \citenamefont
  {Sesana}(2020)}]{Bonetti:2020jku}%
  \BibitemOpen
  \bibfield  {author} {\bibinfo {author} {\bibfnamefont {M.}~\bibnamefont
  {Bonetti}}\ and\ \bibinfo {author} {\bibfnamefont {A.}~\bibnamefont
  {Sesana}},\ }\href {https://doi.org/10.1103/PhysRevD.102.103023} {\bibfield
  {journal} {\bibinfo  {journal} {Phys. Rev. D}\ }\textbf {\bibinfo {volume}
  {102}},\ \bibinfo {pages} {103023} (\bibinfo {year} {2020})},\ \Eprint
  {https://arxiv.org/abs/2007.14403} {arXiv:2007.14403 [astro-ph.GA]}
  \BibitemShut {NoStop}%
\bibitem [{\citenamefont {{Pozzoli}}\ \emph {et~al.}(2023)\citenamefont
  {{Pozzoli}}, \citenamefont {{Babak}}, \citenamefont {{Sesana}}, \citenamefont
  {{Bonetti}},\ and\ \citenamefont {{Karnesis}}}]{2023PhRvD.108j3039P}%
  \BibitemOpen
  \bibfield  {author} {\bibinfo {author} {\bibfnamefont {F.}~\bibnamefont
  {{Pozzoli}}}, \bibinfo {author} {\bibfnamefont {S.}~\bibnamefont {{Babak}}},
  \bibinfo {author} {\bibfnamefont {A.}~\bibnamefont {{Sesana}}}, \bibinfo
  {author} {\bibfnamefont {M.}~\bibnamefont {{Bonetti}}},\ and\ \bibinfo
  {author} {\bibfnamefont {N.}~\bibnamefont {{Karnesis}}},\ }\href
  {https://doi.org/10.1103/PhysRevD.108.103039} {\bibfield  {journal} {\bibinfo
   {journal} {\prd}\ }\textbf {\bibinfo {volume} {108}},\ \bibinfo {eid}
  {103039} (\bibinfo {year} {2023})},\ \Eprint
  {https://arxiv.org/abs/2302.07043} {arXiv:2302.07043 [astro-ph.GA]}
  \BibitemShut {NoStop}%
\bibitem [{\citenamefont {Team}(2018)}]{scirdv}%
  \BibitemOpen
  \bibfield  {author} {\bibinfo {author} {\bibfnamefont {L.~S.~S.}\
  \bibnamefont {Team}},\ }\href {https://doi.org/10.2140/camcos.2010.5.65}
  {\bibfield  {journal} {\bibinfo  {journal} {ESA-L3-EST-SCI-RS-001}\ }\textbf
  {\bibinfo {volume} {5}},\ \bibinfo {pages} {65} (\bibinfo {year}
  {2018})}\BibitemShut {NoStop}%
\bibitem [{\citenamefont {Caprini}(2015)}]{Caprini:2015tfa}%
  \BibitemOpen
  \bibfield  {author} {\bibinfo {author} {\bibfnamefont {C.}~\bibnamefont
  {Caprini}},\ }\href {https://doi.org/10.1088/1742-6596/610/1/012004}
  {\bibfield  {journal} {\bibinfo  {journal} {J. Phys. Conf. Ser.}\ }\textbf
  {\bibinfo {volume} {610}},\ \bibinfo {pages} {012004} (\bibinfo {year}
  {2015})},\ \Eprint {https://arxiv.org/abs/1501.01174} {arXiv:1501.01174
  [gr-qc]} \BibitemShut {NoStop}%
\bibitem [{\citenamefont {Caprini}\ \emph {et~al.}(2020)\citenamefont {Caprini}
  \emph {et~al.}}]{Caprini:2019egz}%
  \BibitemOpen
  \bibfield  {author} {\bibinfo {author} {\bibfnamefont {C.}~\bibnamefont
  {Caprini}} \emph {et~al.},\ }\href
  {https://doi.org/10.1088/1475-7516/2020/03/024} {\bibfield  {journal}
  {\bibinfo  {journal} {JCAP}\ }\textbf {\bibinfo {volume} {03}},\ \bibinfo
  {pages} {024}},\ \Eprint {https://arxiv.org/abs/1910.13125} {arXiv:1910.13125
  [astro-ph.CO]} \BibitemShut {NoStop}%
\bibitem [{\citenamefont {Caprini}\ \emph {et~al.}(2019)\citenamefont
  {Caprini}, \citenamefont {Figueroa}, \citenamefont {Flauger}, \citenamefont
  {Nardini}, \citenamefont {Peloso}, \citenamefont {Pieroni}, \citenamefont
  {Ricciardone},\ and\ \citenamefont {Tasinato}}]{Caprini:2019pxz}%
  \BibitemOpen
  \bibfield  {author} {\bibinfo {author} {\bibfnamefont {C.}~\bibnamefont
  {Caprini}}, \bibinfo {author} {\bibfnamefont {D.~G.}\ \bibnamefont
  {Figueroa}}, \bibinfo {author} {\bibfnamefont {R.}~\bibnamefont {Flauger}},
  \bibinfo {author} {\bibfnamefont {G.}~\bibnamefont {Nardini}}, \bibinfo
  {author} {\bibfnamefont {M.}~\bibnamefont {Peloso}}, \bibinfo {author}
  {\bibfnamefont {M.}~\bibnamefont {Pieroni}}, \bibinfo {author} {\bibfnamefont
  {A.}~\bibnamefont {Ricciardone}},\ and\ \bibinfo {author} {\bibfnamefont
  {G.}~\bibnamefont {Tasinato}},\ }\href
  {https://doi.org/10.1088/1475-7516/2019/11/017} {\bibfield  {journal}
  {\bibinfo  {journal} {JCAP}\ }\textbf {\bibinfo {volume} {11}},\ \bibinfo
  {pages} {017}},\ \Eprint {https://arxiv.org/abs/1906.09244} {arXiv:1906.09244
  [astro-ph.CO]} \BibitemShut {NoStop}%
\bibitem [{\citenamefont {Caprini}\ \emph {et~al.}(2016)\citenamefont {Caprini}
  \emph {et~al.}}]{Caprini:2015zlo}%
  \BibitemOpen
  \bibfield  {author} {\bibinfo {author} {\bibfnamefont {C.}~\bibnamefont
  {Caprini}} \emph {et~al.},\ }\href
  {https://doi.org/10.1088/1475-7516/2016/04/001} {\bibfield  {journal}
  {\bibinfo  {journal} {JCAP}\ }\textbf {\bibinfo {volume} {04}},\ \bibinfo
  {pages} {001}},\ \Eprint {https://arxiv.org/abs/1512.06239} {arXiv:1512.06239
  [astro-ph.CO]} \BibitemShut {NoStop}%
\bibitem [{\citenamefont {{Auclair}}\ \emph {et~al.}(2020)\citenamefont
  {{Auclair}}, \citenamefont {{Blanco-Pillado}}, \citenamefont {{Figueroa}},
  \citenamefont {{Jenkins}}, \citenamefont {{Lewicki}}, \citenamefont
  {{Sakellariadou}}, \citenamefont {{Sanidas}}, \citenamefont {{Sousa}},
  \citenamefont {{Steer}}, \citenamefont {{Wachter}}, \citenamefont
  {{Kuroyanagi}},\ and\ \citenamefont {{LISA Cosmology Working
  Group}}}]{2020JCAP...04..034A}%
  \BibitemOpen
  \bibfield  {author} {\bibinfo {author} {\bibfnamefont {P.}~\bibnamefont
  {{Auclair}}}, \bibinfo {author} {\bibfnamefont {J.~J.}\ \bibnamefont
  {{Blanco-Pillado}}}, \bibinfo {author} {\bibfnamefont {D.~G.}\ \bibnamefont
  {{Figueroa}}}, \bibinfo {author} {\bibfnamefont {A.~C.}\ \bibnamefont
  {{Jenkins}}}, \bibinfo {author} {\bibfnamefont {M.}~\bibnamefont
  {{Lewicki}}}, \bibinfo {author} {\bibfnamefont {M.}~\bibnamefont
  {{Sakellariadou}}}, \bibinfo {author} {\bibfnamefont {S.}~\bibnamefont
  {{Sanidas}}}, \bibinfo {author} {\bibfnamefont {L.}~\bibnamefont {{Sousa}}},
  \bibinfo {author} {\bibfnamefont {D.~A.}\ \bibnamefont {{Steer}}}, \bibinfo
  {author} {\bibfnamefont {J.~M.}\ \bibnamefont {{Wachter}}}, \bibinfo {author}
  {\bibfnamefont {S.}~\bibnamefont {{Kuroyanagi}}},\ and\ \bibinfo {author}
  {\bibnamefont {{LISA Cosmology Working Group}}},\ }\href
  {https://doi.org/10.1088/1475-7516/2020/04/034} {\bibfield  {journal}
  {\bibinfo  {journal} {\jcap}\ }\textbf {\bibinfo {volume} {2020}},\ \bibinfo
  {eid} {034} (\bibinfo {year} {2020})},\ \Eprint
  {https://arxiv.org/abs/1909.00819} {arXiv:1909.00819 [astro-ph.CO]}
  \BibitemShut {NoStop}%
\bibitem [{\citenamefont {Ricciardone}(2017)}]{Ricciardone:2016ddg}%
  \BibitemOpen
  \bibfield  {author} {\bibinfo {author} {\bibfnamefont {A.}~\bibnamefont
  {Ricciardone}},\ }\href {https://doi.org/10.1088/1742-6596/840/1/012030}
  {\bibfield  {journal} {\bibinfo  {journal} {J. Phys. Conf. Ser.}\ }\textbf
  {\bibinfo {volume} {840}},\ \bibinfo {pages} {012030} (\bibinfo {year}
  {2017})},\ \Eprint {https://arxiv.org/abs/1612.06799} {arXiv:1612.06799
  [astro-ph.CO]} \BibitemShut {NoStop}%
\bibitem [{\citenamefont {Cai}\ \emph {et~al.}(2019)\citenamefont {Cai},
  \citenamefont {Pi},\ and\ \citenamefont {Sasaki}}]{Cai:2018dig}%
  \BibitemOpen
  \bibfield  {author} {\bibinfo {author} {\bibfnamefont {R.-g.}\ \bibnamefont
  {Cai}}, \bibinfo {author} {\bibfnamefont {S.}~\bibnamefont {Pi}},\ and\
  \bibinfo {author} {\bibfnamefont {M.}~\bibnamefont {Sasaki}},\ }\href
  {https://doi.org/10.1103/PhysRevLett.122.201101} {\bibfield  {journal}
  {\bibinfo  {journal} {Phys. Rev. Lett.}\ }\textbf {\bibinfo {volume} {122}},\
  \bibinfo {pages} {201101} (\bibinfo {year} {2019})},\ \Eprint
  {https://arxiv.org/abs/1810.11000} {arXiv:1810.11000 [astro-ph.CO]}
  \BibitemShut {NoStop}%
\bibitem [{\citenamefont {Bartolo}\ \emph {et~al.}(2019)\citenamefont
  {Bartolo}, \citenamefont {De~Luca}, \citenamefont {Franciolini},
  \citenamefont {Lewis}, \citenamefont {Peloso},\ and\ \citenamefont
  {Riotto}}]{Bartolo:2018evs}%
  \BibitemOpen
  \bibfield  {author} {\bibinfo {author} {\bibfnamefont {N.}~\bibnamefont
  {Bartolo}}, \bibinfo {author} {\bibfnamefont {V.}~\bibnamefont {De~Luca}},
  \bibinfo {author} {\bibfnamefont {G.}~\bibnamefont {Franciolini}}, \bibinfo
  {author} {\bibfnamefont {A.}~\bibnamefont {Lewis}}, \bibinfo {author}
  {\bibfnamefont {M.}~\bibnamefont {Peloso}},\ and\ \bibinfo {author}
  {\bibfnamefont {A.}~\bibnamefont {Riotto}},\ }\href
  {https://doi.org/10.1103/PhysRevLett.122.211301} {\bibfield  {journal}
  {\bibinfo  {journal} {Phys. Rev. Lett.}\ }\textbf {\bibinfo {volume} {122}},\
  \bibinfo {pages} {211301} (\bibinfo {year} {2019})},\ \Eprint
  {https://arxiv.org/abs/1810.12218} {arXiv:1810.12218 [astro-ph.CO]}
  \BibitemShut {NoStop}%
\bibitem [{\citenamefont {Auclair}\ \emph {et~al.}(2023)\citenamefont {Auclair}
  \emph {et~al.}}]{LISACosmologyWorkingGroup:2022jok}%
  \BibitemOpen
  \bibfield  {author} {\bibinfo {author} {\bibfnamefont {P.}~\bibnamefont
  {Auclair}} \emph {et~al.} (\bibinfo {collaboration} {LISA Cosmology Working
  Group}),\ }\href {https://doi.org/10.1007/s41114-023-00045-2} {\bibfield
  {journal} {\bibinfo  {journal} {Living Rev. Rel.}\ }\textbf {\bibinfo
  {volume} {26}},\ \bibinfo {pages} {5} (\bibinfo {year} {2023})},\ \Eprint
  {https://arxiv.org/abs/2204.05434} {arXiv:2204.05434 [astro-ph.CO]}
  \BibitemShut {NoStop}%
\bibitem [{\citenamefont {Littenberg}\ and\ \citenamefont
  {Cornish}(2023)}]{Littenberg:2023xpl}%
  \BibitemOpen
  \bibfield  {author} {\bibinfo {author} {\bibfnamefont {T.~B.}\ \bibnamefont
  {Littenberg}}\ and\ \bibinfo {author} {\bibfnamefont {N.~J.}\ \bibnamefont
  {Cornish}},\ }\href {https://doi.org/10.1103/PhysRevD.107.063004} {\bibfield
  {journal} {\bibinfo  {journal} {Phys. Rev. D}\ }\textbf {\bibinfo {volume}
  {107}},\ \bibinfo {pages} {063004} (\bibinfo {year} {2023})},\ \Eprint
  {https://arxiv.org/abs/2301.03673} {arXiv:2301.03673 [gr-qc]} \BibitemShut
  {NoStop}%
\bibitem [{\citenamefont {{Katz}}\ \emph {et~al.}(2024)\citenamefont {{Katz}},
  \citenamefont {{Karnesis}}, \citenamefont {{Korsakova}}, \citenamefont
  {{Gair}},\ and\ \citenamefont {{Stergioulas}}}]{2024arXiv240504690K}%
  \BibitemOpen
  \bibfield  {author} {\bibinfo {author} {\bibfnamefont {M.~L.}\ \bibnamefont
  {{Katz}}}, \bibinfo {author} {\bibfnamefont {N.}~\bibnamefont {{Karnesis}}},
  \bibinfo {author} {\bibfnamefont {N.}~\bibnamefont {{Korsakova}}}, \bibinfo
  {author} {\bibfnamefont {J.~R.}\ \bibnamefont {{Gair}}},\ and\ \bibinfo
  {author} {\bibfnamefont {N.}~\bibnamefont {{Stergioulas}}},\ }\href
  {https://doi.org/10.48550/arXiv.2405.04690} {\bibfield  {journal} {\bibinfo
  {journal} {arXiv e-prints}\ ,\ \bibinfo {eid} {arXiv:2405.04690}} (\bibinfo
  {year} {2024})},\ \Eprint {https://arxiv.org/abs/2405.04690}
  {arXiv:2405.04690 [gr-qc]} \BibitemShut {NoStop}%
\bibitem [{\citenamefont {Deng}\ \emph {et~al.}(2025)\citenamefont {Deng},
  \citenamefont {Babak}, \citenamefont {Le~Jeune}, \citenamefont {Marsat},
  \citenamefont {Plagnol},\ and\ \citenamefont {Sartirana}}]{Deng:2025wgk}%
  \BibitemOpen
  \bibfield  {author} {\bibinfo {author} {\bibfnamefont {S.}~\bibnamefont
  {Deng}}, \bibinfo {author} {\bibfnamefont {S.}~\bibnamefont {Babak}},
  \bibinfo {author} {\bibfnamefont {M.}~\bibnamefont {Le~Jeune}}, \bibinfo
  {author} {\bibfnamefont {S.}~\bibnamefont {Marsat}}, \bibinfo {author}
  {\bibfnamefont {E.}~\bibnamefont {Plagnol}},\ and\ \bibinfo {author}
  {\bibfnamefont {A.}~\bibnamefont {Sartirana}},\ }\href
  {https://doi.org/10.1103/PhysRevD.111.103014} {\bibfield  {journal} {\bibinfo
   {journal} {Phys. Rev. D}\ }\textbf {\bibinfo {volume} {111}},\ \bibinfo
  {pages} {103014} (\bibinfo {year} {2025})},\ \Eprint
  {https://arxiv.org/abs/2501.10277} {arXiv:2501.10277 [gr-qc]} \BibitemShut
  {NoStop}%
\bibitem [{\citenamefont {Strub}\ \emph {et~al.}(2024)\citenamefont {Strub},
  \citenamefont {Ferraioli}, \citenamefont {Schmelzbach}, \citenamefont
  {St\"ahler},\ and\ \citenamefont {Giardini}}]{Strub:2024kbe}%
  \BibitemOpen
  \bibfield  {author} {\bibinfo {author} {\bibfnamefont {S.~H.}\ \bibnamefont
  {Strub}}, \bibinfo {author} {\bibfnamefont {L.}~\bibnamefont {Ferraioli}},
  \bibinfo {author} {\bibfnamefont {C.}~\bibnamefont {Schmelzbach}}, \bibinfo
  {author} {\bibfnamefont {S.~C.}\ \bibnamefont {St\"ahler}},\ and\ \bibinfo
  {author} {\bibfnamefont {D.}~\bibnamefont {Giardini}},\ }\href
  {https://doi.org/10.1103/PhysRevD.110.024005} {\bibfield  {journal} {\bibinfo
   {journal} {Phys. Rev. D}\ }\textbf {\bibinfo {volume} {110}},\ \bibinfo
  {pages} {024005} (\bibinfo {year} {2024})},\ \Eprint
  {https://arxiv.org/abs/2403.15318} {arXiv:2403.15318 [gr-qc]} \BibitemShut
  {NoStop}%
\bibitem [{\citenamefont {Boileau}\ \emph {et~al.}(2023)\citenamefont
  {Boileau}, \citenamefont {Christensen}, \citenamefont {Gowling},
  \citenamefont {Hindmarsh},\ and\ \citenamefont {Meyer}}]{Boileau:2022ter}%
  \BibitemOpen
  \bibfield  {author} {\bibinfo {author} {\bibfnamefont {G.}~\bibnamefont
  {Boileau}}, \bibinfo {author} {\bibfnamefont {N.}~\bibnamefont
  {Christensen}}, \bibinfo {author} {\bibfnamefont {C.}~\bibnamefont
  {Gowling}}, \bibinfo {author} {\bibfnamefont {M.}~\bibnamefont {Hindmarsh}},\
  and\ \bibinfo {author} {\bibfnamefont {R.}~\bibnamefont {Meyer}},\ }\href
  {https://doi.org/10.1088/1475-7516/2023/02/056} {\bibfield  {journal}
  {\bibinfo  {journal} {JCAP}\ }\textbf {\bibinfo {volume} {02}},\ \bibinfo
  {pages} {056}},\ \Eprint {https://arxiv.org/abs/2209.13277} {arXiv:2209.13277
  [gr-qc]} \BibitemShut {NoStop}%
\bibitem [{\citenamefont {Adams}\ and\ \citenamefont
  {Cornish}(2014)}]{Adams:2013qma}%
  \BibitemOpen
  \bibfield  {author} {\bibinfo {author} {\bibfnamefont {M.~R.}\ \bibnamefont
  {Adams}}\ and\ \bibinfo {author} {\bibfnamefont {N.~J.}\ \bibnamefont
  {Cornish}},\ }\href {https://doi.org/10.1103/PhysRevD.89.022001} {\bibfield
  {journal} {\bibinfo  {journal} {Phys. Rev. D}\ }\textbf {\bibinfo {volume}
  {89}},\ \bibinfo {pages} {022001} (\bibinfo {year} {2014})},\ \Eprint
  {https://arxiv.org/abs/1307.4116} {arXiv:1307.4116 [gr-qc]} \BibitemShut
  {NoStop}%
\bibitem [{\citenamefont {Hartwig}\ \emph {et~al.}(2023)\citenamefont
  {Hartwig}, \citenamefont {Lilley}, \citenamefont {Muratore},\ and\
  \citenamefont {Pieroni}}]{Hartwig:2023pft}%
  \BibitemOpen
  \bibfield  {author} {\bibinfo {author} {\bibfnamefont {O.}~\bibnamefont
  {Hartwig}}, \bibinfo {author} {\bibfnamefont {M.}~\bibnamefont {Lilley}},
  \bibinfo {author} {\bibfnamefont {M.}~\bibnamefont {Muratore}},\ and\
  \bibinfo {author} {\bibfnamefont {M.}~\bibnamefont {Pieroni}},\ }\href
  {https://doi.org/10.1103/PhysRevD.107.123531} {\bibfield  {journal} {\bibinfo
   {journal} {Phys. Rev. D}\ }\textbf {\bibinfo {volume} {107}},\ \bibinfo
  {pages} {123531} (\bibinfo {year} {2023})},\ \Eprint
  {https://arxiv.org/abs/2303.15929} {arXiv:2303.15929 [gr-qc]} \BibitemShut
  {NoStop}%
\bibitem [{\citenamefont {Criswell}\ \emph {et~al.}(2025)\citenamefont
  {Criswell}, \citenamefont {Rieck},\ and\ \citenamefont
  {Mandic}}]{Criswell:2024hfn}%
  \BibitemOpen
  \bibfield  {author} {\bibinfo {author} {\bibfnamefont {A.~W.}\ \bibnamefont
  {Criswell}}, \bibinfo {author} {\bibfnamefont {S.}~\bibnamefont {Rieck}},\
  and\ \bibinfo {author} {\bibfnamefont {V.}~\bibnamefont {Mandic}},\ }\href
  {https://doi.org/10.1103/PhysRevD.111.023025} {\bibfield  {journal} {\bibinfo
   {journal} {Phys. Rev. D}\ }\textbf {\bibinfo {volume} {111}},\ \bibinfo
  {pages} {023025} (\bibinfo {year} {2025})},\ \Eprint
  {https://arxiv.org/abs/2410.23260} {arXiv:2410.23260 [astro-ph.IM]}
  \BibitemShut {NoStop}%
\bibitem [{\citenamefont {Armano}\ \emph {et~al.}(2016)\citenamefont {Armano}
  \emph {et~al.}}]{PhysRevLett.116.231101}%
  \BibitemOpen
  \bibfield  {author} {\bibinfo {author} {\bibfnamefont {M.}~\bibnamefont
  {Armano}} \emph {et~al.},\ }\href
  {https://doi.org/10.1103/PhysRevLett.116.231101} {\bibfield  {journal}
  {\bibinfo  {journal} {Phys. Rev. Lett.}\ }\textbf {\bibinfo {volume} {116}},\
  \bibinfo {pages} {231101} (\bibinfo {year} {2016})}\BibitemShut {NoStop}%
\bibitem [{\citenamefont {{Armano}}\ \emph {et~al.}(2024)\citenamefont
  {{Armano}}, \citenamefont {{LISA Pathfinder Collaboration}} \emph
  {et~al.}}]{2024PhRvD.110d2004A}%
  \BibitemOpen
  \bibfield  {author} {\bibinfo {author} {\bibfnamefont {M.}~\bibnamefont
  {{Armano}}}, \bibinfo {author} {\bibnamefont {{LISA Pathfinder
  Collaboration}}}, \emph {et~al.},\ }\href
  {https://doi.org/10.1103/PhysRevD.110.042004} {\bibfield  {journal} {\bibinfo
   {journal} {\prd}\ }\textbf {\bibinfo {volume} {110}},\ \bibinfo {eid}
  {042004} (\bibinfo {year} {2024})},\ \Eprint
  {https://arxiv.org/abs/2405.05207} {arXiv:2405.05207 [astro-ph.IM]}
  \BibitemShut {NoStop}%
\bibitem [{\citenamefont {{Muratore}}\ \emph {et~al.}(2024)\citenamefont
  {{Muratore}}, \citenamefont {{Gair}},\ and\ \citenamefont
  {{Speri}}}]{2024PhRvD.109d2001M}%
  \BibitemOpen
  \bibfield  {author} {\bibinfo {author} {\bibfnamefont {M.}~\bibnamefont
  {{Muratore}}}, \bibinfo {author} {\bibfnamefont {J.}~\bibnamefont {{Gair}}},\
  and\ \bibinfo {author} {\bibfnamefont {L.}~\bibnamefont {{Speri}}},\ }\href
  {https://doi.org/10.1103/PhysRevD.109.042001} {\bibfield  {journal} {\bibinfo
   {journal} {\prd}\ }\textbf {\bibinfo {volume} {109}},\ \bibinfo {eid}
  {042001} (\bibinfo {year} {2024})},\ \Eprint
  {https://arxiv.org/abs/2308.01056} {arXiv:2308.01056 [gr-qc]} \BibitemShut
  {NoStop}%
\bibitem [{\citenamefont {{Caprini}}\ \emph {et~al.}(2019)\citenamefont
  {{Caprini}}, \citenamefont {{Figueroa}}, \citenamefont {{Flauger}},
  \citenamefont {{Nardini}}, \citenamefont {{Peloso}}, \citenamefont
  {{Pieroni}}, \citenamefont {{Ricciardone}},\ and\ \citenamefont
  {{Tasinato}}}]{2019JCAP...11..017C}%
  \BibitemOpen
  \bibfield  {author} {\bibinfo {author} {\bibfnamefont {C.}~\bibnamefont
  {{Caprini}}}, \bibinfo {author} {\bibfnamefont {D.~G.}\ \bibnamefont
  {{Figueroa}}}, \bibinfo {author} {\bibfnamefont {R.}~\bibnamefont
  {{Flauger}}}, \bibinfo {author} {\bibfnamefont {G.}~\bibnamefont
  {{Nardini}}}, \bibinfo {author} {\bibfnamefont {M.}~\bibnamefont {{Peloso}}},
  \bibinfo {author} {\bibfnamefont {M.}~\bibnamefont {{Pieroni}}}, \bibinfo
  {author} {\bibfnamefont {A.}~\bibnamefont {{Ricciardone}}},\ and\ \bibinfo
  {author} {\bibfnamefont {G.}~\bibnamefont {{Tasinato}}},\ }\href
  {https://doi.org/10.1088/1475-7516/2019/11/017} {\bibfield  {journal}
  {\bibinfo  {journal} {\jcap}\ }\textbf {\bibinfo {volume} {2019}},\ \bibinfo
  {eid} {017} (\bibinfo {year} {2019})},\ \Eprint
  {https://arxiv.org/abs/1906.09244} {arXiv:1906.09244 [astro-ph.CO]}
  \BibitemShut {NoStop}%
\bibitem [{\citenamefont {Flauger}\ \emph {et~al.}(2021)\citenamefont
  {Flauger}, \citenamefont {Karnesis}, \citenamefont {Nardini}, \citenamefont
  {Pieroni}, \citenamefont {Ricciardone},\ and\ \citenamefont
  {Torrado}}]{Flauger:2020qyi}%
  \BibitemOpen
  \bibfield  {author} {\bibinfo {author} {\bibfnamefont {R.}~\bibnamefont
  {Flauger}}, \bibinfo {author} {\bibfnamefont {N.}~\bibnamefont {Karnesis}},
  \bibinfo {author} {\bibfnamefont {G.}~\bibnamefont {Nardini}}, \bibinfo
  {author} {\bibfnamefont {M.}~\bibnamefont {Pieroni}}, \bibinfo {author}
  {\bibfnamefont {A.}~\bibnamefont {Ricciardone}},\ and\ \bibinfo {author}
  {\bibfnamefont {J.}~\bibnamefont {Torrado}},\ }\href
  {https://doi.org/10.1088/1475-7516/2021/01/059} {\bibfield  {journal}
  {\bibinfo  {journal} {JCAP}\ }\textbf {\bibinfo {volume} {01}},\ \bibinfo
  {pages} {059}},\ \Eprint {https://arxiv.org/abs/2009.11845} {arXiv:2009.11845
  [astro-ph.CO]} \BibitemShut {NoStop}%
\bibitem [{\citenamefont {{Baghi}}\ \emph {et~al.}(2023)\citenamefont
  {{Baghi}}, \citenamefont {{Karnesis}}, \citenamefont {{Bayle}}, \citenamefont
  {{Besan{\c{c}}on}},\ and\ \citenamefont
  {{Inchausp{\'e}}}}]{2023JCAP...04..066B}%
  \BibitemOpen
  \bibfield  {author} {\bibinfo {author} {\bibfnamefont {Q.}~\bibnamefont
  {{Baghi}}}, \bibinfo {author} {\bibfnamefont {N.}~\bibnamefont {{Karnesis}}},
  \bibinfo {author} {\bibfnamefont {J.-B.}\ \bibnamefont {{Bayle}}}, \bibinfo
  {author} {\bibfnamefont {M.}~\bibnamefont {{Besan{\c{c}}on}}},\ and\ \bibinfo
  {author} {\bibfnamefont {H.}~\bibnamefont {{Inchausp{\'e}}}},\ }\href
  {https://doi.org/10.1088/1475-7516/2023/04/066} {\bibfield  {journal}
  {\bibinfo  {journal} {\jcap}\ }\textbf {\bibinfo {volume} {2023}},\ \bibinfo
  {eid} {066} (\bibinfo {year} {2023})},\ \Eprint
  {https://arxiv.org/abs/2302.12573} {arXiv:2302.12573 [gr-qc]} \BibitemShut
  {NoStop}%
\bibitem [{\citenamefont {{Pozzoli}}\ \emph {et~al.}(2024)\citenamefont
  {{Pozzoli}}, \citenamefont {{Buscicchio}}, \citenamefont {{Moore}},
  \citenamefont {{Haardt}},\ and\ \citenamefont
  {{Sesana}}}]{2024PhRvD.109h3029P}%
  \BibitemOpen
  \bibfield  {author} {\bibinfo {author} {\bibfnamefont {F.}~\bibnamefont
  {{Pozzoli}}}, \bibinfo {author} {\bibfnamefont {R.}~\bibnamefont
  {{Buscicchio}}}, \bibinfo {author} {\bibfnamefont {C.~J.}\ \bibnamefont
  {{Moore}}}, \bibinfo {author} {\bibfnamefont {F.}~\bibnamefont {{Haardt}}},\
  and\ \bibinfo {author} {\bibfnamefont {A.}~\bibnamefont {{Sesana}}},\ }\href
  {https://doi.org/10.1103/PhysRevD.109.083029} {\bibfield  {journal} {\bibinfo
   {journal} {\prd}\ }\textbf {\bibinfo {volume} {109}},\ \bibinfo {eid}
  {083029} (\bibinfo {year} {2024})},\ \Eprint
  {https://arxiv.org/abs/2311.12111} {arXiv:2311.12111 [astro-ph.CO]}
  \BibitemShut {NoStop}%
\bibitem [{\citenamefont {Rasmussen}\ and\ \citenamefont
  {Williams}(2005)}]{10.7551/mitpress/3206.001.0001}%
  \BibitemOpen
  \bibfield  {author} {\bibinfo {author} {\bibfnamefont {C.~E.}\ \bibnamefont
  {Rasmussen}}\ and\ \bibinfo {author} {\bibfnamefont {C.~K.~I.}\ \bibnamefont
  {Williams}},\ }\href {https://doi.org/10.7551/mitpress/3206.001.0001} {\emph
  {\bibinfo {title} {Gaussian Processes for Machine Learning}}}\ (\bibinfo
  {publisher} {The MIT Press},\ \bibinfo {year} {2005})\ \Eprint
  {https://arxiv.org/abs/https://direct.mit.edu/book-pdf/2514321/book\_9780262256834.pdf}
  {https://direct.mit.edu/book-pdf/2514321/book\_9780262256834.pdf}
  \BibitemShut {NoStop}%
\bibitem [{\citenamefont {Tinto}\ and\ \citenamefont
  {Dhurandhar}(2005)}]{Tinto:2004wu}%
  \BibitemOpen
  \bibfield  {author} {\bibinfo {author} {\bibfnamefont {M.}~\bibnamefont
  {Tinto}}\ and\ \bibinfo {author} {\bibfnamefont {S.~V.}\ \bibnamefont
  {Dhurandhar}},\ }\href {https://doi.org/10.12942/lrr-2005-4} {\bibfield
  {journal} {\bibinfo  {journal} {Living Rev. Rel.}\ }\textbf {\bibinfo
  {volume} {8}},\ \bibinfo {pages} {4} (\bibinfo {year} {2005})},\ \Eprint
  {https://arxiv.org/abs/gr-qc/0409034} {arXiv:gr-qc/0409034} \BibitemShut
  {NoStop}%
\bibitem [{\citenamefont {Cutler}\ and\ \citenamefont
  {Flanagan}(1994)}]{Cutler:1994ys}%
  \BibitemOpen
  \bibfield  {author} {\bibinfo {author} {\bibfnamefont {C.}~\bibnamefont
  {Cutler}}\ and\ \bibinfo {author} {\bibfnamefont {E.~E.}\ \bibnamefont
  {Flanagan}},\ }\href {https://doi.org/10.1103/PhysRevD.49.2658} {\bibfield
  {journal} {\bibinfo  {journal} {Phys. Rev. D}\ }\textbf {\bibinfo {volume}
  {49}},\ \bibinfo {pages} {2658} (\bibinfo {year} {1994})},\ \Eprint
  {https://arxiv.org/abs/gr-qc/9402014} {arXiv:gr-qc/9402014} \BibitemShut
  {NoStop}%
\bibitem [{\citenamefont {Vallisneri}(2008)}]{Vallisneri:2007ev}%
  \BibitemOpen
  \bibfield  {author} {\bibinfo {author} {\bibfnamefont {M.}~\bibnamefont
  {Vallisneri}},\ }\href {https://doi.org/10.1103/PhysRevD.77.042001}
  {\bibfield  {journal} {\bibinfo  {journal} {Phys. Rev. D}\ }\textbf {\bibinfo
  {volume} {77}},\ \bibinfo {pages} {042001} (\bibinfo {year} {2008})},\
  \Eprint {https://arxiv.org/abs/gr-qc/0703086} {arXiv:gr-qc/0703086}
  \BibitemShut {NoStop}%
\bibitem [{\citenamefont {{Amaro-Seoane}}\ \emph {et~al.}(2017)\citenamefont
  {{Amaro-Seoane}} \emph {et~al.}}]{2017arXiv170200786A}%
  \BibitemOpen
  \bibfield  {author} {\bibinfo {author} {\bibfnamefont {P.}~\bibnamefont
  {{Amaro-Seoane}}} \emph {et~al.},\ }\href
  {https://doi.org/10.48550/arXiv.1702.00786} {\bibfield  {journal} {\bibinfo
  {journal} {arXiv e-prints}\ ,\ \bibinfo {eid} {arXiv:1702.00786}} (\bibinfo
  {year} {2017})},\ \Eprint {https://arxiv.org/abs/1702.00786}
  {arXiv:1702.00786 [astro-ph.IM]} \BibitemShut {NoStop}%
\bibitem [{\citenamefont {Tinto}\ \emph {et~al.}(1994)\citenamefont {Tinto},
  \citenamefont {Giampieri}, \citenamefont {Hellings}, \citenamefont {Bender},\
  and\ \citenamefont {Faller}}]{Tinto:1994kg}%
  \BibitemOpen
  \bibfield  {author} {\bibinfo {author} {\bibfnamefont {M.}~\bibnamefont
  {Tinto}}, \bibinfo {author} {\bibfnamefont {G.}~\bibnamefont {Giampieri}},
  \bibinfo {author} {\bibfnamefont {R.~W.}\ \bibnamefont {Hellings}}, \bibinfo
  {author} {\bibfnamefont {P.~L.}\ \bibnamefont {Bender}},\ and\ \bibinfo
  {author} {\bibfnamefont {J.~E.}\ \bibnamefont {Faller}},\ }in\ \href@noop {}
  {\emph {\bibinfo {booktitle} {{7th Marcel Grossmann Meeting on General
  Relativity (MG 7)}}}}\ (\bibinfo {year} {1994})\ pp.\ \bibinfo {pages}
  {1668--1670}\BibitemShut {NoStop}%
\bibitem [{\citenamefont {Armstrong}\ \emph {et~al.}(1999)\citenamefont
  {Armstrong}, \citenamefont {Estabrook},\ and\ \citenamefont
  {Tinto}}]{Armstrong_1999}%
  \BibitemOpen
  \bibfield  {author} {\bibinfo {author} {\bibfnamefont {J.~W.}\ \bibnamefont
  {Armstrong}}, \bibinfo {author} {\bibfnamefont {F.~B.}\ \bibnamefont
  {Estabrook}},\ and\ \bibinfo {author} {\bibfnamefont {M.}~\bibnamefont
  {Tinto}},\ }\href {https://doi.org/10.1086/308110} {\bibfield  {journal}
  {\bibinfo  {journal} {The Astrophysical Journal}\ }\textbf {\bibinfo {volume}
  {527}},\ \bibinfo {pages} {814} (\bibinfo {year} {1999})}\BibitemShut
  {NoStop}%
\bibitem [{\citenamefont {Shaddock}\ \emph {et~al.}(2004)\citenamefont
  {Shaddock}, \citenamefont {Ware}, \citenamefont {Spero},\ and\ \citenamefont
  {Vallisneri}}]{Shaddock:2004ua}%
  \BibitemOpen
  \bibfield  {author} {\bibinfo {author} {\bibfnamefont {D.~A.}\ \bibnamefont
  {Shaddock}}, \bibinfo {author} {\bibfnamefont {B.}~\bibnamefont {Ware}},
  \bibinfo {author} {\bibfnamefont {R.~E.}\ \bibnamefont {Spero}},\ and\
  \bibinfo {author} {\bibfnamefont {M.}~\bibnamefont {Vallisneri}},\ }\href
  {https://doi.org/10.1103/PhysRevD.70.081101} {\bibfield  {journal} {\bibinfo
  {journal} {Phys. Rev. D}\ }\textbf {\bibinfo {volume} {70}},\ \bibinfo
  {pages} {081101} (\bibinfo {year} {2004})},\ \Eprint
  {https://arxiv.org/abs/gr-qc/0406106} {arXiv:gr-qc/0406106} \BibitemShut
  {NoStop}%
\bibitem [{\citenamefont {Shaddock}\ \emph {et~al.}(2003)\citenamefont
  {Shaddock}, \citenamefont {Tinto}, \citenamefont {Estabrook},\ and\
  \citenamefont {Armstrong}}]{Shaddock:2003dj}%
  \BibitemOpen
  \bibfield  {author} {\bibinfo {author} {\bibfnamefont {D.~A.}\ \bibnamefont
  {Shaddock}}, \bibinfo {author} {\bibfnamefont {M.}~\bibnamefont {Tinto}},
  \bibinfo {author} {\bibfnamefont {F.~B.}\ \bibnamefont {Estabrook}},\ and\
  \bibinfo {author} {\bibfnamefont {J.~W.}\ \bibnamefont {Armstrong}},\ }\href
  {https://doi.org/10.1103/PhysRevD.68.061303} {\bibfield  {journal} {\bibinfo
  {journal} {Phys. Rev. D}\ }\textbf {\bibinfo {volume} {68}},\ \bibinfo
  {pages} {061303} (\bibinfo {year} {2003})},\ \Eprint
  {https://arxiv.org/abs/gr-qc/0307080} {arXiv:gr-qc/0307080} \BibitemShut
  {NoStop}%
\bibitem [{\citenamefont {Vallisneri}(2005)}]{Vallisneri:2005ji}%
  \BibitemOpen
  \bibfield  {author} {\bibinfo {author} {\bibfnamefont {M.}~\bibnamefont
  {Vallisneri}},\ }\href {https://doi.org/10.1103/PhysRevD.76.109903}
  {\bibfield  {journal} {\bibinfo  {journal} {Phys. Rev. D}\ }\textbf {\bibinfo
  {volume} {72}},\ \bibinfo {pages} {042003} (\bibinfo {year} {2005})},\
  \bibinfo {note} {[Erratum: Phys.Rev.D 76, 109903 (2007)]},\ \Eprint
  {https://arxiv.org/abs/gr-qc/0504145} {arXiv:gr-qc/0504145} \BibitemShut
  {NoStop}%
\bibitem [{\citenamefont {Paczkowski}\ \emph {et~al.}(2022)\citenamefont
  {Paczkowski}, \citenamefont {Giusteri}, \citenamefont {Hewitson},
  \citenamefont {Karnesis}, \citenamefont {Fitzsimons}, \citenamefont
  {Wanner},\ and\ \citenamefont {Heinzel}}]{Paczkowski:2022nrt}%
  \BibitemOpen
  \bibfield  {author} {\bibinfo {author} {\bibfnamefont {S.}~\bibnamefont
  {Paczkowski}}, \bibinfo {author} {\bibfnamefont {R.}~\bibnamefont
  {Giusteri}}, \bibinfo {author} {\bibfnamefont {M.}~\bibnamefont {Hewitson}},
  \bibinfo {author} {\bibfnamefont {N.}~\bibnamefont {Karnesis}}, \bibinfo
  {author} {\bibfnamefont {E.~D.}\ \bibnamefont {Fitzsimons}}, \bibinfo
  {author} {\bibfnamefont {G.}~\bibnamefont {Wanner}},\ and\ \bibinfo {author}
  {\bibfnamefont {G.}~\bibnamefont {Heinzel}},\ }\href
  {https://doi.org/10.1103/PhysRevD.106.042005} {\bibfield  {journal} {\bibinfo
   {journal} {Phys. Rev. D}\ }\textbf {\bibinfo {volume} {106}},\ \bibinfo
  {pages} {042005} (\bibinfo {year} {2022})}\BibitemShut {NoStop}%
\bibitem [{\citenamefont {Hartig}\ and\ \citenamefont
  {Wanner}(2023)}]{Hartig:2023ofu}%
  \BibitemOpen
  \bibfield  {author} {\bibinfo {author} {\bibfnamefont {M.-S.}\ \bibnamefont
  {Hartig}}\ and\ \bibinfo {author} {\bibfnamefont {G.}~\bibnamefont
  {Wanner}},\ }\href {https://doi.org/10.1103/PhysRevD.108.022008} {\bibfield
  {journal} {\bibinfo  {journal} {Phys. Rev. D}\ }\textbf {\bibinfo {volume}
  {108}},\ \bibinfo {pages} {022008} (\bibinfo {year} {2023})},\ \Eprint
  {https://arxiv.org/abs/2305.03667} {arXiv:2305.03667 [gr-qc]} \BibitemShut
  {NoStop}%
\bibitem [{\citenamefont {Wanner}\ \emph {et~al.}(2024)\citenamefont {Wanner},
  \citenamefont {Shah}, \citenamefont {Staab}, \citenamefont {Wegener},\ and\
  \citenamefont {Paczkowski}}]{Wanner:2024eoa}%
  \BibitemOpen
  \bibfield  {author} {\bibinfo {author} {\bibfnamefont {G.}~\bibnamefont
  {Wanner}}, \bibinfo {author} {\bibfnamefont {S.}~\bibnamefont {Shah}},
  \bibinfo {author} {\bibfnamefont {M.}~\bibnamefont {Staab}}, \bibinfo
  {author} {\bibfnamefont {H.}~\bibnamefont {Wegener}},\ and\ \bibinfo {author}
  {\bibfnamefont {S.}~\bibnamefont {Paczkowski}},\ }\href
  {https://doi.org/10.1103/PhysRevD.110.022003} {\bibfield  {journal} {\bibinfo
   {journal} {Phys. Rev. D}\ }\textbf {\bibinfo {volume} {110}},\ \bibinfo
  {pages} {022003} (\bibinfo {year} {2024})},\ \Eprint
  {https://arxiv.org/abs/2403.06526} {arXiv:2403.06526 [astro-ph.IM]}
  \BibitemShut {NoStop}%
\bibitem [{\citenamefont {Hartig}\ \emph {et~al.}(2022)\citenamefont {Hartig},
  \citenamefont {Schuster},\ and\ \citenamefont {Wanner}}]{Hartig_2022}%
  \BibitemOpen
  \bibfield  {author} {\bibinfo {author} {\bibfnamefont {M.-S.}\ \bibnamefont
  {Hartig}}, \bibinfo {author} {\bibfnamefont {S.}~\bibnamefont {Schuster}},\
  and\ \bibinfo {author} {\bibfnamefont {G.}~\bibnamefont {Wanner}},\ }\href
  {https://doi.org/10.1088/2040-8986/ac675e} {\bibfield  {journal} {\bibinfo
  {journal} {Journal of Optics}\ }\textbf {\bibinfo {volume} {24}},\ \bibinfo
  {pages} {065601} (\bibinfo {year} {2022})}\BibitemShut {NoStop}%
\bibitem [{\citenamefont {Chwalla}\ \emph {et~al.}(2020)\citenamefont
  {Chwalla}, \citenamefont {Danzmann}, \citenamefont {\'Alvarez}, \citenamefont
  {Delgado}, \citenamefont {Fern\'andez~Barranco}, \citenamefont {Fitzsimons},
  \citenamefont {Gerberding}, \citenamefont {Heinzel}, \citenamefont {Killow},
  \citenamefont {Lieser}, \citenamefont {Perreur-Lloyd}, \citenamefont
  {Robertson}, \citenamefont {Rohr}, \citenamefont {Schuster}, \citenamefont
  {Schwarze}, \citenamefont {Tr\"obs}, \citenamefont {Wanner},\ and\
  \citenamefont {Ward}}]{PhysRevApplied.14.014030}%
  \BibitemOpen
  \bibfield  {author} {\bibinfo {author} {\bibfnamefont {M.}~\bibnamefont
  {Chwalla}}, \bibinfo {author} {\bibfnamefont {K.}~\bibnamefont {Danzmann}},
  \bibinfo {author} {\bibfnamefont {M.~D.}\ \bibnamefont {\'Alvarez}}, \bibinfo
  {author} {\bibfnamefont {J.~E.}\ \bibnamefont {Delgado}}, \bibinfo {author}
  {\bibfnamefont {G.}~\bibnamefont {Fern\'andez~Barranco}}, \bibinfo {author}
  {\bibfnamefont {E.}~\bibnamefont {Fitzsimons}}, \bibinfo {author}
  {\bibfnamefont {O.}~\bibnamefont {Gerberding}}, \bibinfo {author}
  {\bibfnamefont {G.}~\bibnamefont {Heinzel}}, \bibinfo {author} {\bibfnamefont
  {C.}~\bibnamefont {Killow}}, \bibinfo {author} {\bibfnamefont
  {M.}~\bibnamefont {Lieser}}, \bibinfo {author} {\bibfnamefont
  {M.}~\bibnamefont {Perreur-Lloyd}}, \bibinfo {author} {\bibfnamefont
  {D.}~\bibnamefont {Robertson}}, \bibinfo {author} {\bibfnamefont
  {J.}~\bibnamefont {Rohr}}, \bibinfo {author} {\bibfnamefont {S.}~\bibnamefont
  {Schuster}}, \bibinfo {author} {\bibfnamefont {T.}~\bibnamefont {Schwarze}},
  \bibinfo {author} {\bibfnamefont {M.}~\bibnamefont {Tr\"obs}}, \bibinfo
  {author} {\bibfnamefont {G.}~\bibnamefont {Wanner}},\ and\ \bibinfo {author}
  {\bibfnamefont {H.}~\bibnamefont {Ward}},\ }\href
  {https://doi.org/10.1103/PhysRevApplied.14.014030} {\bibfield  {journal}
  {\bibinfo  {journal} {Phys. Rev. Appl.}\ }\textbf {\bibinfo {volume} {14}},\
  \bibinfo {pages} {014030} (\bibinfo {year} {2020})}\BibitemShut {NoStop}%
\bibitem [{\citenamefont {Hartwig}\ and\ \citenamefont
  {Bayle}(2021)}]{PhysRevD.103.123027}%
  \BibitemOpen
  \bibfield  {author} {\bibinfo {author} {\bibfnamefont {O.}~\bibnamefont
  {Hartwig}}\ and\ \bibinfo {author} {\bibfnamefont {J.-B.}\ \bibnamefont
  {Bayle}},\ }\href {https://doi.org/10.1103/PhysRevD.103.123027} {\bibfield
  {journal} {\bibinfo  {journal} {Phys. Rev. D}\ }\textbf {\bibinfo {volume}
  {103}},\ \bibinfo {pages} {123027} (\bibinfo {year} {2021})}\BibitemShut
  {NoStop}%
\bibitem [{\citenamefont {Hartwig}\ and\ \citenamefont
  {Muratore}(2022)}]{PhysRevD.105.062006}%
  \BibitemOpen
  \bibfield  {author} {\bibinfo {author} {\bibfnamefont {O.}~\bibnamefont
  {Hartwig}}\ and\ \bibinfo {author} {\bibfnamefont {M.}~\bibnamefont
  {Muratore}},\ }\href {https://doi.org/10.1103/PhysRevD.105.062006} {\bibfield
   {journal} {\bibinfo  {journal} {Phys. Rev. D}\ }\textbf {\bibinfo {volume}
  {105}},\ \bibinfo {pages} {062006} (\bibinfo {year} {2022})}\BibitemShut
  {NoStop}%
\bibitem [{\citenamefont {Muratore}(2021)}]{Muratore:2021rwq}%
  \BibitemOpen
  \bibfield  {author} {\bibinfo {author} {\bibfnamefont {M.}~\bibnamefont
  {Muratore}},\ }\emph {\bibinfo {title} {{Time delay interferometry for LISA
  science and instrument characterization}}},\ \href
  {https://doi.org/10.15168/11572_312487} {Ph.D. thesis},\ \bibinfo  {school}
  {Trento U.} (\bibinfo {year} {2021})\BibitemShut {NoStop}%
\bibitem [{\citenamefont {Hartwig}(2021)}]{Hartwig:2021dlc}%
  \BibitemOpen
  \bibfield  {author} {\bibinfo {author} {\bibfnamefont {O.}~\bibnamefont
  {Hartwig}},\ }\emph {\bibinfo {title} {{Instrumental modelling and noise
  reduction algorithms for the Laser Interferometer Space Antenna}}},\ \href
  {https://doi.org/10.15488/11372} {Ph.D. thesis},\ \bibinfo  {school} {Leibniz
  U., Hannover} (\bibinfo {year} {2021})\BibitemShut {NoStop}%
\bibitem [{\citenamefont {Staab}(2023)}]{Staab:2023idg}%
  \BibitemOpen
  \bibfield  {author} {\bibinfo {author} {\bibfnamefont {M.~B.}\ \bibnamefont
  {Staab}},\ }\emph {\bibinfo {title} {{Time-delay interferometric ranging for
  LISA: Statistical analysis of bias-free ranging using laser noise
  minimization}}},\ \href {https://doi.org/10.15488/15739} {Ph.D. thesis},\
  \bibinfo  {school} {Leibniz U., Hannover} (\bibinfo {year}
  {2023})\BibitemShut {NoStop}%
\bibitem [{\citenamefont {Reinhardt}(2025)}]{Reinhardt:2025}%
  \BibitemOpen
  \bibfield  {author} {\bibinfo {author} {\bibfnamefont {J.~N.}\ \bibnamefont
  {Reinhardt}},\ }\emph {\bibinfo {title} {{Intersatellite clock
  synchronization and absolute ranging for space-based gravitational-wave
  detectors}}},\ \href {https://doi.org/10.15488/19219} {Ph.D. thesis},\
  \bibinfo  {school} {Leibniz U., Hannover} (\bibinfo {year}
  {2025})\BibitemShut {NoStop}%
\bibitem [{\citenamefont {{Muratore}}\ \emph {et~al.}(2023)\citenamefont
  {{Muratore}}, \citenamefont {{Hartwig}}, \citenamefont {{Vetrugno}},
  \citenamefont {{Vitale}},\ and\ \citenamefont
  {{Weber}}}]{2023PhRvD.107h2004M}%
  \BibitemOpen
  \bibfield  {author} {\bibinfo {author} {\bibfnamefont {M.}~\bibnamefont
  {{Muratore}}}, \bibinfo {author} {\bibfnamefont {O.}~\bibnamefont
  {{Hartwig}}}, \bibinfo {author} {\bibfnamefont {D.}~\bibnamefont
  {{Vetrugno}}}, \bibinfo {author} {\bibfnamefont {S.}~\bibnamefont
  {{Vitale}}},\ and\ \bibinfo {author} {\bibfnamefont {W.~J.}\ \bibnamefont
  {{Weber}}},\ }\href {https://doi.org/10.1103/PhysRevD.107.082004} {\bibfield
  {journal} {\bibinfo  {journal} {\prd}\ }\textbf {\bibinfo {volume} {107}},\
  \bibinfo {eid} {082004} (\bibinfo {year} {2023})}\BibitemShut {NoStop}%
\bibitem [{\citenamefont {Muratore}\ \emph {et~al.}(2020)\citenamefont
  {Muratore}, \citenamefont {Vetrugno},\ and\ \citenamefont
  {Vitale}}]{muratore2020revisitation}%
  \BibitemOpen
  \bibfield  {author} {\bibinfo {author} {\bibfnamefont {M.}~\bibnamefont
  {Muratore}}, \bibinfo {author} {\bibfnamefont {D.}~\bibnamefont {Vetrugno}},\
  and\ \bibinfo {author} {\bibfnamefont {S.}~\bibnamefont {Vitale}},\ }\href
  {https://api.semanticscholar.org/CorpusID:210966231} {\bibfield  {journal}
  {\bibinfo  {journal} {Classical and Quantum Gravity}\ }\textbf {\bibinfo
  {volume} {37}} (\bibinfo {year} {2020})}\BibitemShut {NoStop}%
\bibitem [{\citenamefont {{LISA Data Challenges Working Group}}(2022)}]{ldc}%
  \BibitemOpen
  \bibfield  {author} {\bibinfo {author} {\bibnamefont {{LISA Data Challenges
  Working Group}}},\ }\href {https://lisa-ldc.lal.in2p3.fr/} {\bibinfo {title}
  {{LISA Data Challenges}}} (\bibinfo {year} {2022})\BibitemShut {NoStop}%
\bibitem [{\citenamefont {{LISA Data Challenges Working
  Group}}({\natexlab{a}})}]{radler_doc}%
  \BibitemOpen
  \bibfield  {author} {\bibinfo {author} {\bibnamefont {{LISA Data Challenges
  Working Group}}},\ }\href {https://lisa-ldc.lal.in2p3.fr/challenge1a}
  {\bibinfo {title} {{Radler dataset description}}}
  ({\natexlab{a}})\BibitemShut {NoStop}%
\bibitem [{\citenamefont {{LISA Data Challenges Working
  Group}}({\natexlab{b}})}]{sangria_doc}%
  \BibitemOpen
  \bibfield  {author} {\bibinfo {author} {\bibnamefont {{LISA Data Challenges
  Working Group}}},\ }\href {https://lisa-ldc.lal.in2p3.fr/challenge2a}
  {\bibinfo {title} {{Sangria dataset description}}}
  ({\natexlab{b}})\BibitemShut {NoStop}%
\bibitem [{\citenamefont {{LISA Data Challenges Working
  Group}}({\natexlab{c}})}]{spritz_doc}%
  \BibitemOpen
  \bibfield  {author} {\bibinfo {author} {\bibnamefont {{LISA Data Challenges
  Working Group}}},\ }\href {https://lisa-ldc.lal.in2p3.fr/challenge2b}
  {\bibinfo {title} {{Spritz dataset description}}}
  ({\natexlab{c}})\BibitemShut {NoStop}%
\bibitem [{\citenamefont {Prince}\ \emph {et~al.}(2002)\citenamefont {Prince},
  \citenamefont {Tinto}, \citenamefont {Larson},\ and\ \citenamefont
  {Armstrong}}]{PhysRevD.66.122002}%
  \BibitemOpen
  \bibfield  {author} {\bibinfo {author} {\bibfnamefont {T.~A.}\ \bibnamefont
  {Prince}}, \bibinfo {author} {\bibfnamefont {M.}~\bibnamefont {Tinto}},
  \bibinfo {author} {\bibfnamefont {S.~L.}\ \bibnamefont {Larson}},\ and\
  \bibinfo {author} {\bibfnamefont {J.~W.}\ \bibnamefont {Armstrong}},\ }\href
  {https://doi.org/10.1103/PhysRevD.66.122002} {\bibfield  {journal} {\bibinfo
  {journal} {Phys. Rev. D}\ }\textbf {\bibinfo {volume} {66}},\ \bibinfo
  {pages} {122002} (\bibinfo {year} {2002})}\BibitemShut {NoStop}%
\bibitem [{\citenamefont {{Babak}}\ \emph {et~al.}(2021)\citenamefont
  {{Babak}}, \citenamefont {{Hewitson}},\ and\ \citenamefont
  {{Petiteau}}}]{2021arXiv210801167B}%
  \BibitemOpen
  \bibfield  {author} {\bibinfo {author} {\bibfnamefont {S.}~\bibnamefont
  {{Babak}}}, \bibinfo {author} {\bibfnamefont {M.}~\bibnamefont
  {{Hewitson}}},\ and\ \bibinfo {author} {\bibfnamefont {A.}~\bibnamefont
  {{Petiteau}}},\ }\href {https://doi.org/10.48550/arXiv.2108.01167} {\bibfield
   {journal} {\bibinfo  {journal} {arXiv e-prints}\ ,\ \bibinfo {eid}
  {arXiv:2108.01167}} (\bibinfo {year} {2021})},\ \Eprint
  {https://arxiv.org/abs/2108.01167} {arXiv:2108.01167 [astro-ph.IM]}
  \BibitemShut {NoStop}%
\bibitem [{\citenamefont {Caprini}\ and\ \citenamefont
  {Figueroa}(2018)}]{Caprini_2018}%
  \BibitemOpen
  \bibfield  {author} {\bibinfo {author} {\bibfnamefont {C.}~\bibnamefont
  {Caprini}}\ and\ \bibinfo {author} {\bibfnamefont {D.~G.}\ \bibnamefont
  {Figueroa}},\ }\href {https://doi.org/10.1088/1361-6382/aac608} {\bibfield
  {journal} {\bibinfo  {journal} {Classical and Quantum Gravity}\ }\textbf
  {\bibinfo {volume} {35}},\ \bibinfo {pages} {163001} (\bibinfo {year}
  {2018})}\BibitemShut {NoStop}%
\bibitem [{\citenamefont {{Aghanim}}\ \emph {et~al.}(2020)\citenamefont
  {{Aghanim}} \emph {et~al.}}]{2020A&A...641A...6P}%
  \BibitemOpen
  \bibfield  {author} {\bibinfo {author} {\bibfnamefont {N.}~\bibnamefont
  {{Aghanim}}} \emph {et~al.},\ }\href
  {https://doi.org/10.1051/0004-6361/201833910} {\bibfield  {journal} {\bibinfo
   {journal} {\aap}\ }\textbf {\bibinfo {volume} {641}},\ \bibinfo {eid} {A6}
  (\bibinfo {year} {2020})},\ \Eprint {https://arxiv.org/abs/1807.06209}
  {arXiv:1807.06209 [astro-ph.CO]} \BibitemShut {NoStop}%
\bibitem [{\citenamefont {{Lehoucq}}\ \emph {et~al.}(2023)\citenamefont
  {{Lehoucq}}, \citenamefont {{Dvorkin}}, \citenamefont {{Srinivasan}},
  \citenamefont {{Pellouin}},\ and\ \citenamefont
  {{Lamberts}}}]{2023MNRAS.526.4378L}%
  \BibitemOpen
  \bibfield  {author} {\bibinfo {author} {\bibfnamefont {L.}~\bibnamefont
  {{Lehoucq}}}, \bibinfo {author} {\bibfnamefont {I.}~\bibnamefont
  {{Dvorkin}}}, \bibinfo {author} {\bibfnamefont {R.}~\bibnamefont
  {{Srinivasan}}}, \bibinfo {author} {\bibfnamefont {C.}~\bibnamefont
  {{Pellouin}}},\ and\ \bibinfo {author} {\bibfnamefont {A.}~\bibnamefont
  {{Lamberts}}},\ }\href {https://doi.org/10.1093/mnras/stad2917} {\bibfield
  {journal} {\bibinfo  {journal} {\mnras}\ }\textbf {\bibinfo {volume} {526}},\
  \bibinfo {pages} {4378} (\bibinfo {year} {2023})},\ \Eprint
  {https://arxiv.org/abs/2306.09861} {arXiv:2306.09861 [astro-ph.HE]}
  \BibitemShut {NoStop}%
\bibitem [{\citenamefont {Binetruy}\ \emph {et~al.}(2012)\citenamefont
  {Binetruy}, \citenamefont {Bohe}, \citenamefont {Caprini},\ and\
  \citenamefont {Dufaux}}]{Binetruy:2012ze}%
  \BibitemOpen
  \bibfield  {author} {\bibinfo {author} {\bibfnamefont {P.}~\bibnamefont
  {Binetruy}}, \bibinfo {author} {\bibfnamefont {A.}~\bibnamefont {Bohe}},
  \bibinfo {author} {\bibfnamefont {C.}~\bibnamefont {Caprini}},\ and\ \bibinfo
  {author} {\bibfnamefont {J.-F.}\ \bibnamefont {Dufaux}},\ }\href
  {https://doi.org/10.1088/1475-7516/2012/06/027} {\bibfield  {journal}
  {\bibinfo  {journal} {JCAP}\ }\textbf {\bibinfo {volume} {06}},\ \bibinfo
  {pages} {027}},\ \Eprint {https://arxiv.org/abs/1201.0983} {arXiv:1201.0983
  [gr-qc]} \BibitemShut {NoStop}%
\bibitem [{Note1()}]{Note1}%
  \BibitemOpen
  \bibinfo {note} {\protect \url
  {https://github.com/martinaAEI/noise_knowledge_uncertainty}}\BibitemShut
  {NoStop}%
\bibitem [{\citenamefont {Thrane}\ and\ \citenamefont
  {Romano}(2013)}]{Thrane:2013oya}%
  \BibitemOpen
  \bibfield  {author} {\bibinfo {author} {\bibfnamefont {E.}~\bibnamefont
  {Thrane}}\ and\ \bibinfo {author} {\bibfnamefont {J.~D.}\ \bibnamefont
  {Romano}},\ }\href {https://doi.org/10.1103/PhysRevD.88.124032} {\bibfield
  {journal} {\bibinfo  {journal} {Phys. Rev. D}\ }\textbf {\bibinfo {volume}
  {88}},\ \bibinfo {pages} {124032} (\bibinfo {year} {2013})},\ \Eprint
  {https://arxiv.org/abs/1310.5300} {arXiv:1310.5300 [astro-ph.IM]}
  \BibitemShut {NoStop}%
\bibitem [{\citenamefont {{Janssen}}\ \emph {et~al.}(2015)\citenamefont
  {{Janssen}}, \citenamefont {{Hobbs}}, \citenamefont {{McLaughlin}},
  \citenamefont {{Bassa}}, \citenamefont {{Deller}}, \citenamefont {{Kramer}},
  \citenamefont {{Lee}}, \citenamefont {{Mingarelli}}, \citenamefont
  {{Rosado}}, \citenamefont {{Sanidas}}, \citenamefont {{Sesana}},
  \citenamefont {{Shao}}, \citenamefont {{Stairs}}, \citenamefont
  {{Stappers}},\ and\ \citenamefont {{Verbiest}}}]{2015aska.confE..37J}%
  \BibitemOpen
  \bibfield  {author} {\bibinfo {author} {\bibfnamefont {G.}~\bibnamefont
  {{Janssen}}}, \bibinfo {author} {\bibfnamefont {G.}~\bibnamefont {{Hobbs}}},
  \bibinfo {author} {\bibfnamefont {M.}~\bibnamefont {{McLaughlin}}}, \bibinfo
  {author} {\bibfnamefont {C.}~\bibnamefont {{Bassa}}}, \bibinfo {author}
  {\bibfnamefont {A.}~\bibnamefont {{Deller}}}, \bibinfo {author}
  {\bibfnamefont {M.}~\bibnamefont {{Kramer}}}, \bibinfo {author}
  {\bibfnamefont {K.}~\bibnamefont {{Lee}}}, \bibinfo {author} {\bibfnamefont
  {C.}~\bibnamefont {{Mingarelli}}}, \bibinfo {author} {\bibfnamefont
  {P.}~\bibnamefont {{Rosado}}}, \bibinfo {author} {\bibfnamefont
  {S.}~\bibnamefont {{Sanidas}}}, \bibinfo {author} {\bibfnamefont
  {A.}~\bibnamefont {{Sesana}}}, \bibinfo {author} {\bibfnamefont
  {L.}~\bibnamefont {{Shao}}}, \bibinfo {author} {\bibfnamefont
  {I.}~\bibnamefont {{Stairs}}}, \bibinfo {author} {\bibfnamefont
  {B.}~\bibnamefont {{Stappers}}},\ and\ \bibinfo {author} {\bibfnamefont
  {J.~P.~W.}\ \bibnamefont {{Verbiest}}},\ }in\ \href
  {https://doi.org/10.22323/1.215.0037} {\emph {\bibinfo {booktitle} {Advancing
  Astrophysics with the Square Kilometre Array (AASKA14)}}}\ (\bibinfo {year}
  {2015})\ p.~\bibinfo {pages} {37},\ \Eprint
  {https://arxiv.org/abs/1501.00127} {arXiv:1501.00127 [astro-ph.IM]}
  \BibitemShut {NoStop}%
\bibitem [{\citenamefont {{Edwards}}\ \emph {et~al.}(2020)\citenamefont
  {{Edwards}}, \citenamefont {{Maturana-Russel}}, \citenamefont {{Meyer}},
  \citenamefont {{Gair}}, \citenamefont {{Korsakova}},\ and\ \citenamefont
  {{Christensen}}}]{2020PhRvD.102h4062E}%
  \BibitemOpen
  \bibfield  {author} {\bibinfo {author} {\bibfnamefont {M.~C.}\ \bibnamefont
  {{Edwards}}}, \bibinfo {author} {\bibfnamefont {P.}~\bibnamefont
  {{Maturana-Russel}}}, \bibinfo {author} {\bibfnamefont {R.}~\bibnamefont
  {{Meyer}}}, \bibinfo {author} {\bibfnamefont {J.}~\bibnamefont {{Gair}}},
  \bibinfo {author} {\bibfnamefont {N.}~\bibnamefont {{Korsakova}}},\ and\
  \bibinfo {author} {\bibfnamefont {N.}~\bibnamefont {{Christensen}}},\ }\href
  {https://doi.org/10.1103/PhysRevD.102.084062} {\bibfield  {journal} {\bibinfo
   {journal} {\prd}\ }\textbf {\bibinfo {volume} {102}},\ \bibinfo {eid}
  {084062} (\bibinfo {year} {2020})},\ \Eprint
  {https://arxiv.org/abs/2004.07515} {arXiv:2004.07515 [gr-qc]} \BibitemShut
  {NoStop}%
\bibitem [{\citenamefont {Quang~Nam}\ \emph {et~al.}(2023)\citenamefont
  {Quang~Nam}, \citenamefont {Lemi\`ere}, \citenamefont {Petiteau},
  \citenamefont {Bayle}, \citenamefont {Hartwig}, \citenamefont {Martino},\
  and\ \citenamefont {Staab}}]{QuangNam:2022gjz}%
  \BibitemOpen
  \bibfield  {author} {\bibinfo {author} {\bibfnamefont {D.}~\bibnamefont
  {Quang~Nam}}, \bibinfo {author} {\bibfnamefont {Y.}~\bibnamefont
  {Lemi\`ere}}, \bibinfo {author} {\bibfnamefont {A.}~\bibnamefont {Petiteau}},
  \bibinfo {author} {\bibfnamefont {J.-B.}\ \bibnamefont {Bayle}}, \bibinfo
  {author} {\bibfnamefont {O.}~\bibnamefont {Hartwig}}, \bibinfo {author}
  {\bibfnamefont {J.}~\bibnamefont {Martino}},\ and\ \bibinfo {author}
  {\bibfnamefont {M.}~\bibnamefont {Staab}},\ }\href
  {https://doi.org/10.1103/PhysRevD.108.082004} {\bibfield  {journal} {\bibinfo
   {journal} {Phys. Rev. D}\ }\textbf {\bibinfo {volume} {108}},\ \bibinfo
  {pages} {082004} (\bibinfo {year} {2023})},\ \Eprint
  {https://arxiv.org/abs/2211.02539} {arXiv:2211.02539 [gr-qc]} \BibitemShut
  {NoStop}%
\bibitem [{\citenamefont {{Quang Nam}}\ \emph {et~al.}(2023)\citenamefont
  {{Quang Nam}}, \citenamefont {{Martino}}, \citenamefont {{Lemi{\`e}re}},
  \citenamefont {{Petiteau}}, \citenamefont {{Bayle}}, \citenamefont
  {{Hartwig}},\ and\ \citenamefont {{Staab}}}]{2023PhRvD.108h2004Q}%
  \BibitemOpen
  \bibfield  {author} {\bibinfo {author} {\bibfnamefont {D.}~\bibnamefont
  {{Quang Nam}}}, \bibinfo {author} {\bibfnamefont {J.}~\bibnamefont
  {{Martino}}}, \bibinfo {author} {\bibfnamefont {Y.}~\bibnamefont
  {{Lemi{\`e}re}}}, \bibinfo {author} {\bibfnamefont {A.}~\bibnamefont
  {{Petiteau}}}, \bibinfo {author} {\bibfnamefont {J.-B.}\ \bibnamefont
  {{Bayle}}}, \bibinfo {author} {\bibfnamefont {O.}~\bibnamefont {{Hartwig}}},\
  and\ \bibinfo {author} {\bibfnamefont {M.}~\bibnamefont {{Staab}}},\ }\href
  {https://doi.org/10.1103/PhysRevD.108.082004} {\bibfield  {journal} {\bibinfo
   {journal} {\prd}\ }\textbf {\bibinfo {volume} {108}},\ \bibinfo {eid}
  {082004} (\bibinfo {year} {2023})},\ \Eprint
  {https://arxiv.org/abs/2211.02539} {arXiv:2211.02539 [gr-qc]} \BibitemShut
  {NoStop}%
\bibitem [{\citenamefont {{Hartwig}}\ \emph {et~al.}(2023)\citenamefont
  {{Hartwig}}, \citenamefont {{Lilley}}, \citenamefont {{Muratore}},\ and\
  \citenamefont {{Pieroni}}}]{2023PhRvD.107l3531H}%
  \BibitemOpen
  \bibfield  {author} {\bibinfo {author} {\bibfnamefont {O.}~\bibnamefont
  {{Hartwig}}}, \bibinfo {author} {\bibfnamefont {M.}~\bibnamefont {{Lilley}}},
  \bibinfo {author} {\bibfnamefont {M.}~\bibnamefont {{Muratore}}},\ and\
  \bibinfo {author} {\bibfnamefont {M.}~\bibnamefont {{Pieroni}}},\ }\href
  {https://doi.org/10.1103/PhysRevD.107.123531} {\bibfield  {journal} {\bibinfo
   {journal} {\prd}\ }\textbf {\bibinfo {volume} {107}},\ \bibinfo {eid}
  {123531} (\bibinfo {year} {2023})},\ \Eprint
  {https://arxiv.org/abs/2303.15929} {arXiv:2303.15929 [gr-qc]} \BibitemShut
  {NoStop}%
\bibitem [{\citenamefont {Akima}(1970)}]{10.1145/321607.321609}%
  \BibitemOpen
  \bibfield  {author} {\bibinfo {author} {\bibfnamefont {H.}~\bibnamefont
  {Akima}},\ }\href {https://doi.org/10.1145/321607.321609} {\bibfield
  {journal} {\bibinfo  {journal} {J. ACM}\ }\textbf {\bibinfo {volume} {17}},\
  \bibinfo {pages} {589–602} (\bibinfo {year} {1970})}\BibitemShut {NoStop}%
\bibitem [{\citenamefont {{Virtanen, Pauli and
  others}}(2020)}]{2020SciPy-NMeth}%
  \BibitemOpen
  \bibfield  {author} {\bibinfo {author} {\bibnamefont {{Virtanen, Pauli and
  others}}},\ }\href {https://doi.org/10.1038/s41592-019-0686-2} {\bibfield
  {journal} {\bibinfo  {journal} {Nature Methods}\ }\textbf {\bibinfo {volume}
  {17}},\ \bibinfo {pages} {261} (\bibinfo {year} {2020})}\BibitemShut
  {NoStop}%
\bibitem [{\citenamefont {Katz}\ \emph {et~al.}(2021)\citenamefont {Katz},
  \citenamefont {Chua}, \citenamefont {Speri}, \citenamefont {Warburton},\ and\
  \citenamefont {Hughes}}]{Katz:2021yft}%
  \BibitemOpen
  \bibfield  {author} {\bibinfo {author} {\bibfnamefont {M.~L.}\ \bibnamefont
  {Katz}}, \bibinfo {author} {\bibfnamefont {A.~J.~K.}\ \bibnamefont {Chua}},
  \bibinfo {author} {\bibfnamefont {L.}~\bibnamefont {Speri}}, \bibinfo
  {author} {\bibfnamefont {N.}~\bibnamefont {Warburton}},\ and\ \bibinfo
  {author} {\bibfnamefont {S.~A.}\ \bibnamefont {Hughes}},\ }\href
  {https://doi.org/10.1103/PhysRevD.104.064047} {\bibfield  {journal} {\bibinfo
   {journal} {Phys. Rev. D}\ }\textbf {\bibinfo {volume} {104}},\ \bibinfo
  {pages} {064047} (\bibinfo {year} {2021})},\ \Eprint
  {https://arxiv.org/abs/2104.04582} {arXiv:2104.04582 [gr-qc]} \BibitemShut
  {NoStop}%
\bibitem [{\citenamefont {Chapman-Bird}\ \emph {et~al.}(2025)\citenamefont
  {Chapman-Bird} \emph {et~al.}}]{Chapman-Bird:2025xtd}%
  \BibitemOpen
  \bibfield  {author} {\bibinfo {author} {\bibfnamefont {C.~E.~A.}\
  \bibnamefont {Chapman-Bird}} \emph {et~al.},\ }\href@noop {} {\bibfield
  {journal} {\bibinfo  {journal} {ArXiv preprint}\ } (\bibinfo {year}
  {2025})},\ \Eprint {https://arxiv.org/abs/2506.09470} {arXiv:2506.09470
  [gr-qc]} \BibitemShut {NoStop}%
\bibitem [{\citenamefont {{Green}}(1995)}]{10.1093/biomet/82.4.711}%
  \BibitemOpen
  \bibfield  {author} {\bibinfo {author} {\bibfnamefont {P.~J.}\ \bibnamefont
  {{Green}}},\ }\href {https://doi.org/10.1093/biomet/82.4.711} {\bibfield
  {journal} {\bibinfo  {journal} {Biometrika}\ }\textbf {\bibinfo {volume}
  {82}},\ \bibinfo {pages} {711} (\bibinfo {year} {1995})},\ \Eprint
  {https://arxiv.org/abs/https://academic.oup.com/biomet/article-pdf/82/4/711/699533/82-4-711.pdf}
  {https://academic.oup.com/biomet/article-pdf/82/4/711/699533/82-4-711.pdf}
  \BibitemShut {NoStop}%
\bibitem [{\citenamefont {{Foreman-Mackey}}\ \emph {et~al.}(2013)\citenamefont
  {{Foreman-Mackey}}, \citenamefont {{Hogg}}, \citenamefont {{Lang}},\ and\
  \citenamefont {{Goodman}}}]{2013PASP..125..306F}%
  \BibitemOpen
  \bibfield  {author} {\bibinfo {author} {\bibfnamefont {D.}~\bibnamefont
  {{Foreman-Mackey}}}, \bibinfo {author} {\bibfnamefont {D.~W.}\ \bibnamefont
  {{Hogg}}}, \bibinfo {author} {\bibfnamefont {D.}~\bibnamefont {{Lang}}},\
  and\ \bibinfo {author} {\bibfnamefont {J.}~\bibnamefont {{Goodman}}},\ }\href
  {https://doi.org/10.1086/670067} {\bibfield  {journal} {\bibinfo  {journal}
  {\pasp}\ }\textbf {\bibinfo {volume} {125}},\ \bibinfo {pages} {306}
  (\bibinfo {year} {2013})},\ \Eprint {https://arxiv.org/abs/1202.3665}
  {arXiv:1202.3665 [astro-ph.IM]} \BibitemShut {NoStop}%
\bibitem [{\citenamefont {{Karnesis}}\ \emph {et~al.}(2023)\citenamefont
  {{Karnesis}}, \citenamefont {{Katz}}, \citenamefont {{Korsakova}},
  \citenamefont {{Gair}},\ and\ \citenamefont
  {{Stergioulas}}}]{2023MNRAS.526.4814K}%
  \BibitemOpen
  \bibfield  {author} {\bibinfo {author} {\bibfnamefont {N.}~\bibnamefont
  {{Karnesis}}}, \bibinfo {author} {\bibfnamefont {M.~L.}\ \bibnamefont
  {{Katz}}}, \bibinfo {author} {\bibfnamefont {N.}~\bibnamefont {{Korsakova}}},
  \bibinfo {author} {\bibfnamefont {J.~R.}\ \bibnamefont {{Gair}}},\ and\
  \bibinfo {author} {\bibfnamefont {N.}~\bibnamefont {{Stergioulas}}},\ }\href
  {https://doi.org/10.1093/mnras/stad2939} {\bibfield  {journal} {\bibinfo
  {journal} {\mnras}\ }\textbf {\bibinfo {volume} {526}},\ \bibinfo {pages}
  {4814} (\bibinfo {year} {2023})},\ \Eprint {https://arxiv.org/abs/2303.02164}
  {arXiv:2303.02164 [astro-ph.IM]} \BibitemShut {NoStop}%
\bibitem [{\citenamefont {Katz}\ \emph {et~al.}(2023)\citenamefont {Katz},
  \citenamefont {Karnesis},\ and\ \citenamefont
  {Korsakova}}]{michael_katz_2023_7705496}%
  \BibitemOpen
  \bibfield  {author} {\bibinfo {author} {\bibfnamefont {M.}~\bibnamefont
  {Katz}}, \bibinfo {author} {\bibfnamefont {N.}~\bibnamefont {Karnesis}},\
  and\ \bibinfo {author} {\bibfnamefont {N.}~\bibnamefont {Korsakova}},\ }\href
  {https://doi.org/10.5281/zenodo.7705496} {\bibinfo {title} {mikekatz04/eryn:
  first full release}} (\bibinfo {year} {2023})\BibitemShut {NoStop}%
\bibitem [{\citenamefont {Muratore}\ \emph {et~al.}(2025)\citenamefont
  {Muratore}, \citenamefont {Gair}, \citenamefont {Hartwig}, \citenamefont
  {Katz},\ and\ \citenamefont
  {Toubiana}}]{muratore2025pipelinesearchingfittinginstrumental}%
  \BibitemOpen
  \bibfield  {author} {\bibinfo {author} {\bibfnamefont {M.}~\bibnamefont
  {Muratore}}, \bibinfo {author} {\bibfnamefont {J.}~\bibnamefont {Gair}},
  \bibinfo {author} {\bibfnamefont {O.}~\bibnamefont {Hartwig}}, \bibinfo
  {author} {\bibfnamefont {M.~L.}\ \bibnamefont {Katz}},\ and\ \bibinfo
  {author} {\bibfnamefont {A.}~\bibnamefont {Toubiana}},\ }\href
  {https://arxiv.org/abs/2505.19870} {\bibinfo {title} {A pipeline for
  searching and fitting instrumental glitches in lisa data}} (\bibinfo {year}
  {2025}),\ \Eprint {https://arxiv.org/abs/2505.19870} {arXiv:2505.19870
  [gr-qc]} \BibitemShut {NoStop}%
\bibitem [{\citenamefont
  {Santini}(2024{\natexlab{a}})}]{cudakima_2024_13919394}%
  \BibitemOpen
  \bibfield  {author} {\bibinfo {author} {\bibfnamefont {A.}~\bibnamefont
  {Santini}},\ }\href {https://doi.org/10.5281/zenodo.13919394} {\bibinfo
  {title} {asantini29/cudakima: First official release}} (\bibinfo {year}
  {2024}{\natexlab{a}})\BibitemShut {NoStop}%
\bibitem [{\citenamefont {Okuta}\ \emph {et~al.}(2017)\citenamefont {Okuta},
  \citenamefont {Unno}, \citenamefont {Nishino}, \citenamefont {Hido},\ and\
  \citenamefont {Loomis}}]{cupy_learningsys2017}%
  \BibitemOpen
  \bibfield  {author} {\bibinfo {author} {\bibfnamefont {R.}~\bibnamefont
  {Okuta}}, \bibinfo {author} {\bibfnamefont {Y.}~\bibnamefont {Unno}},
  \bibinfo {author} {\bibfnamefont {D.}~\bibnamefont {Nishino}}, \bibinfo
  {author} {\bibfnamefont {S.}~\bibnamefont {Hido}},\ and\ \bibinfo {author}
  {\bibfnamefont {C.}~\bibnamefont {Loomis}},\ }in\ \href
  {http://learningsys.org/nips17/assets/papers/paper_16.pdf} {\emph {\bibinfo
  {booktitle} {Proceedings of Workshop on Machine Learning Systems
  (LearningSys) in The Thirty-first Annual Conference on Neural Information
  Processing Systems (NIPS)}}}\ (\bibinfo {year} {2017})\BibitemShut {NoStop}%
\bibitem [{\citenamefont {Bradbury}\ \emph {et~al.}(2018)\citenamefont
  {Bradbury}, \citenamefont {Frostig}, \citenamefont {Hawkins}, \citenamefont
  {Johnson}, \citenamefont {Leary}, \citenamefont {Maclaurin}, \citenamefont
  {Necula}, \citenamefont {Paszke}, \citenamefont {Vander{P}las}, \citenamefont
  {Wanderman-{M}ilne},\ and\ \citenamefont {Zhang}}]{jax2018github}%
  \BibitemOpen
  \bibfield  {author} {\bibinfo {author} {\bibfnamefont {J.}~\bibnamefont
  {Bradbury}}, \bibinfo {author} {\bibfnamefont {R.}~\bibnamefont {Frostig}},
  \bibinfo {author} {\bibfnamefont {P.}~\bibnamefont {Hawkins}}, \bibinfo
  {author} {\bibfnamefont {M.~J.}\ \bibnamefont {Johnson}}, \bibinfo {author}
  {\bibfnamefont {C.}~\bibnamefont {Leary}}, \bibinfo {author} {\bibfnamefont
  {D.}~\bibnamefont {Maclaurin}}, \bibinfo {author} {\bibfnamefont
  {G.}~\bibnamefont {Necula}}, \bibinfo {author} {\bibfnamefont
  {A.}~\bibnamefont {Paszke}}, \bibinfo {author} {\bibfnamefont
  {J.}~\bibnamefont {Vander{P}las}}, \bibinfo {author} {\bibfnamefont
  {S.}~\bibnamefont {Wanderman-{M}ilne}},\ and\ \bibinfo {author}
  {\bibfnamefont {Q.}~\bibnamefont {Zhang}},\ }\href
  {http://github.com/google/jax} {\bibinfo {title} {{JAX}: composable
  transformations of {P}ython+{N}um{P}y programs}} (\bibinfo {year}
  {2018})\BibitemShut {NoStop}%
\bibitem [{\citenamefont {{Goodman}}\ and\ \citenamefont
  {{Weare}}(2010)}]{2010CAMCS...5...65G}%
  \BibitemOpen
  \bibfield  {author} {\bibinfo {author} {\bibfnamefont {J.}~\bibnamefont
  {{Goodman}}}\ and\ \bibinfo {author} {\bibfnamefont {J.}~\bibnamefont
  {{Weare}}},\ }\href {https://doi.org/10.2140/camcos.2010.5.65} {\bibfield
  {journal} {\bibinfo  {journal} {Communications in Applied Mathematics and
  Computational Science}\ }\textbf {\bibinfo {volume} {5}},\ \bibinfo {pages}
  {65} (\bibinfo {year} {2010})}\BibitemShut {NoStop}%
\bibitem [{\citenamefont {Xie}\ \emph {et~al.}(2010)\citenamefont {Xie},
  \citenamefont {Lewis}, \citenamefont {Fan}, \citenamefont {Kuo},\ and\
  \citenamefont {Chen}}]{10.1093/sysbio/syq085}%
  \BibitemOpen
  \bibfield  {author} {\bibinfo {author} {\bibfnamefont {W.}~\bibnamefont
  {Xie}}, \bibinfo {author} {\bibfnamefont {P.~O.}\ \bibnamefont {Lewis}},
  \bibinfo {author} {\bibfnamefont {Y.}~\bibnamefont {Fan}}, \bibinfo {author}
  {\bibfnamefont {L.}~\bibnamefont {Kuo}},\ and\ \bibinfo {author}
  {\bibfnamefont {M.-H.}\ \bibnamefont {Chen}},\ }\href
  {https://doi.org/10.1093/sysbio/syq085} {\bibfield  {journal} {\bibinfo
  {journal} {Systematic Biology}\ }\textbf {\bibinfo {volume} {60}},\ \bibinfo
  {pages} {150} (\bibinfo {year} {2010})},\ \Eprint
  {https://arxiv.org/abs/https://academic.oup.com/sysbio/article-pdf/60/2/150/24552358/syq085.pdf}
  {https://academic.oup.com/sysbio/article-pdf/60/2/150/24552358/syq085.pdf}
  \BibitemShut {NoStop}%
\bibitem [{\citenamefont {{Del Pozzo}}\ \emph {et~al.}(2011)\citenamefont {{Del
  Pozzo}}, \citenamefont {{Veitch}},\ and\ \citenamefont
  {{Vecchio}}}]{2011PhRvD..83h2002D}%
  \BibitemOpen
  \bibfield  {author} {\bibinfo {author} {\bibfnamefont {W.}~\bibnamefont {{Del
  Pozzo}}}, \bibinfo {author} {\bibfnamefont {J.}~\bibnamefont {{Veitch}}},\
  and\ \bibinfo {author} {\bibfnamefont {A.}~\bibnamefont {{Vecchio}}},\ }\href
  {https://doi.org/10.1103/PhysRevD.83.082002} {\bibfield  {journal} {\bibinfo
  {journal} {\prd}\ }\textbf {\bibinfo {volume} {83}},\ \bibinfo {eid} {082002}
  (\bibinfo {year} {2011})},\ \Eprint {https://arxiv.org/abs/1101.1391}
  {arXiv:1101.1391 [gr-qc]} \BibitemShut {NoStop}%
\bibitem [{\citenamefont {{Toubiana}}\ \emph {et~al.}(2021)\citenamefont
  {{Toubiana}}, \citenamefont {{Wong}}, \citenamefont {{Babak}}, \citenamefont
  {{Barausse}}, \citenamefont {{Berti}}, \citenamefont {{Gair}}, \citenamefont
  {{Marsat}},\ and\ \citenamefont {{Taylor}}}]{2021PhRvD.104h3027T}%
  \BibitemOpen
  \bibfield  {author} {\bibinfo {author} {\bibfnamefont {A.}~\bibnamefont
  {{Toubiana}}}, \bibinfo {author} {\bibfnamefont {K.~W.~K.}\ \bibnamefont
  {{Wong}}}, \bibinfo {author} {\bibfnamefont {S.}~\bibnamefont {{Babak}}},
  \bibinfo {author} {\bibfnamefont {E.}~\bibnamefont {{Barausse}}}, \bibinfo
  {author} {\bibfnamefont {E.}~\bibnamefont {{Berti}}}, \bibinfo {author}
  {\bibfnamefont {J.~R.}\ \bibnamefont {{Gair}}}, \bibinfo {author}
  {\bibfnamefont {S.}~\bibnamefont {{Marsat}}},\ and\ \bibinfo {author}
  {\bibfnamefont {S.~R.}\ \bibnamefont {{Taylor}}},\ }\href
  {https://doi.org/10.1103/PhysRevD.104.083027} {\bibfield  {journal} {\bibinfo
   {journal} {\prd}\ }\textbf {\bibinfo {volume} {104}},\ \bibinfo {eid}
  {083027} (\bibinfo {year} {2021})},\ \Eprint
  {https://arxiv.org/abs/2106.13819} {arXiv:2106.13819 [gr-qc]} \BibitemShut
  {NoStop}%
\bibitem [{\citenamefont {{Toubiana}}\ \emph {et~al.}(2024)\citenamefont
  {{Toubiana}}, \citenamefont {{Pompili}}, \citenamefont {{Buonanno}},
  \citenamefont {{Gair}},\ and\ \citenamefont {{Katz}}}]{2024PhRvD.109j4019T}%
  \BibitemOpen
  \bibfield  {author} {\bibinfo {author} {\bibfnamefont {A.}~\bibnamefont
  {{Toubiana}}}, \bibinfo {author} {\bibfnamefont {L.}~\bibnamefont
  {{Pompili}}}, \bibinfo {author} {\bibfnamefont {A.}~\bibnamefont
  {{Buonanno}}}, \bibinfo {author} {\bibfnamefont {J.~R.}\ \bibnamefont
  {{Gair}}},\ and\ \bibinfo {author} {\bibfnamefont {M.~L.}\ \bibnamefont
  {{Katz}}},\ }\href {https://doi.org/10.1103/PhysRevD.109.104019} {\bibfield
  {journal} {\bibinfo  {journal} {\prd}\ }\textbf {\bibinfo {volume} {109}},\
  \bibinfo {eid} {104019} (\bibinfo {year} {2024})},\ \Eprint
  {https://arxiv.org/abs/2307.15086} {arXiv:2307.15086 [gr-qc]} \BibitemShut
  {NoStop}%
\bibitem [{\citenamefont {Maturana-Russel}\ \emph {et~al.}(2019)\citenamefont
  {Maturana-Russel}, \citenamefont {Meyer}, \citenamefont {Veitch},\ and\
  \citenamefont {Christensen}}]{Maturana-Russel:2018yos}%
  \BibitemOpen
  \bibfield  {author} {\bibinfo {author} {\bibfnamefont {P.}~\bibnamefont
  {Maturana-Russel}}, \bibinfo {author} {\bibfnamefont {R.}~\bibnamefont
  {Meyer}}, \bibinfo {author} {\bibfnamefont {J.}~\bibnamefont {Veitch}},\ and\
  \bibinfo {author} {\bibfnamefont {N.}~\bibnamefont {Christensen}},\ }\href
  {https://doi.org/10.1103/PhysRevD.99.084006} {\bibfield  {journal} {\bibinfo
  {journal} {Phys. Rev. D}\ }\textbf {\bibinfo {volume} {99}},\ \bibinfo
  {pages} {084006} (\bibinfo {year} {2019})},\ \Eprint
  {https://arxiv.org/abs/1810.04488} {arXiv:1810.04488 [physics.data-an]}
  \BibitemShut {NoStop}%
\bibitem [{\citenamefont {{Zahraoui}}\ \emph {et~al.}(2025)\citenamefont
  {{Zahraoui}}, \citenamefont {{Maturana-Russel}}, \citenamefont {{van
  Straten}}, \citenamefont {{Meyer}},\ and\ \citenamefont
  {{Gulyaev}}}]{2025MNRAS.540.3818Z}%
  \BibitemOpen
  \bibfield  {author} {\bibinfo {author} {\bibfnamefont {E.~M.}\ \bibnamefont
  {{Zahraoui}}}, \bibinfo {author} {\bibfnamefont {P.}~\bibnamefont
  {{Maturana-Russel}}}, \bibinfo {author} {\bibfnamefont {W.}~\bibnamefont
  {{van Straten}}}, \bibinfo {author} {\bibfnamefont {R.}~\bibnamefont
  {{Meyer}}},\ and\ \bibinfo {author} {\bibfnamefont {S.}~\bibnamefont
  {{Gulyaev}}},\ }\href {https://doi.org/10.1093/mnras/staf953} {\bibfield
  {journal} {\bibinfo  {journal} {\mnras}\ }\textbf {\bibinfo {volume} {540}},\
  \bibinfo {pages} {3818} (\bibinfo {year} {2025})},\ \Eprint
  {https://arxiv.org/abs/2411.14736} {arXiv:2411.14736 [astro-ph.IM]}
  \BibitemShut {NoStop}%
\bibitem [{\citenamefont
  {Whittle}(1953)}]{faaafff7-3364-33b0-82ea-e107a78593d3}%
  \BibitemOpen
  \bibfield  {author} {\bibinfo {author} {\bibfnamefont {P.}~\bibnamefont
  {Whittle}},\ }\href {http://www.jstor.org/stable/2983728} {\bibfield
  {journal} {\bibinfo  {journal} {Journal of the Royal Statistical Society.
  Series B (Methodological)}\ }\textbf {\bibinfo {volume} {15}},\ \bibinfo
  {pages} {125} (\bibinfo {year} {1953})}\BibitemShut {NoStop}%
\bibitem [{\citenamefont {Franciolini}\ \emph {et~al.}()\citenamefont
  {Franciolini}, \citenamefont {Pieroni}, \citenamefont {Ricciardone},\ and\
  \citenamefont {Romano}}]{Franciolini:2025leq}%
  \BibitemOpen
  \bibfield  {author} {\bibinfo {author} {\bibfnamefont {G.}~\bibnamefont
  {Franciolini}}, \bibinfo {author} {\bibfnamefont {M.}~\bibnamefont
  {Pieroni}}, \bibinfo {author} {\bibfnamefont {A.}~\bibnamefont
  {Ricciardone}},\ and\ \bibinfo {author} {\bibfnamefont {J.~D.}\ \bibnamefont
  {Romano}},\ }\href {https://arxiv.org/abs/2505.24695} {\bibinfo {title}
  {{Likelihoods for Stochastic Gravitational Wave Background Data Analysis}}},\
  \Eprint {https://arxiv.org/abs/2505.24695} {arXiv:2505.24695 [gr-qc]}
  \BibitemShut {NoStop}%
\bibitem [{\citenamefont {Muratore}\ \emph {et~al.}(2022)\citenamefont
  {Muratore}, \citenamefont {Vetrugno}, \citenamefont {Vitale},\ and\
  \citenamefont {Hartwig}}]{Muratore:2021uqj}%
  \BibitemOpen
  \bibfield  {author} {\bibinfo {author} {\bibfnamefont {M.}~\bibnamefont
  {Muratore}}, \bibinfo {author} {\bibfnamefont {D.}~\bibnamefont {Vetrugno}},
  \bibinfo {author} {\bibfnamefont {S.}~\bibnamefont {Vitale}},\ and\ \bibinfo
  {author} {\bibfnamefont {O.}~\bibnamefont {Hartwig}},\ }\href
  {https://doi.org/10.1103/PhysRevD.105.023009} {\bibfield  {journal} {\bibinfo
   {journal} {Phys. Rev. D}\ }\textbf {\bibinfo {volume} {105}},\ \bibinfo
  {pages} {023009} (\bibinfo {year} {2022})},\ \Eprint
  {https://arxiv.org/abs/2108.02738} {arXiv:2108.02738 [gr-qc]} \BibitemShut
  {NoStop}%
\bibitem [{\citenamefont {Kass}\ and\ \citenamefont
  {Raftery}(1995)}]{doi:10.1080/01621459.1995.10476572}%
  \BibitemOpen
  \bibfield  {author} {\bibinfo {author} {\bibfnamefont {R.~E.}\ \bibnamefont
  {Kass}}\ and\ \bibinfo {author} {\bibfnamefont {A.~E.}\ \bibnamefont
  {Raftery}},\ }\href {https://doi.org/10.1080/01621459.1995.10476572}
  {\bibfield  {journal} {\bibinfo  {journal} {Journal of the American
  Statistical Association}\ }\textbf {\bibinfo {volume} {90}},\ \bibinfo
  {pages} {773} (\bibinfo {year} {1995})}\BibitemShut {NoStop}%
\bibitem [{\citenamefont {Pozzoli}\ \emph {et~al.}(2024)\citenamefont
  {Pozzoli}, \citenamefont {Gair}, \citenamefont {Buscicchio},\ and\
  \citenamefont {Speri}}]{Pozzoli:2024hkt}%
  \BibitemOpen
  \bibfield  {author} {\bibinfo {author} {\bibfnamefont {F.}~\bibnamefont
  {Pozzoli}}, \bibinfo {author} {\bibfnamefont {J.}~\bibnamefont {Gair}},
  \bibinfo {author} {\bibfnamefont {R.}~\bibnamefont {Buscicchio}},\ and\
  \bibinfo {author} {\bibfnamefont {L.}~\bibnamefont {Speri}},\ }\href@noop {}
  {\bibfield  {journal} {\bibinfo  {journal} {arXiv e-prints}\ } (\bibinfo
  {year} {2024})},\ \Eprint {https://arxiv.org/abs/2412.10468}
  {arXiv:2412.10468 [astro-ph.IM]} \BibitemShut {NoStop}%
\bibitem [{\citenamefont {Spiegelhalter}\ \emph {et~al.}(2002)\citenamefont
  {Spiegelhalter}, \citenamefont {Best}, \citenamefont {Carlin},\ and\
  \citenamefont {Van Der~Linde}}]{https://doi.org/10.1111/1467-9868.00353}%
  \BibitemOpen
  \bibfield  {author} {\bibinfo {author} {\bibfnamefont {D.~J.}\ \bibnamefont
  {Spiegelhalter}}, \bibinfo {author} {\bibfnamefont {N.~G.}\ \bibnamefont
  {Best}}, \bibinfo {author} {\bibfnamefont {B.~P.}\ \bibnamefont {Carlin}},\
  and\ \bibinfo {author} {\bibfnamefont {A.}~\bibnamefont {Van Der~Linde}},\
  }\href {https://doi.org/https://doi.org/10.1111/1467-9868.00353} {\bibfield
  {journal} {\bibinfo  {journal} {Journal of the Royal Statistical Society:
  Series B (Statistical Methodology)}\ }\textbf {\bibinfo {volume} {64}},\
  \bibinfo {pages} {583} (\bibinfo {year} {2002})},\ \Eprint
  {https://arxiv.org/abs/https://rss.onlinelibrary.wiley.com/doi/pdf/10.1111/1467-9868.00353}
  {https://rss.onlinelibrary.wiley.com/doi/pdf/10.1111/1467-9868.00353}
  \BibitemShut {NoStop}%
\bibitem [{\citenamefont {Gelman}\ \emph {et~al.}(2004)\citenamefont {Gelman},
  \citenamefont {Carlin}, \citenamefont {Stern},\ and\ \citenamefont
  {Rubin}}]{MR2027492}%
  \BibitemOpen
  \bibfield  {author} {\bibinfo {author} {\bibfnamefont {A.}~\bibnamefont
  {Gelman}}, \bibinfo {author} {\bibfnamefont {J.~B.}\ \bibnamefont {Carlin}},
  \bibinfo {author} {\bibfnamefont {H.~S.}\ \bibnamefont {Stern}},\ and\
  \bibinfo {author} {\bibfnamefont {D.~B.}\ \bibnamefont {Rubin}},\ }\href@noop
  {} {\emph {\bibinfo {title} {Bayesian data analysis}}},\ \bibinfo {edition}
  {2nd}\ ed.,\ Texts in Statistical Science Series\ (\bibinfo  {publisher}
  {Chapman \& Hall/CRC, Boca Raton, FL},\ \bibinfo {year} {2004})\ pp.\
  \bibinfo {pages} {xxvi+668}\BibitemShut {NoStop}%
\bibitem [{\citenamefont {MacKay}(2002)}]{10.5555/971143}%
  \BibitemOpen
  \bibfield  {author} {\bibinfo {author} {\bibfnamefont {D.~J.~C.}\
  \bibnamefont {MacKay}},\ }\href@noop {} {\emph {\bibinfo {title} {Information
  Theory, Inference \& Learning Algorithms}}}\ (\bibinfo  {publisher}
  {Cambridge University Press},\ \bibinfo {address} {USA},\ \bibinfo {year}
  {2002})\BibitemShut {NoStop}%
\bibitem [{\citenamefont {Santini}(2025)}]{lisaps}%
  \BibitemOpen
  \bibfield  {author} {\bibinfo {author} {\bibfnamefont {A.}~\bibnamefont
  {Santini}},\ }\href {https://doi.org/10.5281/zenodo.16323271} {\bibinfo
  {title} {asantini29/lisa-ps: First release}} (\bibinfo {year}
  {2025})\BibitemShut {NoStop}%
\bibitem [{\citenamefont {Harris}\ \emph {et~al.}(2020)\citenamefont {Harris}
  \emph {et~al.}}]{harris2020array}%
  \BibitemOpen
  \bibfield  {author} {\bibinfo {author} {\bibfnamefont {C.~R.}\ \bibnamefont
  {Harris}} \emph {et~al.},\ }\href {https://doi.org/10.1038/s41586-020-2649-2}
  {\bibfield  {journal} {\bibinfo  {journal} {Nature}\ }\textbf {\bibinfo
  {volume} {585}},\ \bibinfo {pages} {357} (\bibinfo {year}
  {2020})}\BibitemShut {NoStop}%
\bibitem [{\citenamefont {Hunter}(2007)}]{Hunter:2007}%
  \BibitemOpen
  \bibfield  {author} {\bibinfo {author} {\bibfnamefont {J.~D.}\ \bibnamefont
  {Hunter}},\ }\href {https://doi.org/10.1109/MCSE.2007.55} {\bibfield
  {journal} {\bibinfo  {journal} {Computing in Science \& Engineering}\
  }\textbf {\bibinfo {volume} {9}},\ \bibinfo {pages} {90} (\bibinfo {year}
  {2007})}\BibitemShut {NoStop}%
\bibitem [{\citenamefont {Lam}\ \emph {et~al.}(2015)\citenamefont {Lam},
  \citenamefont {Pitrou},\ and\ \citenamefont {Seibert}}]{lam2015numba}%
  \BibitemOpen
  \bibfield  {author} {\bibinfo {author} {\bibfnamefont {S.~K.}\ \bibnamefont
  {Lam}}, \bibinfo {author} {\bibfnamefont {A.}~\bibnamefont {Pitrou}},\ and\
  \bibinfo {author} {\bibfnamefont {S.}~\bibnamefont {Seibert}},\ }in\
  \href@noop {} {\emph {\bibinfo {booktitle} {Proceedings of the Second
  Workshop on the LLVM Compiler Infrastructure in HPC}}}\ (\bibinfo {year}
  {2015})\ pp.\ \bibinfo {pages} {1--6}\BibitemShut {NoStop}%
\bibitem [{\citenamefont {Santini}(2024{\natexlab{b}})}]{pysco_2024_13930440}%
  \BibitemOpen
  \bibfield  {author} {\bibinfo {author} {\bibfnamefont {A.}~\bibnamefont
  {Santini}},\ }\href {https://doi.org/10.5281/zenodo.13930440} {\bibinfo
  {title} {asantini29/pysco: First release}} (\bibinfo {year}
  {2024}{\natexlab{b}})\BibitemShut {NoStop}%
\end{thebibliography}%

\end{document}